\documentclass[aps,prc,twocolumn,superscriptaddress,showpacs,nofootinbib]{revtex4-1} 

\usepackage{graphicx,amsmath,amssymb,bm,color}

\DeclareGraphicsExtensions{ps,eps,ps.gz,frh.eps,ml.eps}

\usepackage{fmtcount} 
\usepackage{graphicx}
\usepackage{amsmath, amsfonts, amssymb, amsthm, amscd, amsbsy, amstext} 
\usepackage{marvosym} 
\usepackage{yfonts}  
\usepackage{braket}
\usepackage{mathrsfs} 
\usepackage{mathcomp} 
\usepackage{upgreek}
\usepackage{array}
\usepackage{slashed}
\usepackage{enumitem}

\usepackage{txfonts}

\usepackage{enumitem}

\usepackage{cancel}

\usepackage{url}
\usepackage{hyperref}%

\DeclareMathOperator{\e}{\operatorname{e}}

\DeclareMathOperator{\Tr}{\operatorname{Tr}}

\newcommand{\en}{\varepsilon}  
\newcommand{\En}{D}
\newcommand{\Ens}{D}
\newcommand{\im}{\text{i}}    

\newcommand{\breq}{\nonumber \\}  
\newcommand{\bs}{\textbf}  
\newcommand{\F}{\text{F}} 
 
\newcommand{\eps}{\varepsilon}

\def\XXint#1#2#3{{\setbox0=\hbox{$#1{#2#3}{\int}$ }
\vcenter{\hbox{$#2#3$ }}\kern-.6\wd0}}

\makeatletter

\newcommand{\Rmnum}[1]{\expandafter\@slowromancap\romannumeral #1@}
\makeatother

\makeatletter

\def\env@blscases{
  \let\@ifnextchar\new@ifnextchar
  \left.
  \def\arraystretch{1.2}
  \array{@{}l@{\quad}l@{}}
}
\makeatother

\makeatletter

\def\env@rcases{
  \let\@ifnextchar\new@ifnextchar
  \left.
  \def\arraystretch{1.2}
  \array{@{}l@{\quad}l@{}}
}
\makeatother

\begin {document}
\title{Zero-Temperature Limit and Statistical Quasiparticles in Many-Body Perturbation Theory}
\author{Corbinian Wellenhofer}
\email[E-mail:~]{wellenhofer@theorie.ikp.physik.tu-darmstadt.de}
\affiliation{Institut f\"{u}r Kernphysik, Technische Universit\"{a}t Darmstadt, D-64289 Darmstadt, Germany}  
\affiliation{ExtreMe Matter Institute EMMI, GSI Helmholtzzentrum f\"{u}r Schwerionenforschung GmbH, 64291 Darmstadt, Germany}

\begin{abstract}
The order-by-order renormalization of the self-consistent mean-field potential 
in many-body perturbation theory for normal Fermi systems is investigated in detail. 
Building on previous work mainly by 
Balian and de Dominicis, as a key result we
derive a thermodynamic 
perturbation series that manifests the consistency of the adiabatic zero-temperature formalism with
perturbative statistical mechanics---for both isotropic and anisotropic systems---and
satisfies 
at each order and for all temperatures the thermodynamic relations associated with Fermi-liquid theory. 
These properties are proved to all orders.
\end{abstract}

\maketitle
\section{Introduction}
Many-body perturbation theory (MBPT) represents the elementary framework for calculations aimed at the properties of nonrelativistic many-fermion systems at zero and finite temperature.
In general, for Fermi systems the correct ground-state is not a normal state but involves Cooper 
pairs~\cite{PhysRev.150.202,PhysRevLett.15.524,doi:10.1142/S0217979292001249,RevModPhys.66.129,Salmhofer:1999uq}.
However, pairing effects can often be neglected for approximative calculations of thermodynamic properties close to zero temperature.
For such calculations there are 
two formalisms: first, there is grand-canonical perturbation theory, and second, 
the zero-temperature formalism based on the adiabatic continuation of the ground state \cite{PhysRev.84.350,Goldstone:1957zz,Nozbook,Runge,negele,Fetter,Abrikosov}. 
In their time-dependent (i.e., in frequency space) formulations, these two formalisms give matching results
if all quantities are derived from
the exact Green's functions, i.e., from the
self-consistently renormalized propagators~\cite{Luttinger:1960ua,Fetter,Abrikosov,Feldman1999}. 
The renormalization of MBPT in frequency space can be generalized to vertex functions~\cite{dommar2,dommar1,Hausmann,Rossi:2015lda,PhysRevB.93.161102,DICKHOFF2004377,VanHoucke:2011ux},
and is essential to obtain a fully consistent framework for calculating transport properties \cite{Baym:1961zz,Baym:1962sx,stefanuc}.

Nevertheless, the use of bare propagators has the benefit that in that case the time integrals can be performed analytically.
With bare propagators, MBPT in its most basic form corresponds to a perturbative expansion
in terms of the interaction Hamiltonian $V$ 
about the noninteracting system with Hamiltonian $H_0$, where $H=H_0+V$ is the full Hamiltonian.
First-order self-energy effects can be included to all orders in bare MBPT
by expanding instead about a reference Hamiltonian $H_\text{ref}=H_0+U_1$, 
where $U_1$ includes the first-order contribution to the (frequency-space) self-energy $\Sigma_{1,{\bs{k}}}$ as a 
self-consistent single-particle potential (mean field).
The renormalization of $H_\text{ref}$ in terms of $U_1$ has the effect that all
two-particle reducible diagrams with first-order pieces (single-vertex loops) are canceled.
At second order the self-energy becomes frequency dependent and complex, 
so the equivalence between 
the propagator renormalization in frequency space and the renormalization of the mean-field part of $H_\text{ref}$ in bare MBPT is restricted to the Hartree-Fock level.

Zero-temperature MBPT calculations with bare propagators and a Hartree-Fock reference Hamiltonian ${H_\text{ref}=H_0+U_1}$ are common in quantum chemistry
and nuclear physics.
With a Hartree-Fock reference Hamiltonian (or, with ${H_\text{ref}=H_0}$), however, the adiabatic zero-temperature formalism 
is inconsistent with
the zero-temperature limit (${T\rightarrow 0}$) of grand-canonical MBPT.
The (main) fault however lies not with zero-temperature MBPT, but with the grand-canonical perturbation series: 
in the bare grand-canonical formalism (with $H_\text{ref}\in\{H_0,H_0+U_1\}$) there is a mismatch in the Fermi-Dirac distribution functions caused by using
the \textit{reference} spectrum $\varepsilon_{\bs{k}}$ together with the \textit{true} chemical potential $\mu$, 
and in general this leads to deficient results \cite{Fritsch:2002hp,PhysRevC.89.064009,Wellenhofer:2017qla}.
The adiabatic formalism on the other hand uses the reference chemical potential, i.e., the reference Fermi energy $\eps_\F$.
Related to this is the presence of additional contributions from two-particle reducible diagrams, the so-called anomalous contributions, in the grand-canonical formalism.

This issue is usually dealt with by modifying 
the grand-canonical perturbation series for the free energy
in terms of an expansion about the 
chemical potential $\mu_\text{ref}\xrightarrow{T\rightarrow 0}\eps_\F$ of the reference system~\cite{Kohn:1960zz,brout2} (see also Sec.~\ref{sec42}).
This expansion introduces additional anomalous contributions, and for isotropic systems these can be seen to cancel the old ones for ${T\rightarrow 0}$~\cite{Luttinger:1960ua}. 
Thus, the modified perturbation series for the free energy $F(T,\mu_\text{ref})$ reproduces the adiabatic series in the isotropic case. 
For anisotropic systems, however, the anomalous contributions persist at ${T=0}$ (for ${H_\text{ref}=H_0+U_1}$, at fourth order and beyond). 
Negele and Orland~\cite{negele} interpret this feature as follows: 
there is nothing fundamentally wrong with the bare zero-temperature formalism, but for anisotropic systems the adiabatic continuation must be based on a better reference Hamiltonian 
$H_\text{ref}$.
Since the convergence rate\footnote{In general, MBPT corresponds to 
divergent asymptotic series~\cite{negele,PhysRevLett.121.130405,rossi,refId0,Marino:2019wra}, so convergence rate should be understood in terms of the result at 
optimal truncation.} of MBPT depends on the choice of $H_\text{ref}$, this issue is relevant also for finite-temperature calculations, 
and for isotropic systems.

Recently, 
Holt and Kaiser~\cite{PhysRevC.95.034326} have shown that including
the real part of the bare second-order contribution to the (on-shell) self-energy, $\text{Re}\,[\Sigma_{2,{\bs{k}}}(\varepsilon_{\bs{k}})]$,
as the second-order contribution to the self-consistent mean field 
has a significant effect in perturbative
nuclear matter calculations with modern two- and three-nucleon potentials (see, e.g., Refs.~\cite{RevModPhys.81.1773,MACHLEIDT20111,Bogner:2009bt}).
However, 
a formal 
clarification for the renormalization of $H_\text{ref}$ in terms of  $\text{Re}\,[\Sigma_{2,{\bs{k}}}(\varepsilon_{\bs{k}})]$ was not 
included in Ref.~\cite{PhysRevC.95.034326}. In particular, from the discussion of Ref.~\cite{PhysRevC.95.034326}
it is not clear whether the use of this second-order mean field should be considered an improvement or not, compared to calculations with a Hartree-Fock mean field.\footnote{To 
be precise,
since Ref.~\cite{PhysRevC.95.034326} uses the adiabatic formalism, the considered self-energy is not 
the frequency-space self-energy 
of the imaginary-time formalism,
$\Sigma_{{\bs{k}}}(z)$,
but the collisional one $\Sigma^\text{coll}_{\bs{k}}(\omega)$, 
which however satisfies $\text{Re}\,[\Sigma^\text{coll}_{{\bs{k}}}(\omega)]=\text{Re}\,[\Sigma_{{\bs{k}}}(\omega\pm \im\eta)]$.
We use the notion frequency 
somewhat generalized, i.e., by frequency we refer mostly
to the argument $z\in \mathbb{C}$ of the (usual) 
analytic continuation $\Sigma_{{\bs{k}}}(z)$ of the Matsubara self-energy $\Xi_{{\bs{k}}}(z_l)$. See Appendix \ref{app21} for details.}

A general scheme where the reference Hamiltonian is renormalized at each order in grand-canonical MBPT was introduced by
Balian, Bloch, and de Dominicis \cite{Balian1961529} (see also Refs.~\cite{Balian1961529b,Balianph,BlochImperFerm,boer,1964mbpdedom}). 
This scheme however leads to a mean field 
whose functional form is given by $U[n_\bs{k},T]$, where $n_\bs{k}(T,\mu)$ is the Fermi-Dirac distribution and  
the explicit temperature dependence
involves factors $\e^{\pm(\en_\bs{k}-\mu)/T}$.
Because of the $\e^{\pm(\en_\bs{k}-\mu)/T}$ factors, the resulting perturbation series 
is well-behaved only at sufficiently large temperatures, 
and its ${T\rightarrow 0}$ limit does not exist.\footnote{Note that in Luttinger's analysis~\cite{PhysRev.174.263} 
of the scheme by Balian, Bloch, and de Dominicis
it is incorrectly assumed that the mean field has the form $U[n_\bs{k}]$. 
This is why in Ref.~\cite{Fermiliq}
Luttinger's paper has been (incorrectly) associated with statistical quasiparticles.}

A different renormalization scheme
was outlined by Balian and de Dominicis (BdD) in Refs.~\cite{statquasi3,statquasi1} (see also Refs.~\cite{1964mbpdedom,boer}).
At second order, this scheme leads to the mean field employed by Holt and Kaiser~\cite{PhysRevC.95.034326}.
The outline given in Refs.~\cite{statquasi3,statquasi1} indicates the following results:
\begin{enumerate}[start=0,label={(\arabic*)}]
\item The functional form of the mean field is to all orders given by $U[n_\bs{k}]$, i.e., there 
is no explicit temperature dependence (apart from the one given by the Fermi-Dirac distributions), so the ${T\rightarrow 0}$ limit exists.
\item The zero-temperature limit of 
the renormalized grand-canonical perturbation series for the free energy $F(T,\mu)$ reproduces the (correspondingly renormalized) adiabatic series for the 
ground-state energy $E^{(0)}(\eps_\F)$ to all orders; i.e., 
the reference spectrum $\eps_\bs{k}$ 
has been adjusted to the true chemical potential $\mu$, with $\eps_\F=\mu$ at ${T=0}$.
\item One obtains
at each perturbative order and for all temperatures the thermodynamic relations associated with Fermi-liquid theory~\cite{Landau}. 
This result corresponds to the notion statistical quasiparticles \cite{PhysRevA.7.304,pethick2,Fermiliq}.
\end{enumerate}
The most intricate part in establishing these results is as follows. 
For ${T\neq 0}$, there are no energy denominator poles in the 
(proper)
expressions for the perturbative contributions to the grand-canonical potential.
The BdD renormalization scheme however introduces such
poles, and therefore a regularization procedure is required to apply the scheme.
So far, this issue has been studied in more detail only for the case of impurity systems~\cite{BALIAN1971229,LUTTINGER19731,keitermorandi}.

Motivated by this situation, in the present paper
we revisit the order-by-order renormalization of the reference Hamiltonian in bare MBPT.\footnote{Several of 
the results presented here are also discussed in the authors dissertation~\cite{Wellenhofer:2017qla}, but note that some 
technical details have been missed and several typos appear there.}
First, in Sec.~\ref{sec2} we give a short review of grand-canonical perturbation theory with bare propagators 
and introduce the various order-by-order renormalizations of the reference Hamiltonian.
We also discuss how dynamical quasiparticles arise in (frequency-space) MBPT, and show that their energies are distinguished from the ones of 
the statistical quasiparticles associated with result (2).
In Sec.~\ref{sec3} we discuss the 
regularization procedure for the BdD renormalization scheme, and analyze the resulting 
expressions for the second- and third-order contributions to the grand-canonical potential and the BdD mean field. 
In Sec.~\ref{sec4} we prove to all orders that the BdD
renormalized perturbation series 
satisfies the Fermi-liquid relations (2) and, as a consequence,
manifests the consistency of the adiabatic zero-temperature formalism (1).
The paper is concluded in Sec.~\ref{summary}. 
In Appendix~\ref{app1},
we derive explicitly the renormalized contribution from two-particle reducible diagrams at fourth order. In Appendix~\ref{app2},
we discuss in more detail the various forms of the self-energy, derive various expressions for the mean occupation numbers, 
and examine the functional relations between the grand-canonical potential and the (various forms of the) self-energy in bare MBPT.

\section{Grand-Canonical Perturbation Theory} \label{sec2}

\subsection{Setup}\label{sec21}

We consider a homogeneous but not necessarily isotropic system of nonrelativistic fermions in thermodynamic equilibrium.
The Hamiltonian is given by ${H= H_0+V}$, where $V$ is a two-body operator representing pair interactions. 
Multi-fermion interactions do not raise any new formal or conceptual issues, and are therefore neglected.
For notational simplicity and without loss of generality we assume a single species of spinless fermions. 
If there is no external potential, then $H_0$ is the kinetic energy operator.
We now introduce an additional one-body operator $U$, and write 
\begin{align}\label{Eq0}
H= \underbrace{(H_0+U)}_{H_\text{ref}}+(V- U).
\end{align}
The operator $U$ represents a mean field, i.e., an effective one-body potential 
which allows to define a solvable reference system 
that 
includes the effects of pair interactions in the system to a certain degree.
For a homogeneous system the mean field should preserve translational invariance, so the eigenstates $\ket{\psi_{\bs{k}}}$
of the momentum operator are eigenstates of $H_\text{ref}=H_0+U$, i.e., 
\begin{align}
H_{\text{ref}} \ket{\psi_{\bs{k}}} = \eps_{\bs{k}}\ket{\psi_{\bs{k}}}.
\end{align}
Because the mean field is supposed to include interaction effects self-consistently, the single-particle energies $\eps_{\bs{k}}$ are determined by the self-consistent equation
\begin{align} 
\eps_{\bs{k}} = \eps_{0,\bs{k}} + U_{\bs{k}}[\eps_{\bs{k}}],
\end{align}
where $\eps_{0,\bs{k}}=\braket{\psi_{\bs{k}}|H_0|\psi_{\bs{k}}}$ and $U_{\bs{k}}[\eps_{\bs{k}}]=\braket{\psi_{\bs{k}}|U|\psi_{\bs{k}}}$.\footnote{In the 
Hartree-Fock case the self-consistency requirement 
can be evaded for isotropic systems at ${T=0}$ by replacing in the expression for $U_{\bs{k}}=U_{1,k}$
the distribution functions $n_k=\theta(\mu-\en_k)$
by $\theta(k_{\F,\text{ref}}-k)$, where the unperturbed Fermi momentum $k_{\F,\text{ref}}$ 
is defined via ${\en_{k_{\F,\text{ref}}}=\mu}$. In that case, first-order MBPT is identical for ${U=0}$ and ${U=U_1}$ (more generally, ${U\propto U_1}$).}
The occupation number representation of the
reference Hamiltonian $H_{\text{ref}}=H_0+ U$ is then given by
\begin{align}
\mathcal{H}_{\text{ref}} &= \sum_{\bs{k}} \braket{\psi_{\bs{k}}|H_{\text{ref}}|\psi_{\bs{k}} }  a^\dagger_{\bs{k}} a_{\bs{k}},
\end{align}
where $a^\dagger_\bs{k}$ and $a_\bs{k}$ 
are creation and annihiliation operators with respect to momentum eigenstates.
If not indicated explicitly otherwise, we assume the thermodynamic limit where $\sum_\bs{k}\rightarrow\int \! d^3 k/(2\pi)^3$.\footnote{In the thermodynamic limit the expressions for all size
extensive quantities scale linearly with the confining volume. 
For notational simplicity, we neglect the scale factors. For discussions regarding our choice of basis states, see Refs.~\cite{2016arXiv160900014C,PhysRevA.95.062124}.}
The occupation number representation of the perturbation Hamiltonian ${H-H_\text {ref}=V-U}$ is given by
\begin{align}\label{VHamilton}
\mathcal{V}&= \frac{1}{2!}\sum_{\bs{k}_1,\bs{k}_2,\bs{k}_3,\bs{k}_4} 
\braket{\psi_{\bs{k}_1}\psi_{\bs{k}_2}| V|\psi_{\bs{k}_3}\psi_{\bs{k}_4} } a^\dagger_{\bs{k}_1} a^\dagger_{\bs{k}_2} a_{\bs{k}_4} a_{\bs{k}_3}
\breq & \quad
-\sum_{\bs{k}}
\braket{\psi_{\bs{k}}|  U|\psi_{\bs{k}} } a^\dagger_{\bs{k}}  a_{\bs{k}},
\end{align}
where momentum conservation is implied, i.e., $\bs{k}_1+\bs{k}_2$\break$=\bs{k}_3+\bs{k}_4$.
We assume that the potential $V$ is sufficiently regular(ized) such that no ultraviolet~\cite{Hammer:2000xg,Wellenhofer:2018dwh} or infrared~\cite{PhysRev.106.364} divergences appear in 
perturbation theory.\footnote{At ${T=0}$, in MBPT there 
are still divergences due to vanishing energy denominators, but these cancel each other at each order~\cite{Wellenhofer:2018dwh} (see also Sec.~\ref{sec44}).}
Further, we require that $V$ has a form (e.g., finite-ranged\break interactions) for which the thermodynamic limit exists; see, e.g., Refs.~\cite{haagbook,ruelle,lieb}.

\subsection{Perturbation series and diagrammatic analysis}\label{sec22}

\subsubsection{Grand-canonical perturbation series}

For truncation order $N$, the perturbation series for the grand-canonical potential $\Omega(T,\mu)$ is 
given by
\begin{align} \label{MBPTgc0}
\Omega(T,\mu) = \Omega_\text{ref}(T,\mu) 
+\Omega_{U}(T,\mu) 
+  \sum_{n=1}^{N} \Omega_n(T,\mu),
\end{align}
where
\begin{align} 
\Omega_\text{ref}(T,\mu) &= T\sum_{\bs{k}}\ln(\bar n_\bs{k}),
\breq
\Omega_{U}(T,\mu)&= -
\sum_{\bs{k}} U_{\bs{k}} n_\bs{k}.
\end{align}
Here, $\bar n_\bs{k}=1-n_\bs{k}$, with $n_\bs{k}=[1+\e^{\beta(\en_\bs{k}-\mu}]^{-1}$ the Fermi-Dirac distribution function, and $\beta=1/T$.
From the 
grand-canonical version of Wick's theorem one obtains the following
formula~\cite{Fetter,BDDnuclphys7} for 
$\Omega_{n}(T,\mu)$:
\begin{align} \label{OmegaT}  
\Omega_n^{\text{direct}[P]}=
-\frac{1}{\beta}\frac{(-1)^{n}}{n!} 
\int\limits_{0}^\beta \!
d \tau_n \cdots d \tau_1 \;
\Braket{  \mathcal{T}\big[
\mathcal{V}(\tau_n) \cdots
\mathcal{V}(\tau_1) \big]
}_{L},
\end{align}
where $\mathcal{T}$ is the time-ordering operator and
$\mathcal{V}(\tau)=\e^{\mathcal{H}_\text{ref} \tau}\mathcal{V}\e^{-\mathcal{H}_\text{ref} \tau}$ 
is the interaction picture representation (in imaginary time) of the 
perturbation operator $\mathcal{V}$ given by Eq.~\eqref{VHamilton}.

\subsubsection{Classification of diagrams}

The various ways the Wick contractions in the unperturbed ensemble average $\braket{\ldots}$ can be performed can be represented by Hugenholtz diagrams, i.e., diagrams composed of $V$ and $-U$ vertices,\footnote{The diagram composed of a single $-U$ vertex corresponds to $\Omega_{U}(T,\mu)$ and is excluded here. For the diagrammatic rules, see, e.g., Refs.~\cite{szabo,Runge}.}
and directed lines attached to vertices at both ends.
Left-pointing lines are called holes and correspond to factors $n_\bs{k}$, right-pointing lines are called particles and have factors $\bar n_\bs{k}$.
In the case of two-particle reducible diagrams, momentum conservation implies that
there are two or more lines with identical three-momenta. We refer to these lines as articulation lines.
The diagrammatic parts connected via articulation lines are referred to as pieces.
Two-particle irreducible diagrams have only $V$ vertices.
Two-particle reducible diagrams
where at least one set of lines with identical three-momenta includes both holes and particles are called anomalous, with the
indicative lines referred to as anomalous articulation lines.
All other (two-particle reducible or irreducible) diagrams are called normal. 
The parts of anomalous diagrams connected via anomalous articulation lines are called normal pieces.\footnote{That is, normal pieces correspond to the linked normal subdiagrams of the normal unlinked diagram generated by 
cutting all anomalous articulation lines and closing them in each separated part.}
In general, normal two-particle reducible diagrams transform into anomalous diagrams under vertex permutations, see 
Figs.~\ref{fig3red}, \ref{figx} and \ref{fig4}.

In Eq.~\eqref{OmegaT}, the subscript $L$ means that only linked diagrams are taken into account.
By virtue of the time integration and the time-ordering operator, 
in Eq.~\eqref{OmegaT} there is no distinction between diagrams connected via vertex permutations; 
in particular, there is no distinction between normal and anomalous two-particle reducible diagrams.
The distinction between the different diagrams in the permutation invariant sets of diagrams
is however
relevant for the time-independent formulas discussed below.

\subsubsection{Time-independent formulas}

From Eq.~\eqref{OmegaT}, Bloch and de Dominicis \cite{BDDnuclphys7} (see also 
Refs.~\cite{1964mbpdedom0,boer,keitermorandi}) have derived several time-independent formulas for $\Omega_{n}(T,\mu)$.
One of them, here referred to as the \text{direct formula}, is given by
\begin{align} \label{direct}
\Omega_{n}^\text{direct}&=\frac{1}{\beta} \frac{(-1)^{n}}{2\pi \im} 
\oint_{C}  dz \frac{\e^{-\beta z}}{z^2}
\Braket{ 
\mathcal{V} \frac{1}{\En_n-z} \cdots
\mathcal{V} \frac{1}{\En_1-z}
\mathcal{V}
}_{\!L},
\end{align}
where the contour $C$ encloses all the poles ${z=0,\En_1,\ldots,\En_n}$, with $\En_{\nu\in\{1,\ldots,n\}}$ the energy denominators for the respective diagrams.
Furthermore, in Eq.~\eqref{direct}, it is implied that
the contributions from all poles are summed before the momentum integration, i.e., the $z$ integral is performed 
inside the momentum integrals.
This has the consequence that the integrands of the momentum integrals
have no poles (for ${T\neq 0}$, see below) from vanishing energy denominators.
The expressions obtained from the direct formula
deviate from the ones obtained from the time-dependent formula Eq.~\eqref{OmegaT}, but---as evident from the derivation of direct formula~\cite{BDDnuclphys7}---the sum of the direct expressions obtained for a set of diagrams that is closed under vertex permutations
is equivalent (but not identical) to the expression obtained from Eq.~\eqref{OmegaT}.

From the cyclic property of the trace, another time-independent formula can be derived~\cite{BDDnuclphys7}, here referred to as the cyclic formula, i.e.,
\begin{align} \label{cyclic}
\Omega_{n}^\text{cyclic}&=\frac{1}{n} \frac{(-1)^{n+1}}{2\pi \im} 
\oint_{C}  dz \frac{\e^{-\beta z}}{z}
\Braket{ 
\mathcal{V} \frac{1}{\En_n-z} \cdots
\mathcal{V} \frac{1}{\En_1-z}
\mathcal{V}
}_{\!L},
\end{align}
where again it is implied that the $z$ integral is performed 
inside the momentum integrals; again, 
this has the consequence that the integrands have no poles (for ${T\neq 0}$).
The direct and the cyclic formula give equivalent (but not identical) expressions only for the sums of diagrams connected via cyclic vertex permutations, and the cyclic expressions for the individual diagrams in these cyclic groups are equivalent.

Finally, from the analysis of the contributions from the different poles in Eq.~\eqref{cyclic}
one can \textit{formally} write down a reduced form of the cyclic formula~\cite{BDDnuclphys7},
here referred to as
the reduced formula, i.e.,
\begin{align} \label{reduced}
\Omega_{n}^\text{reduced}&=  \frac{(-1)^{n+1}}{\mathcal{O}}
\underset{z=0}{\text{Res}}  \frac{\e^{-\beta z}}{z}
\Braket{ 
\mathcal{V} \frac{1}{\En_n-z} \cdots
\mathcal{V} \frac{1}{\En_1-z}
\mathcal{V}
}_{\!L},
\end{align}
where $\mathcal{O}$ is the order of the pole at $z=0$. 
The reduced expressions for normal diagrams 
are identical to the usual expressions of zero-temperature MBPT, except that the step functions 
are replaced by Fermi-Dirac distributions.
As a consequence, while at ${T=0}$ the energy denominator poles in these expressions are at the integration boundary, for ${T\neq 0}$ they are in the interior.
This entails
that the reduced expressions for individual diagrams are not well-defined for ${T\neq 0}$. 

Last, we note that each of the time-independent formulas 
can be applied also to unlinked diagrams (the only change being the omission of the subscript $L$); 
this will become relevant in Sec.~\ref{sec4}.

\subsubsection{Classification of perturbative contributions}

Anomalous  diagrams give no contribution in zero-temperature MBPT.
However, the contributions from anomalous diagrams in grand-canonical MBPT do \textit{not} vanish for ${T\rightarrow 0}$ (in the thermodynamic limit).
The reduced integrands (which are well-defined at ${T=0}$) for diagrams with identically vanishing energy denominators\footnote{That is, diagrams
with energy denominators involving only articulation lines with identical three-momenta. Such diagrams are 
anomalous. Note that one must distinguish between anomalous (normal) diagrams and anomalous (normal) contributions.} have terms of the form 
\begin{align} \label{anomT0lim}
\frac{\partial^\nu n_\bs{k}}{\partial \mu^\nu} \xrightarrow{T\rightarrow 0}
\delta^{(\nu)}(\mu-\en_\bs{k}),
\end{align}
e.g., $\beta  n_\bs{k} \bar n_\bs{k}=\partial n_\bs{k}/\partial \mu  \xrightarrow{T\rightarrow 0}
\delta(\mu-\en_\bs{k})$.
Contributions with such terms are called anomalous contributions.
There are also contributions that vanish for ${T\rightarrow 0}$, e.g.,
\begin{align}
n_\bs{k} \bar n_\bs{k}=T\frac{\partial n_\bs{k}}{\partial \mu} \xrightarrow{T\rightarrow 0}
0.
\end{align}
Such pseudoanomalous contributions can be
associated also with \textit{normal} two-particle reducible diagrams via the relation
\begin{align}\label{ndouble}
\bar n_\bs{k}=1-n_\bs{k},
\end{align}
i.e.,
\begin{align}\label{ndouble1}
n_\bs{k} n_\bs{k} &= n_\bs{k} -n_\bs{k} \bar n_\bs{k}, 
\\ \label{ndouble2}
\bar n_\bs{k} \bar n_\bs{k} &= \bar n_\bs{k} -n_\bs{k} \bar n_\bs{k},
\\ \label{ndouble3}
n_\bs{k} n_\bs{k} n_\bs{k} &= n_\bs{k} -2 n_\bs{k} \bar n_\bs{k}+n_\bs{k} \bar n_\bs{k} \bar n_\bs{k},
\\ \label{ndouble4}
\bar n_\bs{k} \bar n_\bs{k} \bar n_\bs{k} &= \bar n_\bs{k} -2 n_\bs{k} \bar n_\bs{k}+n_\bs{k} n_\bs{k} \bar n_\bs{k},
\end{align}
etc.\footnote{In the case of the direct formula there are also pseudoanomalous contributions (of a different kind, i.e., terms $\sim T$) from the pole at $z=0$.  
Furthermore, in both the direct and the cyclic case the expressions for diagrams with 
several identical energy denominators involve terms
$\sim T^{-\nu}$ with $\nu\geq 1$.
In Hartree-Fock MBPT such diagrams appear first at sixth order, i.e., normal two-particle reducible diagrams 
composed of three second-order pieces.
Because terms $\sim T^{-\nu}$ with $\nu\geq 1$ do not appear in the reduced formula, 
for the cyclic sums of diagrams these terms cancel each other in the ${T\rightarrow 0}$ limit.}
Contributions which are not anomalous or pseudoanomalous are referred to as normal contributions.
Following loosely Balian, Bloch, and de Dominicis~\cite{Balian1961529}, we
refer to the application of Eq.~\eqref{ndouble} 
according to Eqs.~\eqref{ndouble1}--\eqref{ndouble4}, etc. as disentanglement, denoted symbolically by $\div$.

For the ${T\rightarrow 0}$ limit, the energy denominator exponentials present in the 
direct and cyclic formula all have 
to be evaluated via
\begin{align} \label{Dexp}
\bar n_\bs{k} \e^{-\beta (\en_\bs{k}-\mu)} = n_\bs{k}.
\end{align}
The simple relations given by Eqs.~(\ref{ndouble}) and (\ref{Dexp}) play a crucial role in many of the issues and results discussed in the present paper.

\subsection{Discrete spectrum inconsistency and anomalous contributions}\label{sec23a}

Apart from being essential 
for practical many-body calculations, the thermodynamic limit is in fact essential for the thermodynamic consistency of 
the grand-canonical perturbation series at low $T$, in particular for ${T\rightarrow 0}$, in the general case (see below).
For a finite system with a discrete spectrum at ${T=0}$
one has either $\mu\in\{\en_\bs{k}\}$  or $\mu\not\in\{\en_\bs{k}\}$. Both cases are 
inconsistent.

In the first case the ${T\rightarrow 0}$ limit is singular,
because in that case
the anomalous contributions diverge. In addition, for discrete systems and
$\mu\in\{\en_\bs{k}\}$ the ${T\rightarrow 0}$ limit is singular due to 
energy denominator singularities.\footnote{These singularities are present also in the adiabatic case for $\en_\F\in\{\en_\bs{k}\}$ (i.e., for open-shell systems).}

In the second case the anomalous contributions vanish for ${T\rightarrow 0}$.
From $F(T,\mu)=\Omega(T,\mu)+\mu\varrho(T,\mu)$ and the fact that all contributions to $\varrho(T,\mu)=-\partial\Omega(T,\mu)/\partial\mu$ 
except the ones from $\Omega_\text{ref}(T,\mu)$
are anomalous, for $\mu\not\in\{\en_\bs{k}\}$ we obtain
\begin{align} \label{gcfinite}
F(T,\mu)\xrightarrow{T\rightarrow 0} E^{(0)}(\mu), 
\;\;\;\;\;\;\;\;\;\;
\varrho(T,\mu)\xrightarrow{T\rightarrow 0}\sum_{\bs{k}}\theta(\mu-\en_\bs{k}),
\end{align}
where
$E^{(0)}(\en_\F)$ corresponds to the adiabatic series. 
As noted by Kohn and Luttinger~\cite{Kohn:1960zz}, the two parts of Eq.~\eqref{gcfinite} are inconsistent
with each other. A possible definition of the 
chemical potential at ${T=0}$ in the finite case is
\begin{align} 
\mu(T=0,\varrho)=\frac{E^{(0)}(\varrho+1)+E^{(0)}(\varrho)}{2}.
\end{align}
The second part of Eq.~\eqref{gcfinite} however is equivalent to
\begin{align} 
\mu(T=0,\varrho)=\frac{E_\text{ref}^{(0)}(\varrho+1)+E_\text{ref}^{(0)}(\varrho)}{2} \equiv \mu_\text{ref}(T=0,\varrho),
\end{align}
which contradicts the previous equation. For a given particle number the true chemical potential deviates from the chemical potential of the 
reference system, 
and Eq.~\eqref{gcfinite} would imply that they are equal at $T=0$.
Thus, in the discrete case the ${T\rightarrow 0}$ limit of 
the grand-canonical perturbation series is inconsistent also for $\mu\not\in\{\en_\bs{k}\}$.

The same inconsistency can arise in the thermodynamic limit 
if the reference spectrum has a gap $\Delta$ and $\mu\not\in\{\en_\bs{k}\}$.
The ${T\rightarrow 0}$ limit is still smooth in the discrete (gapped) case, so the inconsistency is still present for nonzero $T$, although
it is washed out at sufficiently high $T$. 
Qualitatively, in the discrete case the inconsistency is relevant if the spectrum does not resolve the anomalous terms $\partial^\nu n_\bs{k}/\partial \mu^\nu$.
As discussed in Sec.~\ref{sec23}, contributions with such terms can be seen to account for the mismatch 
generated by using the reference spectrum together with the true chemical potential.
If the anomalous terms are not sufficiently resolved the information about this mismatch gets lost and one approaches the paradoxical result that $\mu(T,\varrho)= \mu_\text{ref}(T,\varrho)$.\footnote{This inconsistency has been overlooked in Ref.~\cite{SANTRA2017355}.}

There are two ways the discrete (gapped) spectrum inconsistency for $\mu\not\in\{\en_\bs{k}\}$ can be partially resolved, i.e., 
\begin{enumerate}[start=1,label={(\roman*)}]
\item by using the reference chemical potential instead of the true one,
\item by choosing a mean-field that leads to $\mu({T=0},\varrho)= \mu_\text{ref}({T=0},\varrho)$.
\end{enumerate}
Case (i) corresponds to the modified perturbation series $F(T,\mu_\text{ref})$.
The partial resolution of the discrete spectrum inconsistency in that case
is as follows:
\begin{enumerate}[start=1,label={(\roman*)}]
\item
$F(T,\mu_\text{ref})$ involves additional anomalous contributions, and the failure to resolve the old ones  
is balanced (for anisotropic systems, only partially) by not resolving the new ones.
\end{enumerate}
In the gapped case $F(T,\mu_\text{ref})$ reproduces the adiabatic series in the ${T\rightarrow 0}$ limit (if $\mu_\text{ref}\not\in\{\en_\bs{k}\}$).
In the gapless case the adiabatic series is reproduced only for isotropic systems; in that case the old and new anomalous contributions cancel for ${T\rightarrow 0}$. 
Thus, there is still a remainder of the discrete spectrum inconsistency.
The information about anisotropy encoded in the anomalous contributions 
is not resolved in the discrete case at low $T$.
In particular, for $F(T,\mu_\text{ref})$
the thermodynamic limit (and ${\Delta\rightarrow 0}$ limit, respectively) and the ${T\rightarrow 0}$ limit are 
noncommuting limits in the anisotropic case.

Regarding case (ii), there are three mean-field renormalization schemes
that lead to $\mu(T,\varrho)= \mu_\text{ref}(T,\varrho)$, and accordingly, $F(T,\mu)=F(T,\mu_\text{ref})$: 
the direct, the cyclic, and the BdD scheme; see Sec.~\ref{sec23}.
The anomalous diagrams are removed in each scheme, but in the direct and cyclic schemes there are still anomalous contributions. Hence, for 
the direct and cyclic schemes
there is no discrete spectrum inconsistency despite anomalous contributions. However, 
these schemes are well-behaved only at high $T$ where the inconsistency ceases to be relevant.
In particular, the ${T\rightarrow 0}$ limit does not exist for the direct and cyclic scheme.

The ${T\rightarrow 0}$ limit exists for the BdD scheme, 
but for ${N>2}$ this scheme exists
only in the thermodynamic limit.
The commutativity of the ${T\rightarrow 0}$  and ${\Delta \rightarrow 0}$ limits is fully restored in the BdD scheme, 
irrespective of isotropy.
The anomalous contributions can be removed and the result $\mu({T=0},\varrho)= \mu_\text{ref}({T=0},\varrho)$ can be achieved
also for finite systems, via the mean-field renormalization scheme specified by Eq.~\eqref{Uredast} below.\footnote{As discussed in Sec.~\ref{sec23}, the Eq.~\eqref{Uredast} scheme leads to 
$\mu(T,\varrho)= \mu_\text{ref}(T,\varrho)$ for nonzero $T$ if the (pseudoanomalous) contributions from energy denominator poles 
are excluded, but it is not clear whether this is justified.} 
For ${N\leq 2}$ the Eq.~\eqref{Uredast} scheme converges to the BdD scheme, but for ${N>2}$ it becomes ill-defined (singular, for ${N\geq 4}$) in the thermodynamic limit.
Altogether, we have:
\begin{enumerate}[start=2,label={(\roman*)}]
\item
The result $\mu({T=0},\varrho)= \mu_\text{ref}({T=0},\varrho)$ can be achieved for finite systems and in the thermodynamic limit, irrespective of isotropy, but for ${N>2}$
these two cases are not smoothly connected.
\end{enumerate}
The commutativity of limits can however be fully restored for $F(T,\mu_\text{ref})$:
\begin{enumerate}[start=3,label={(\roman*)}]
\item For $F(T,\mu_\text{ref})$ together with the mean-field renormalization scheme given by Eq.~\ref{Ufermi} below the limit ${T\rightarrow 0}$ commutes with both the thermodynamic limit and the 
${\Delta \rightarrow 0}$ limit, irrespective of isotropy. 
This is because for $F(T,\mu_\text{ref})$, at ${T=0}$ the Eq.~\ref{Ufermi} scheme removes the anomalous contributions.
\end{enumerate}
Case (ii) and case (iii) both lead to the adiabatic formalism, irrespective of isotropy.
There are however still anomalous contributions at finite $T$ in case (iii), and 
the reference chemical potential is identified with the true chemical potential only in case (ii).\footnote{Accordingly, for the grand-canonical series the Eq.~\ref{Ufermi} scheme 
does not remove the anomalous contributions to $\rho(T,\mu)$ at ${T=0}$ (or ${T\neq 0}$), so in that case the adiabatic series is not reproduced (in any case) and the discrete spectrum inconsistency persists.}

\subsection{Mean-field renormalization schemes}\label{sec23}

The usual choices for the mean-field potential are ${U=0}$ (free reference spectrum) or ${U=U_1}$ (Hartree-Fock spectrum). 
In general, one expects that the choice ${U=U_1}$ leads to an improved perturbation series, compared to ${U=0}$. 
For first-order MBPT this 
can be seen from the fact that ${U=U_1(T,\mu)}$ and ${U=U_1(\varepsilon_\F)}$, respectively, are stationary points of the right-hand sides of the
inequalities
\begin{align} 
\Omega(T,\mu) &\leq \Omega_\text{ref}(T,\mu) 
+\Omega_{U}(T,\mu) 
+ \Omega_1(T,\mu),
\\
E^{(0)}(\varepsilon_\text{F}) &\leq E^{(0)}_\text{ref}(\varepsilon_\text{F}) 
+E^{(0)}_{U}(\varepsilon_\text{F})
+ E^{(0)}_1(\varepsilon_\text{F}),
\end{align}
for grand-canonical and adiabatic MBPT, respectively.
For truncation orders $N> 1$, however, no similar formal argument is available for ${U=U_1}$ representing the best choice.

For both ${U=0}$ and ${U=U_1}$, 
in the thermodynamic limit
the grand-canonical perturbation series does not reproduce the adiabatic one 
for $T\rightarrow 0$. The adiabatic series is
also not reproduced in 
the discrete case, and in that case
the grand-canonical series is inconsistent, in general (in particular, for ${U\in\{\,0,U_1\}}$);
see Sec.~\ref{sec23a}).

It is now important to note that, at least for ${U=0}$, in general bare grand-canonical MBPT
leads to deficient results also in the thermodynamic limit. This 
is particularly evident for a system with a first-order phase transition: for ${U=0}$ it is impossible to obtain 
the nonconvex single-phase constrained free energy from $\Omega(T,\mu)$,  
since
$\Omega(T,\mu)$ is necessarily a single-valued function of $\mu$ for ${U=0}$; see also Refs.~\cite{Fritsch:2002hp,PhysRevC.89.064009,Wellenhofer:2017qla}.

This deficiency can be repaired by
modifying the expression for $F(T,\mu)$
in terms of a (truncated) formal expansion about the 
chemical potential $\mu_\text{ref}\xrightarrow{T\rightarrow 0}\eps_\F$ of the reference system; see Sec.~\ref{sec42} for details.
This expansion introduces additional contributions, and  
the structure of these contributions is very similar to 
anomalous diagrams. 
For isotropic systems it can be seen that the anomalous parts of these additional contributions cancel the old ones for ${T\rightarrow 0}$, 
leading to 
\begin{align} 
F(T,\mu_\text{ref})\xrightarrow{T\rightarrow 0}E^{(0)}(\varepsilon_\F)
\end{align}
in the isotropic case.\footnote{We note that 
$F(T,\mu_\text{ref})$ with $U\in\{\,0,U_1\}$ and ${N=2}$
has been employed in 
nuclear matter calculations in Refs.~\cite{Tolos:2007bh,Fritsch:2002hp,Fiorilla:2011sr,Holt:2013fwa,PhysRevC.89.064009,PhysRevC.92.015801,PhysRevC.93.055802}.
For nuclear matter calculations with self-consistent propagators, see, e.g., Refs.~\cite{PhysRevC.88.044302,PhysRevC.90.054322,PhysRevC.98.025804}.}

For $\Omega(T,\mu)$ with ${U=U_1(T,\mu)}$ one has $\varrho(T,\mu)=\sum_\bs{k} n_\bs{k}$ at truncation order ${N=1}$. 
Thus, 
for ${N=1}$ (but not for ${N>1}$) it is
\begin{align}
F(T,\mu)=F(T,\mu_\text{ref}),
\end{align}
with $\mu(T,\varrho)=\mu_\text{ref}(T,\varrho)$, 
where $F(T,\mu_\text{ref})$ now
corresponds to the modified series with ${U=U_1(T,\mu_\text{ref})}$.
For $\Omega(T,\mu)$ the change from ${U=0}$ to ${U=U_1}$ removes all anomalous (and normal) diagrams with single-vertex loops. For 
$F(T,\mu_\text{ref})$ both the reference spectrum and the reference chemical potential get renormalized, and
both the anomalous diagrams and the additional ones with single-vertex loops are removed.

Now, these features make evident that there is a deficiency in the grand-canonical series with ${U=0}$ irrespective of the 
presence of a first-order phase transition: 
there is a mismatch in the Fermi-Dirac distribution functions 
generated by using
the spectrum of $H_0$ together with the true chemical potential, leading 
to decreased perturbative convergence, compared to $F(T,\mu_\text{ref})$ with the same setup.\footnote{For additional details and numerical evidence, see Ref.~\cite{Wellenhofer:2017qla}.}
One may interpret the anomalous contributions as a symptom of this mismatch. 
In that sense, the ``expanding away'' of the mismatch, i.e., the construction of $F(T,\mu_\text{ref})$, corresponds to 
a symptomatic treatment that provides as remedy additional anomalous contributions that counteract the old ones.\footnote{At second order the two types 
of anomalous contributions have been found to give individually very large but nearly canceling
contributions in nuclear matter calculations~\cite{Wellenhofer:2017qla,Tolos:2007bh,PhysRevC.89.064009}.}
The mismatch can however be ameliorated (cured, for $N=1$, in the $U=U_1$ case) by
improving the quality of the reference Hamiltonian:
the change from ${U=0}$ to ${U=U_1}$ removes the main symptom (and the corresponding remedy, in the modified case), 
the anomalous diagrams with single-vertex loops. 

Altogether, 
this suggests that one can expect that the convergence behavior of $\Omega(T,\mu)$
is inferior to the one of $F(T,\mu_\text{ref})$ also for ${U=U_1}$.
Moreover, one can suspect that both $\Omega(T,\mu)$ and $F(T,\mu_\text{ref})$ may be further improved by using a mean field beyond Hartree-Fock.
In the best case, the additional mean-field contributions should remove all the remaining anomalous diagrams (and additional diagrams, for the modified series), i.e., the ones with higher-order pieces, and lead to 
$F(T,\mu)=F(T,\mu_\text{ref})$ for truncation orders ${N>1}$.

In the following, we 
introduce three different renormalization schemes where the mean field 
receives additional contributions for each $N$, i.e,
\begin{align} \label{mbptren1}
U^{\aleph,(\ast\ast),\div}=U_1+\sum_{n=2}^N U^{\aleph,(\ast\ast),\div}_n.
\end{align}
Here, $\aleph$ refers to one of the three time-independent formulas (direct, cyclic, or reduced), $\div$ to 
the disentanglement, and $\ast\ast$ to the regularization of energy denominators 
required to make the reduced formula well-defined.
Equation~\eqref{mbptren1} is understood to imply 
a reordering of the perturbation contributions
such that a given order $n \in\{1,\ldots, N\}$ involves
only diagrams
for which 
\begin{align}\label{Ncounting}
\mathscr{N}(V)+\mathscr{N}(U_1)+\sum_{m=2}^N m \mathscr{N}(U_m) =n \in\{1,\ldots, N\} ,
\end{align}
where $\mathscr{N}(V)$ is the number of $V$ vertices, and $\mathscr{N}(U_1)$ and $\mathscr{N}(U_n)$ the number of $-U_1$ and $-U^{\aleph,(\ast\ast),\div}_n$ vertices, respectively.
This 
can be implemented by writing Eq.~\eqref{Eq0} as
\begin{align}
H = \underbrace{(U_0+U^{\aleph,(\ast\ast),\div})}_{H_\text{ref}} + \lambda V - \lambda U_1-\sum_{n=2}^N\lambda^n  U^{\aleph,(\ast\ast),\div}_n
\end{align}
and ordering the perturbation series with respect to powers of $\lambda$ (which is at the end set to $\lambda=1$).

For each truncation order $N$, the three schemes
constitute three (different, for ${N>1}$) stationary points 
of MBPT.
Related to this, 
in each of the three schemes the (direct, cyclic and reduced, respectively) contributions from anomalous diagrams are removed, and 
in each scheme the relation between the particle number and the chemical potential matches the adiabatic relation, i.e., 
\begin{align} 
U=U^{\aleph,(\ast\ast),\div}:\;\;\;\;\; 
\varrho(T,\mu)=
-\frac{\partial \Omega(T,\mu)}{\partial\mu}=\sum_{\bs{k}} n_\bs{k},
\end{align}
so $F(T,\mu)=F(T,\mu_\text{ref})$ holds in each scheme.
The ${T\rightarrow 0}$ limit exists however only for the case where $U=U^{\text{reduced},\ast\ast,\div}$.
In that case, the grand-canonical formalism and zero-temperature MBPT are consistent with each other for both isotropic and anisotropic systems, given that 
the adiabatic continuation is based on $H_\text{ref}=H_0+U^{\text{reduced},\ast\ast,\div}$.

\subsubsection{Scheme by Balian, Bloch, and de Dominicis (direct scheme)}

In the renormalization scheme by Balian, Bloch, and de Dominicis~\cite{Balian1961529}, 
the mean-field potential is, for truncation order $N$, defined as
\begin{align} \label{Udirect}
U^{\text{direct},\div}_{\bs{k}} = 
\sum_{n=1}^{N} 
U^{\text{direct},\div}_{n,\bs{k}}
&=
\sum_{n=1}^{N} \frac{\delta 
\Omega_{n,\text{normal}}^{\text{direct},\div}}{\delta n_\bs{k}}
= 
\frac{\delta 
\mathcal{D}^{\text{direct},\div}}{\delta n_\bs{k}},
\end{align}
i.e., only the direct contributions from normal diagrams are included, 
and $\div$ (disentanglement) means that for each set of (normal) articulation lines with identical three-momenta only one (hole or particle) distribution function appears [i.e., 
only the first term of the right-hand sides of Eqs.~\eqref{ndouble1}--\eqref{ndouble4}, etc., is included].
The ${n=1}$ contribution to the mean field corresponds (as in the other schemes) to the usual Hartree-Fock single-particle potential, i.e.,
\begin{align} \label{UHF}
U_{1,\bs{k}} =
\sum_{\bs{k}'}
\braket{\psi_{\bs{k}}\psi_{\bs{k}'}| V|\psi_{\bs{k}} \psi_{\bs{k}'}}
n_{\bs{k}'},
\end{align}
where antisymmetrization is implied.
For the higher-order contributions, the functional derivative $\delta/\delta n_\bs{k}$ has to be evaluated 
\textit{without} applying Eq.~\eqref{Dexp}, i.e., the energy denominator exponentials have to be kept in the form that 
results from the contour integral. Otherwise, the functional derivative would be ill-defined (due to the emergence of poles).
For ${n=2}$, one finds 
\begin{align} \label{U2dir}
U^{\text{direct},(\div)}_{2,\bs{k}} &= 
\frac{1}{2}
\sum_{\bs{k}_2, \bs{k}_3, \bs{k}_4} \!
|\braket{\psi_{\bs{k}}\psi_{\bs{k}_2}| V|\psi_{\bs{k}_3} \psi_{\bs{k}_4}}|^2 
\breq & \quad \times
\Big[ n_{\bs{k}_2} \bar n_{\bs{k}_3} \bar n_{\bs{k}_4}\mathcal{F}^\text{direct}(D) 
- n_{\bs{k}_3} n_{\bs{k}_4} \bar n_{\bs{k}_2} \mathcal{F}^\text{direct}(-D) \Big] ,
\end{align}
where 
\begin{align}
\label{F2T0lim}
\mathcal{F}^\text{direct}(D)
&= \frac{1-\beta D-\e^{-\beta D}}{\beta D^2} \xrightarrow{D\rightarrow 0} -\frac{\beta}{2},
\end{align}
with $D=\en_{\bs{k}_3}+\en_{\bs{k}_4} - \en_{\bs{k}_2} - \en_{\bs{k}}$.
The ${T\rightarrow 0}$ limit of Eq.~\eqref{U2dir} is singular, due to the energy denominator exponential in $\mathcal{F}^\text{direct}(D)$: 
the functional derivative has removed one distribution function in the integrand, 
inhibiting the complete elimination of the energy denominator exponential via Eq.~\eqref{Dexp}.
Hence, the renormalization scheme of Balian, Bloch, and de Dominicis is of interest only for systems which are sufficiently close to the classical 
limit.\footnote{See also the next paragraph, and Sec.~\ref{summary}.}

The direct
contributions from anomalous diagrams composed of two normal pieces 
that are not (but may involve) $-U$ vertices
have
the factorized form 
\begin{align} \label{directfactorized}
\Omega_{n_1+n_2,\text{anomalous}}^{\text{direct},\div}
&= -\frac{\beta}{2} \sum_\bs{k} 
U^{\text{direct},\div}_{n_1,\bs{k}} n_\bs{k} \bar n_\bs{k} \,
U^{\text{direct},\div}_{n_2,\bs{k}}
(2-\delta_{n_1,n_2})
,
\end{align}
and similar for anomalous diagrams with several normal (non $-U$) pieces; see Sec.~\ref{sec42}. 
Given that for normal diagrams with $-U$ vertices
the functional derivative in Eq.~\eqref{Udirect} acts only on the diagrammatic lines,\footnote{That is, the functional dependence on $n_\bs{k}$ of the $-U$ vertices 
is \textit{not} taken into account in Eq.~\eqref{Udirect}.}
Eq.~\eqref{directfactorized} implies 
that the direct contributions from these diagrams are all canceled 
by the contributions from the corresponding diagrams with $-U$ pieces.
The resulting perturbation series is then given by
\begin{align} \label{MBPTrenorm1a}
\Omega(T,\mu) &= \Omega_\text{ref}(T,\mu) + 
\Omega_{U}(T,\mu) +
\mathcal{D}^{\text{direct},\div}(T,\mu).
\end{align}
Using 
$\en_{\bs{k}}=\en_{0,\bs{k}}+U^{\text{direct},\div}_{\bs{k}}$, Eq.~\eqref{MBPTrenorm1a}
can be written in the equivalent form
\begin{align} \label{MBPTrenorm1}
\Omega[n_\bs{k},T] &=T 
\sum_\bs{k} \big( n_\bs{k}\ln n_\bs{k} +  
\bar n_\bs{k}\ln \bar n_\bs{k}\big)
+\sum_\bs{k} \left(\en_{0,\bs{k}} 
- \mu \right) n_\bs{k} 
\breq &  \quad + \mathcal{D}^{\text{direct},\div}[n_\bs{k},T],
\end{align}
which, using Eqs.~\eqref{Dexp} and \eqref{Udirect}, can be seen to be stationary under variations of the distribution functions, 
$\delta \Omega[n_\bs{k},T]/\delta n_\bs{k} =0$.
From this one readily obtains the following expressions for the 
fermion number $\varrho$, the entropy $S$, and the internal energy 
$E$:
\begin{align} \label{StatQP1a}
\varrho &= 
\sum_\bs{k} n_\bs{k},
\\ \label{StatQP2a}
S &=-\sum_\bs{k} \big( n_\bs{k}\ln n_\bs{k} +  
\bar n_\bs{k}\ln \bar n_\bs{k}\big)
- \frac{\partial 
\mathcal{D}^{\text{direct},\div}[n_\bs{k},T]}{\partial T} , 
\\
E &= \sum_\bs{k} \en_{0,\bs{k}} n_\bs{k} +  
\mathcal{D}^{\text{direct},\div}[n_\bs{k},T] 
- T\frac{\partial \mathcal{D}^{\text{direct},\div}[n_\bs{k},T]}{\partial T}.
\end{align}
The variation of the internal energy 
$\delta E[n_\bs{k},T]/\delta  n_\bs{k}$ is given by
\begin{align} \label{StatQP3a}
\frac{\delta E}{\delta n_\bs{k}} &=  \en_{\bs{k}} 
-T\frac{\partial U^{\text{direct},\div}_{\bs{k}}[n_\bs{k},T]}{\partial T}.
\end{align}
The relations given by Eqs.~\eqref{StatQP1a}, \eqref{StatQP2a} and \eqref{StatQP3a} match those of Fermi-liquid theory~\cite{Landau,Landau2,Landau3}, except 
for the terms due to the explicit temperature dependence of $\mathcal{D}^{\text{direct},\div}[n_\bs{k},T]$ and $U^{\text{direct},\div}_\bs{k}[n_\bs{k},T]$.

\subsubsection{Cyclic scheme}

There is a straightforward variant of the scheme by Balian, Bloch, and de Dominicis: the cyclic scheme, 
with
mean-field potential 
\begin{align} \label{Ucyclic}
U^{\text{cyclic},\div}_{\bs{k}} = 
\sum_{n=1}^{N} 
U^{\text{cyclic},\div}_{n,\bs{k}}
&=
\sum_{n=1}^{N} \frac{\delta 
\Omega_{n,\text{normal}}^{\text{cyclic},\div}}{\delta n_\bs{k}}
= 
\frac{\delta 
\mathcal{D}^{\text{cyclic},\div}}{\delta n_\bs{k}}.
\end{align}
At second order one has
\begin{align} \label{U2cyc}
U^{\text{cyclic},(\div)}_{2,\bs{k}} &= 
-\frac{1}{4}
\sum_{\bs{k}_2, \bs{k}_3, \bs{k}_4} \!
|\braket{\psi_{\bs{k}}\psi_{\bs{k}_2}| V|\psi_{\bs{k}_3} \psi_{\bs{k}_4}}|^2 
\breq & \quad \times
\Big[ n_{\bs{k}_2} \bar n_{\bs{k}_3} \bar n_{\bs{k}_4} \mathcal{F}^\text{cyclic}(D)
-n_{\bs{k}_3} n_{\bs{k}_4} \bar n_{\bs{k}_2} \mathcal{F}^\text{cyclic}(-D) \Big],
\end{align}
where 
\begin{align}
\label{F2cycT0lim}
\mathcal{F}^\text{cyclic}(D)
&= \frac{1-\e^{-\beta D}}{ D} \xrightarrow{D\rightarrow 0} \beta.
\end{align}
In the cyclic scheme, the perturbation series and thermodynamic relations have the 
same structure as in the direct scheme. 
In particular, the same factorization property holds (see Sec.~\ref{sec42}), and 
again the zero-temperature limit does not exist [as evident from Eq.~\eqref{U2cyc}].
The direct scheme is, however, distinguished from the 
cyclic scheme in terms of it leading to the identification 
of the Fermi-Dirac distribution functions with the exact mean occupation numbers~\cite{Balian1961529,Balian1961529b,boer} (see also Appendix~\ref{app22}) 
and (in the classical limit) the virial expansion~\cite{Balian1961529b}.
This indicates that, for calculations close to the classical limit, the 
direct scheme is preferable to the cyclic scheme.

\subsubsection{Reduced scheme(s)}

In the renormalization scheme outlined by Balian and de Dominicis (BdD) \cite{statquasi3,statquasi1}, 
the term $\mathcal{D}^{\text{direct},\div}[n_\bs{k},T]$ (or, $\mathcal{D}^{\text{cyclic},\div}[n_\bs{k},T]$) is replaced by 
a term $\mathcal{D}^\text{BdD}[n_\bs{k}]$ that has no explicit temperature dependence in addition 
to the one given by the functional dependence on $n_\bs{k}(T,\mu)$,
and satisfies
\begin{align} \label{T0BdD}
\mathcal{D}^\text{BdD}(T,\mu) \xrightarrow{T\rightarrow 0} \sum_{n=1}^N E^{(0)}_n(\eps_\F),
\end{align}
where, by Eq.~\eqref{StatQP1a}, $\mu\xrightarrow{T\rightarrow 0}\eps_\F$, and $E^{(0)}_n(\eps_\F)$ corresponds to 
the sum of all contributions of order $n$ in zero-temperature MBPT.
This implies consistency with the adiabatic zero-temperature formalism irrespective of isotropy.
The BdD mean field is given by
\begin{align} \label{UBdDfirst}
U^\text{BdD}_\bs{k}[n_\bs{k}]=\frac{\delta\mathcal{D}^\text{BdD}[n_\bs{k}]}{\delta n_\bs{k}}.
\end{align}
Since $\mathcal{D}^\text{BdD}[n_\bs{k}]$ is supposed to have no explicit temperature dependence, it must be constructed
by eliminating all
energy denominator exponentials via Eq.~\eqref{Dexp}.
But then the functional derivative will lead to poles.
To make the functional derivative well-defined, the energy denominators have to be regularized.

Now, 
as first recognized by Balian and de Dominicis~\cite{BALIAN1960502} as well as Horwitz, Brout and Englert~\cite{brout1},
for a finite system with a discrete spectrum the following renormalized perturbation series can be constructed
\begin{align} \label{MBPTrenorm2a}
\Omega(T,\mu) &= \Omega_\text{ref}(T,\mu) + 
\Omega_{U}(T,\mu) +
\mathcal{D}^{\text{reduced,}\ast,\div}(T,\mu),
\end{align}
with mean field 
\begin{align}  \label{Uredast}
U^{\text{reduced,}\ast\ast,\div}_\bs{k}=\frac{\delta \mathcal{D}^{\text{reduced,}\ast,\div}}{\delta n_\bs{k}},
\end{align}
where
\begin{align}\label{MBPTrenorm2b}
\mathcal{D}^{\text{reduced,}\ast,\div}(T,\mu) 
=
\sum_{n=1}^{N} \frac{\delta 
\Omega_{n,\text{normal}}^{\text{reduced,}\ast,\div}}{\delta n_\bs{k}}\xrightarrow{T\rightarrow 0} \sum_{n=1}^N E^{(0)}_n(\eps_\F),
\end{align}
with $\mu\xrightarrow{T\rightarrow 0}\eps_\F$.
Here, ${\ast}$ means that the energy denominator poles are excluded in the discrete state sums (which makes the reduced formula well-defined, for a finite system).
Equation~\eqref{MBPTrenorm2a} entails another factorization property, i.e., (see Sec.~\ref{sec43})
\begin{align}\label{reducedfactorized1}
\Omega_{n_1+n_2,\text{anomalous}}^{\text{reduced,}\ast,\div}
&=  -\frac{\beta}{2}   \sum_\bs{k} 
U^{\text{reduced,}\ast,\div}_{n_1,\bs{k}} n_\bs{k} \bar n_\bs{k} 
U^{\text{reduced,}\ast,\div}_{n_2,\bs{k}}
\breq &\quad \times
(2-\delta_{n_1,n_2})
.
\end{align}
In Eq.~\eqref{reducedfactorized1}, $\div$ implies that the pseudoanomalous terms from the reduced expressions for normal two-particle reducible diagrams 
with the same pieces
are added (to the reduced expressions for the corresponding anomalous diagrams).

Equations~\eqref{MBPTrenorm2a} and \eqref{MBPTrenorm2b}
lead to the Fermi-liquid relations for $\varrho$, $S$, and $\delta E/\delta n_\bs{k}$.
The validity of the ${\ast}$ prescription for finite systems is however somewhat questionable, since it disregards the contributions from 
the energy denominator poles present in the cyclic and direct case for ${T\neq 0}$ [see Eqs.~\eqref{F2T0lim} and \eqref{F2cycT0lim}].\footnote{If the pole contributions are included for a finite system then Eq.~\eqref{MBPTrenorm2a} is valid only for ${T\rightarrow 0}$ (and the ${T\rightarrow 0}$ limit exists 
only for $\mu\not\in\{\en_\bs{k}\}$). In that sense, 
the construction of the thermodynamic Fermi-liquid relations via MBPT depends on the thermodynamic limit.}
In the thermodynamic limit 
the contributions from energy denominator poles have measure 
zero.
However, the thermodynamic limit of $\mathcal{D}^{\text{reduced,}\ast,\div}$ is singular at ${T\neq 0}$, due to 
terms with energy denominator poles of even degree.\footnote{At ${T=0}$, these singular terms 
cancel each other, see Ref.~\cite{Wellenhofer:2018dwh} and Sec.~\ref{sec45}.} 
In addition, 
there are terms with several (odd) energy denominator poles for which
the thermodynamic limit is not well-defined, as evident from 
the Poincar\'{e}-Bertrand transformation 
formula Eq.~\eqref{bertrand}; this implies that in the thermodynamic limit 
$U^{\text{reduced,}\ast,\div}_\bs{k}$ is ill-defined also at ${T=0}$. 

All in all, Eqs.~\ref{MBPTrenorm2a}, \eqref{MBPTrenorm2b}, and \eqref{reducedfactorized1} indicate that 
the BdD renormalization scheme should correspond to
\begin{align} \label{bddassume}
\mathcal{D}^\text{BdD}[n_\bs{k}] 
=
\mathcal{D}^{\text{reduced,}\ast\ast,\div}[n_\bs{k}],
\end{align}
where ${\ast\ast}$ refers to the energy denominator regularization for infinite systems.

\subsection{Statistical versus dynamical quasiparticles}\label{sec24}

The statistical quasiparticles associated 
with the BdD renormalization scheme
are distinguished from the dynamical quasiparticles~\cite{Noz1,Noz2,Benfatto2006}
associated with the asymptotic stability of the low-lying excited states.
In the following, we examine how dynamical quasiparticles arise in grand-canonical MBPT, and compare their energies to the 
ones of the statistical quasiparticles (i.e., the single-particle energies in the BdD scheme).
More details on the (various forms of the) self-energy are given in Appendix \ref{app2}.
In particular, in Appendix \ref{app22} we show
that (only) in the direct scheme the exact mean occupation numbers $f_{\bs{k}}(T,\mu)$ are identified with the Fermi-Dirac distributions.
Note that since 
the ${T\rightarrow 0}$ limit does not exist for the direct scheme,
this result is consistent with the discontinuity of $f_{\bs{k}}(T,\mu)$ at ${T=0}$.
The consistency of $f_{\bs{k}}({T\neq 0},\mu)=n_{\bs{k}}({T\neq 0},\mu)$ 
with the results discussed below is examined in Appendix~\ref{app22}.

\subsubsection{Dynamical quasiparticles without mean field}

In MBPT (for normal systems), dynamical quasiparticles arise as follows.
The perturbative contributions $\Sigma_{n,\bs{k}}(z,T,\mu)$ to the frequency-space self-energy 
$\Sigma_{\bs{k}}(z,T,\mu)$
are given by 
a specific analytic continuation (see Appendix~\ref{app21})
of the 
perturbative contributions to the Matsubara self-energy $\Xi_{\bs{k}}(z_l,T,\mu)$, 
where 
\begin{align} \label{Matsubarafreq}
z_l=\frac{\im(2l+1)\pi}{\beta}+\mu
\end{align}
are the Matsubara frequencies,
with $l\in\mathbb{Z}$. 
For example, in bare MBPT (with ${U=0}$) the two-particle irreducible
second-order contribution to $\Xi_{\bs{k}}(z_l,T,\mu)$ is given by [see Eq.~\eqref{Xi2calc}]
\begin{align} \label{sigma2}
\Xi_{2,\bs{k}}(z_l,T,\mu) &= 
\frac{1}{2}
\sum_{\bs{k}_2, \bs{k}_3, \bs{k}_4} \!
|\braket{\psi_{\bs{k}}\psi_{\bs{k}_2}| V|\psi_{\bs{k}_3} \psi_{\bs{k}_4}}|^2  n_{\bs{k}_2} \bar n_{\bs{k}_3} \bar n_{\bs{k}_4}
\breq & \quad \times
\frac{e^{-\beta(\en_{\bs{k}_3}+\en_{\bs{k}_4} - \en_{\bs{k}_2} - z_l)}-1}{\en_{\bs{k}_3}+\en_{\bs{k}_4} - \en_{\bs{k}_2} - z_l}.
\end{align}
From this, the expression for $\Sigma_{2,\bs{k}}(z,T,\mu)$ 
is obtained by \textit{first} substituting $\e^{\beta(z_l-\mu)}=-1$ and \textit{then} 
performing the analytic continuation.
Using Eq.~\eqref{Dexp}, one gets
\begin{align} \label{sigma2b}
\Sigma_{2,\bs{k}}(z,T,\mu) &= 
-\frac{1}{2}
\sum_{\bs{k}_2, \bs{k}_3, \bs{k}_4} \!
|\braket{\psi_{\bs{k}}\psi_{\bs{k}_2}| V|\psi_{\bs{k}_3} \psi_{\bs{k}_4}}|^2  
\breq & \quad \times
\frac{
n_{\bs{k}_2} \bar n_{\bs{k}_3} \bar n_{\bs{k}_4} +
n_{\bs{k}_3} n_{\bs{k}_4} \bar n_{\bs{k}_2}}
{\en_{\bs{k}_3}+\en_{\bs{k}_4} - \en_{\bs{k}_2} - z}.
\end{align}
As evident from the second-order contribution,
setting $z = \omega\pm\im\eta$, with $\omega$ real and $\eta$ infinitesimal, leads to the general relation~\cite{kadanoffbaym}
\begin{align} \label{dynquasi1a}
\Sigma_\bs{k}(\omega\pm\im\eta,T,\mu) = \mathcal{S}_\bs{k}(\omega,T,\mu) \mp \im \mathcal{J}_\bs{k}(\omega,T,\mu),
\end{align}
where $\mathcal{S}_\bs{k}$ and $\mathcal{J}_\bs{k}$ are real, and ${\mathcal{J}_\bs{k}\geq 0}$ (see Appendix~\ref{app21}).
From the property that at ${T=0}$ the energy denominators in the expressions for the perturbative contributions to the self-energy, $\Sigma_{n,\bs{k}}(z,T,\mu)$, vanish only 
for $z\rightarrow \mu$, Luttinger~\cite{PhysRev.121.942} showed that
\begin{align} \label{dynquasi1}
\mathcal{J}_\bs{k}(\omega,0,\mu) \xrightarrow{\omega\rightarrow \mu} C_\bs{k}(\mu)\, (\omega-\mu)^2,
\end{align}
with ${C_\bs{k}(\mu)\geq 0}$. 
Crucial for our discussion (i.e., in particular for the next paragraph), this result holds not only if $\Sigma_\bs{k}$ is calculated using self-consistent propagators  but also 
if $\Sigma_\bs{k}$ is calculated using bare propagators.

In Ref.~\cite{PhysRev.119.1153}, Luttinger showed that Eq.~\eqref{dynquasi1} implies a discontinuity at ${T=0}$ and ${\bs{k}=\bs{k}_\F}$ of the 
exact mean occupation numbers $f_{\bs{k}}(T,\mu)$ of the momentum eigenstates $\ket{\psi_{\bs{k}}}$, i.e.,~\cite{Luttinger:1960ua,Parry}
\begin{align}
f_{\bs{k}}(T,\mu)=\braket{\!\braket{a_\bs{k}^\dagger a_\bs{k}}\!} 
= 
\int \limits_{-\infty}^\infty \!\frac{d\omega}{2\pi} \frac{1}{1+\e^{\beta(\omega-\mu)}}\mathcal{A}_\bs{k}(\omega,T,\mu),
\end{align}
where $\braket{\!\braket{\ldots}\!}$ denotes the true ensemble average, and the spectral function $\mathcal{A}_\bs{k}(\omega,T,\mu)$ 
is given by~\cite{kadanoffbaym} (see also Appendix \ref{app21})
\begin{align} \label{spectral}
\mathcal{A}_\bs{k}(\omega,T,\mu)=\frac{2 \mathcal{J}_\bs{k}(\omega,T,\mu)}{\left[\omega-\en_{0,\bs{k}}-\mathcal{S}_\bs{k}(\omega,T,\mu)\right]^2+\left[\mathcal{J}_\bs{k}(\omega,T,\mu)\right]^2}.
\end{align}
The (true) Fermi momentum $\bs{k}_\F$, defined in terms of the discontinuity of $f_{\bs{k}}(0,\mu)$, is determined by~\cite{PhysRev.119.1153}
\begin{align} \label{dynquasi2}
\mu=\en_{0,\bs{k}_\F} + \mathcal{S}_{\bs{k}_\F}(\mu,0,\mu).
\end{align}
The lifetime of a single-mode excitation with momentum $\bs{k}$ 
of the ground state
is determined by 
the width of the spectral function at ${T=0}$~\cite{Fetter,kadanoffbaym}. From Eqs.~\eqref{dynquasi1} and \eqref{dynquasi2}, the width
vanishes (i.e., the excitation becomes stable against decay into collective modes) for ${\omega\rightarrow \mu}$ and ${\bs{k}\rightarrow\bs{k}_\F}$.
The energies $\mathcal{E}_\bs{k}$ of the dynamical quasiparticles are therefore
determined by
\begin{align} \label{dynquasi2b}
\mathcal{E}_\bs{k}(\mu)=\en_{0,\bs{k}} + \mathcal{S}_{\bs{k}}(\mathcal{E}_\bs{k},0,\mu),
\end{align}
where ${\mathcal{E}_\bs{k}\approx \mu}$ and ${\bs{k}\approx\bs{k}_\F}$ (low-lying excitations).\footnote{Note that the 
relation ${\mathcal{E}_{\bs{k}_\F}=\partial E({T=0},\varrho)/\partial \varrho}$ (Hugenholtz-Van Hove theorem~\cite{HUGENHOLTZ1958363,Baym:1962sx}) 
is trivial if $E({T=0},\varrho)$ is derived from $\Omega(T,\mu)$.}

\subsubsection{Dynamical quasiparticles with mean field}

The distinction between the energies
of statistical and dynamical quasiparticles can now be made explicit, in a specific sense.
For bare MBPT with mean field $U_\bs{k}(T,\mu)$
the self-energy is given by
\begin{align}\label{dynquasi3b}
\Sigma_{\bs{k}}(z,T,\mu)
&=
-U_\bs{k}(T,\mu) +\Sigma'_{\bs{k}}(z,T,\mu).
\end{align}
Here, the first term
corresponds to the contribution from the self-energy diagram 
composed of a single $-U$ vertex. Since bare propagators are used, 
$\Sigma'_{\bs{k}}(z,T,\mu)$ involves not only one- and two-particle irreducible  
but also two-particle reducible self-energy diagrams (including diagrams with $-U$ vertices); see, e.g., 
Ref.~\cite{PLATTER2003250}.\footnote{Note that this implies that 
there are diagrams with several identical energy denominators, i.e., the Hadamard finite part appears.}
It can be seen that [see Eq.~\eqref{sigmastarred}]
\begin{align}\label{dynquasi4}
\Sigma'_{n,\bs{k}}(z) =  
\left[\frac{\delta \Omega_n^{\text{reduced}}[n_\bs{k}]}{\delta n_\bs{k}}\bigg|_{\en_\bs{k}= z}
\right]_{\bs{k} \notin \{\text{articulation lines}\}},
\end{align}
(with $\text{Im}[z]\neq 0$).
Instead of Eq.~\eqref{dynquasi1} we have
\begin{align}
\Sigma_{\bs{k}}(z,T,\mu)
&=
-U_\bs{k}(T,\mu) +\mathcal{S}'_{\bs{k}}(z,T,\mu)+\im \mathcal{J}'_{\bs{k}}(z,T,\mu),
\end{align}
with 
\begin{align} \label{dynquasi3d}
\mathcal{J}'_\bs{k}(\omega,0,\mu) = C'_\bs{k} (\omega-\mu)^2.
\end{align}
The spectral function is now given by
\begin{align} \label{spectralSJ}
\mathcal{A}_\bs{k}(\omega,T,\mu)=\frac{2 \mathcal{J}'_\bs{k}
(\omega,T,\mu)}{\Big[\omega-\en_{\bs{k}}-\text{Re}\left[\Sigma_{\bs{k}}(z,T,\mu)\right]\Big]^2+\left[\mathcal{J}'_\bs{k}(\omega,T,\mu)\right]^2}.
\end{align}
Using $\en_{\bs{k}}=\en_{0,\bs{k}}+U_{\bs{k}}(T,\mu)$, this 
becomes
\begin{align} 
\mathcal{A}_\bs{k}(\omega,T,\mu)=\frac{2 \mathcal{J}'_\bs{k}(\omega,T,\mu)}{\left[\omega-\en_{0,\bs{k}}-\mathcal{S}'_\bs{k}(\omega,T,\mu)\right]^2+\left[\mathcal{J}'_\bs{k}(\omega,T,\mu)\right]^2},
\end{align}
so the (true) Fermi-momentum $\bs{k}_\F$ is determined by
\begin{align} \label{kfdef}
\mu=\en_{0,\bs{k}_\F} + \mathcal{S}'_{\bs{k}_\F}(\mu,0,\mu),
\end{align}
and the dynamical quasiparticle energies $\mathcal{E}_\bs{k}$ are given by
\begin{align} \label{quasiU}
\mathcal{E}_\bs{k}(\mu)=\en_{0,\bs{k}} + \mathcal{S}'_{\bs{k}}(\mathcal{E}_\bs{k},0,\mu),
\end{align}
where ${\mathcal{E}_\bs{k}\approx \mu}$ and ${\bs{k}\approx\bs{k}_\F}$. 
It is ${\mathcal{E}_\bs{k} = \en_\bs{k}}$ for ${N\leq 2}$ within the BdD renormalization scheme, but
from Eq.~\eqref{dynquasi4} as well as Eqs.~\eqref{UBdDfirst} and \eqref{bddassume} it is clear that 
this correspondence breaks down for truncation orders ${N>2}$.
To have ${\mathcal{E}_\bs{k} = \en_\bs{k}}$ for ${N>2}$ the mean field must 
satisfy
\begin{align}\label{luttward}
U_\bs{k}(0,\mu) = \mathcal{S}'_{\bs{k}}(\mu,0,\mu),
\end{align}
but then no statistical quasiparticle relations are obtained.
In particular, 
formally extending Eq.~\eqref{quasiU} to momenta $\bs{k}\in[0,\bs{k}_\F]$,
the mean-field renormalization 
specified by Eq.~\eqref{luttward} leads to
\begin{align}
\sum_\bs{k} \theta(\mu-\mathcal{E}_\bs{k})
=
\sum_\bs{k} \theta(\mu-\en_\bs{k}),
\end{align}
but the relation (i.e., Luttinger's theorem~\cite{PhysRev.121.942,stefanuc,Baym:1962sx})
\begin{align}
\sum_\bs{k} \theta(\mu-\mathcal{E}_\bs{k})=\varrho(T=0,\mu)
\end{align}
is satisfied only for truncation orders $N\leq 2$.

\section{Regularization of energy denominators}\label{sec3}

An energy denominator regularization scheme is a procedure that allows to evaluate
the contributions associated with the various parts ${\mathcal{F}_\alpha=f_\alpha/D}$ of
the energy denominator terms ${\mathcal{F}=\sum_\alpha\mathcal{F}_\alpha}$ separately [cf., e.g.,~Eq.~\eqref{F2T0lim}].
The (formal) splitting of the $\mathcal{F}$'s
into parts $\mathcal{F}_\alpha$ introduces poles, so the essence of any regularization scheme must be a change in the way the contributions near the zeros of the 
denominators $D=\prod_\nu D_\nu^{n_\nu}$ of these terms are evaluated (in particular for the case where some $n_\nu$ are even). 
This change must be such that, for a fixed mean field, the same results are
obtained as from the original unregularized expressions 
for the $\mathcal{F}$'s
(e.g.,
the expressions obtained from the direct or cyclic formula).

For the second-order normal contribution the regularization is (essentially) unique and corresponds
to evaluating the two parts of Eq.~\eqref{F2cycT0lim} separately via 
principal value integrals.
For the higher-order contributions, the regularization scheme introduced here 
starts by adding infinitesimal imaginary parts 
to the individual energy denominators $D_\nu$, i.e.,  
$\prod_\nu D_\nu^{n_\nu}\rightarrow \prod_\nu (D_\nu+i\eta_\nu)^{n_\nu}$. 
The regularization then corresponds to evaluating the 
various parts with energy denominator terms
$\mathcal{F}_{\alpha,[\{\eta_\nu\}]}=f_\alpha/[\prod_\nu (D_\nu+i\eta_\nu)]^{n_\nu}$ via the
Sokhotski-Plemelj-Fox formula. 
That this is a valid procedure can be seen from the 
fact that (after adding infinitesimal imaginary parts) the Sokhotski-Plemelj-Fox formula can be applied (formally) also 
to the unsplit expressions with energy denominator terms 
$\mathcal{F}_{[\{\eta_\nu\}]}=\sum_\alpha\mathcal{F}_{\alpha,[\{\eta_\nu\}]}$, and 
after its application the splitting corresponds again (i.e., as in the second-order case) to a separation into
principal value integrals, by virtue of Eq.~\eqref{hadamard} below.

The crucial point of this particular regularization scheme is 
that it allows to separate the normal, anomalous, and pseudoanomalous contributions (at finite $T$) such that 
these contributions have a form that matches the (regularized) disentangled reduced formula.
This feature
is essential for the cancellation of the 
pseudoanomalous contributions and the factorization of the anomalous contributions, and 
these properties lead to    
the thermodynamic Fermi-liquid relations via the BdD scheme.
In other terms, the Fermi-liquid relations uniquely determine the regularization of the energy 
denominators.\footnote{A different regularization scheme can for example be set up via
$\prod_\nu D_\nu^{n_\nu}\rightarrow (\prod_\nu D_\nu)^{n_\nu}+i\eta$. The parts $\mathcal{F}_{\alpha,[\eta]}$ 
then have a form that deviates from the reduced formula (in particular, the pseudoanomalous contributions do not cancel; see also Appendix \ref{app1}), so
the Fermi-liquid relations cannot be obtained in this scheme.}

In Sec.~\ref{sec31} we introduce the formal approach to the energy denominator regularization for the BdD scheme.\footnote{Rules for the formal regularization have been presented also in Refs.~\cite{BALIAN1971229,LUTTINGER19731,keitermorandi} for the case of
impurity systems.}
The numerical evaluation of the resulting expressions is discussed in Sec.~\ref{sec32}.

\subsection{Formal regularization}\label{sec31}

From the cyclic expressions, the regularized (${\ast\ast}$) disentangled ($\div$) reduced expressions are 
obtained by performing the following steps:
\begin{enumerate}[start=1,label={(\roman*)}]
\item add infinitesimal imaginary parts $\eta_\nu$ to the energy denominators $D_\nu$ (where $\eta_1\neq \eta_2\neq \ldots$),
\item eliminate the energy denominator exponentials via Eq.~\eqref{Dexp}, 
\item apply Eq.~\eqref{ndouble}.
\end{enumerate}
Here, the first step is part of $\ast\ast$, the second step is part of the reduction, and the third step is associated with $\div$.
Then
\begin{enumerate}[start=4,label={(\roman*)}]
\item for two-particle reducible diagrams, average over the signs $\text{sgn}(\eta_\nu)$ of the imaginary parts,
\item split the integrals such that the various parts of the cyclic energy denominator terms are integrated separately, then suitably relabel indices in some integrals, and finally recombine the integrals that lead to normal, pseudoanomalous and anomalous contributions,
\item observe that the pseudoanomalous contributions vanish (this is proved to all orders in Sec.~\ref{sec4}),
\item observe that the anomalous contributions factorize (this is proved to all orders in Sec.~\ref{sec4}),
\end{enumerate}
where the first step is part of $\ast\ast$, and the second, third and fourth steps are associated with $\div$ and reduction.
To show how these rules arise we now regularize, disentangle, and reduce the expressions for the 
contributions from the normal
second-order diagram and from selected third-order diagrams.

\vspace*{2mm}
\begin{figure}[h] 
\centering
\vspace*{0mm}
\hspace*{-0.0cm}
\includegraphics[width=0.12\textwidth]{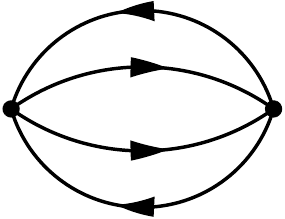} 
\vspace*{-2mm}
\caption{The normal second-order diagram. It is invariant under vertex permutations.}
\label{fig2normal}
\end{figure}
The cyclic expression for the normal 
second-order diagram shown in Fig.~\ref{fig2normal} is given by
\begin{align}\label{Omega2}
\Omega_{2,\text{normal}}^\text{cyclic}= -\frac{1}{8} \sum_{ijab}
\zeta^{ijab} n_{ij}\bar n_{ab} \frac{1-\e^{-\beta D_{ab,ij}}}{D_{ab,ij}},
\end{align}
where $\zeta^{ijab}=V^{ij,ab}V^{ab,ij}$,
with $V^{ij,ab}=
\braket{\psi_{\bs{k}_i}\psi_{\bs{k}_j}|V|\psi_{\bs{k}_a}\psi_{\bs{k}_b} }$.
Moreover, $\sum_i=\int d^3 k_i/(2\pi)^3$, $n_{ij}=n_{\bs{k}_i}n_{\bs{k}_j}$ and 
$\bar n_{ij}=(1-n_{\bs{k}_a})(1-n_{\bs{k}_b})$, 
and $D_{ab,ij}=\en_{\bs{k}_a}+\en_{\bs{k}_b}-\en_{\bs{k}_i}-\en_{\bs{k}_j}$.

In Eq.~\eqref{Omega2}, the term $(1-\e^{-\beta D_{ab,ij}})/D_{ab,ij}$ is regular for ${D_{ab,ij}=0}$.
To evaluate the 
two parts of the numerator of this term separately, we add 
an infinitesimal imaginary term $\im\eta$ to the energy denominator.
This leads to 
\begin{align}\label{Omega2b}
\Omega_{2,\text{normal}}^\text{cyclic}&= -\frac{1}{8} \sum_{ijab}
\zeta^{ijab} n_{ij} \bar n_{ab} \frac{1-\e^{-\beta D_{ab,ij}}}{D_{ab,ij}+\im\eta}.
\breq &=
-\frac{1}{8} \sum_{ijab}
\zeta^{ijab} n_{ij} \bar n_{ab} \frac{1}{D_{ab,ij}+\im\eta}
\breq & \quad +\frac{1}{8} \sum_{ijab}
\zeta^{ijab} n_{ab} \bar n_{ij} \frac{1}{D_{ab,ij}-\im\eta}.
\end{align}
where we have applied Eq.~\eqref{Dexp} to eliminate the energy denominator exponential 
in the second part.
Relabeling indices $(i,j)\leftrightarrow (a,b)$ and recombining the two terms leads to
\begin{align}\label{Omega2c}
\Omega_{2,\text{normal}}^\text{cyclic}&=
-\frac{1}{8} \sum_{ijab}
\zeta^{ijab} n_{ij} \bar n_{ab} \left[\frac{1}{D_{ab,ij}+\im\eta}+\frac{1}{D_{ab,ij}-\im\eta}\right]
\breq & \equiv \Omega_{2,\text{normal}}^{\text{reduced,}\ast\ast,(\div)}=\Omega_{2,\text{normal}}^\text{BdD}.
\end{align}
From this, one obtains for the second-order contribution to the BdD mean field the expression
\begin{align}\label{Omega2d}
U^\text{BdD}_{2,i} = U_{2,i}^{\text{reduced,}\ast\ast,(\div)} &=
-\frac{1}{4} \sum_{jab}
\zeta^{ijab} \big( n_{j} \bar n_{ab}+n_{ab} \bar n_{j}\big)
\breq & \quad \times
\left[\frac{1}{D_{ab,ij}+\im\eta}+\frac{1}{D_{ab,ij}-\im\eta}\right].
\end{align}
Note that the expressions for $\Omega_{2,\text{normal}}^\text{BdD}$ and $U^\text{BdD}_{2,i}$ are real.
Given that the integration variables include $D_{ab,ij}$ [or an equivalent variable, see Eq.~\eqref{Omega2num1}], 
this can be seen explicitly from the Sokhotski-Plemelj theorem
\begin{align} \label{plemelj}
\frac{1}{x+\im \eta}= \frac{P}{x}-\im\pi\,\text{sgn}(\eta)\,\delta(x),
\end{align}
where $P$ refers to the Cauchy principal value. 
For actual numerical calculations it is however more practical not to use 
$D_{ab,ij}$ as an integration variable, and then
the application of the Sokhotski-Plemelj theorem
requires further attention. This issue is discussed in Sec.~\ref{sec32}.

It will be useful now to examine how 
Eq.~\eqref{Omega2c}
can be derived from the direct formula.
The direct expression 
is given by
\begin{align} \label{Omega2dira}
\Omega_{2,\text{normal}}^\text{direct}= \frac{1}{4} \sum_{ijab}
\zeta^{ijab} n_{ij}\bar n_{ab} \frac{1-\e^{-\beta D}-\beta D}{\beta D^2},
\end{align}
where $D=D_{ab,ij}$.
Adding an imaginary part to the energy denominator we have
\begin{align} \label{Omega2dirb}
\Omega_{2,\text{normal}}^\text{direct}= \frac{1}{4} \sum_{ijab}
\zeta^{ijab} n_{ij}\bar n_{ab} \frac{1-\e^{-\beta D}-\beta D}{\beta (D+\im\eta)^2}.
\end{align}
Here, the integral can be evaluated in terms of the 
Sokhotski-Plemelj-Fox formula~\cite{fox}
\begin{align}\label{plemelj2}
\frac{1}{(x+\im\eta)^n}= \frac{P}{x^n} + \im\pi(-1)^{n} \,\text{sgn}(\eta)\,\delta^{(n-1)}(x),
\end{align}
where now $P$ denotes the Hadamard finite part~\cite{hadamard} (see also Refs.~\cite{MONEGATO2009425,galap,dispersions}), i.e.,
\begin{align}\label{hadamard}
\int \!\! dx\, \varphi(x)\frac{P}{x^{n+1}}\equiv
\frac{1}{n!}
\lim_{y\rightarrow 0}
\frac{\partial^{n}}{\partial y^{n}}\!\!
\int \!\! dx\, \varphi(x)\frac{P}{x-y},
\end{align}
and $\delta^{(n-1)}(x)=\partial \delta(x)/\partial x^n$.
Note that this prescription satisfies $x^k/(x+\im\eta)^n=1/(x+\im\eta)^{n-k}$.
Since $\partial(1-\e^{-\beta D}-\beta D)/\partial D=0$ for $D=0$, evaluating Eq.~\eqref{Omega2dirb} with the Sokhotski-Plemelj-Fox formula
gives the same result as Eq.~\eqref{Omega2dira}.
This equivalence is maintained 
if the three parts of the $1-\e^{-\beta D}-\beta D$ are integrated separately (and evaluated with the Sokhotski-Plemelj-Fox formula).
That is, applying \textit{first} the Sokhotski-Plemelj-Fox formula and \textit{then}
Eq.~\eqref{Dexp} and the relabeling of indices we find 
\begin{align} \label{Omega2dirc}
\Omega_{2,\text{normal}}^\text{direct} &= -\frac{1}{4} \sum_{ijab}
\zeta^{ijab} n_{ij}\bar n_{ab} \, D\frac{P}{D^2}
=-\frac{1}{4} \sum_{ijab}
\zeta^{ijab} n_{ij}\bar n_{ab} \, \frac{P}{D}
\breq & \equiv \Omega_{2,\text{normal}}^{\text{reduced,}\ast\ast,(\div)}=\Omega_{2,\text{normal}}^\text{BdD}.
\end{align}
It is now important to note that
applying Eq.~\eqref{Dexp} and relabeling indices in the second part (which implies $D\rightarrow -D$) 
\textit{before} applying the Sokhotski-Plemelj-Fox formula 
would lead to incorrect results, i.e., 
this procedure would leave the real part invariant but produce a finite imaginary part. 
This is because 
\begin{align} \label{Omega2wronga}
f(D)\frac{\e^{-\beta D}}{(D+\im\eta)^2}
&=f(D)\e^{-\beta D}\frac{P}{D^2}+\im\pi \,\text{sgn}(\eta)\,\delta(D)\,\beta f(0)
\breq & \quad
-\im\pi \,\text{sgn}(\eta)\,\delta(D)\,\frac{\partial f(D)}{\partial D},
\end{align}
whereas
\begin{align}
\frac{f(-D)}{(-D+\im\eta)^2}
=f(-D)\frac{P}{(-D)^2}+\im\pi \,\text{sgn}(\eta)\,\delta(D)\,f(0).
\end{align}
In general, for ${n>1}$ it is
\begin{align} 
f(D)\frac{\e^{-\beta D}}{(D+\im\eta)^n} \neq \frac{f(-D)}{(-D+\im\eta)^n}.
\end{align}
However, note that 
$f(-D)\frac{P}{(-D)^n}=f(D)\frac{P}{D^n}$
since $D$ is integrated in the whole real domain, and therefore 
\begin{align} 
\text{Re}\left[f(D)\frac{\e^{-\beta D}}{(D+\im\eta)^n}\right] \equiv \text{Re}\left[\frac{f(-D)}{(-D+\im\eta)^n}\right]
\end{align}
for the considered $f(D)$.
Hence, applying Eq.~\eqref{Dexp} and relabeling indices without first applying the Sokhotski-Plemelj-Fox formula
becomes valid if we average over the sign of $\eta$, i.e.,
\begin{align} 
\frac{1}{2}\sum_{\text{sgn}(\eta)}f(D)\frac{\e^{-\beta D}}{(D+\im\eta)^n} = \frac{1}{2}\sum_{\text{sgn}(\eta)} \frac{f(-D)}{(-D+\im\eta)^n}.
\end{align}
Note that the average has to be taken for all three parts of Eq.~\eqref{Omega2dirb}, otherwise imaginary parts would remain.

\vspace*{0mm}
\begin{figure}[h] 
\centering
\vspace*{0mm}
\hspace*{-0.0cm}
\includegraphics[width=0.45\textwidth]{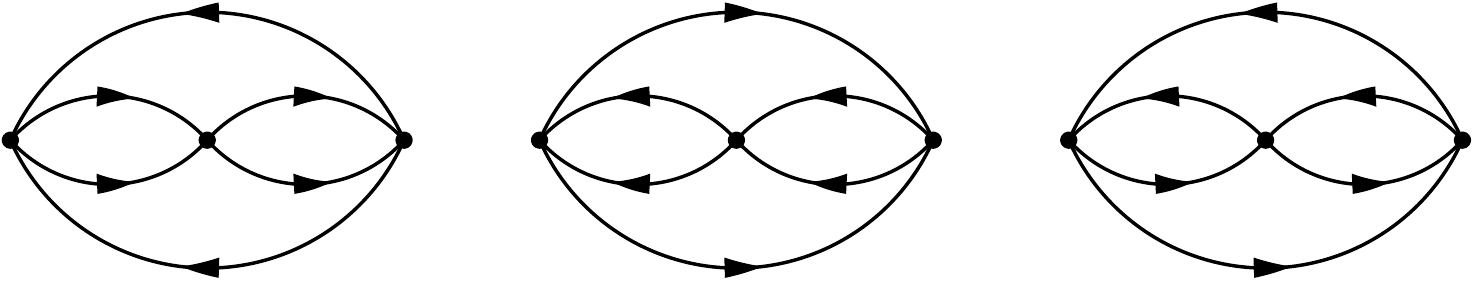} 
\vspace*{-2mm}
\caption{The third-order two-particle irreducible diagrams. Each diagram is invariant under cyclic vertex permutations. 
The first (pp) and second (hh) diagram transform into each other under noncyclic permutations, and the third (ph) diagram is permutation invariant.}
\label{fig3ppph}
\end{figure}

The cyclic expressions for the third-order two-particle irreducible diagrams shown in Fig.~\ref{fig3ppph} are given by
\begin{align} \label{Omega3pp}
\Omega_{3,\text{pp}}^\text{cyclic}&= \frac{1}{24} \sum_{ijabcd}
\zeta_{\text{pp}}^{ijabcd} n_{ij}\bar n_{abcd} 
\mathcal{F}_\text{pp}^\text{cyclic},
\\ \label{Omega3hh}
\Omega_{3,\text{hh}}^\text{cyclic}&= \frac{1}{24} \sum_{ijklab}
\zeta_{\text{hh}}^{ijklcd} n_{ijkl}\bar n_{ab} 
\mathcal{F}_\text{hh}^\text{cyclic},
\\ \label{Omega3ph}
\Omega_{3,\text{ph}}^\text{cyclic}&= \frac{1}{3} \sum_{ijkabc}
\zeta_{\text{ph}}^{ijkabc} n_{ijk}\bar n_{abc} 
\mathcal{F}_\text{ph}^\text{cyclic},
\end{align}

\noindent
where $\zeta_{\text{pp}}^{ijabcd}=V^{ij,ab}V^{ab,cd}V^{cd,ij}$, 
$\zeta_{\text{hh}}^{ijabcd}=V^{ij,ab}V^{kl,ij}V^{ab,kl}$, and
$\zeta_{\text{ph}}^{ijabcd}=V^{ij,ab}V^{kb,ic}V^{ac,jk}$. The energy denominator terms are given by
\begin{align}
\mathcal{F}_\text{pp,hh,ph}^\text{cyclic}
=
\left[\frac{1}{D_1 D_2}
+\frac{\e^{-\beta D_1}}{D_1 (D_1 - D_2)}
-\frac{\e^{-\beta D_2}}{D_2 (D_1 - D_2)}\right],
\end{align}
with $D_1=D_{ab,ij}$ and $D_2=D_{cd,ij}$ for the pp diagram, $D_1=D_{ab,ij}$ and $D_2=D_{ab,kl}$ for the hh diagram, and 
$D_1=D_{ab,ij}$ and $D_2=D_{ac,jk}$ for the ph diagram.
In each case, substituting $D_1\rightarrow D_1+\im\eta_1$ and $D_2\rightarrow D_2+\im\eta_2$, 
with $\eta_1\neq\eta_2$, splitting the integrals, eliminating the energy denominator exponentials and
relabeling indices leads to
\begin{align}
\mathcal{F}_\text{pp,hh,ph}^{\text{cyclic},\ast\ast} 
&=
\left[\frac{1}{(D_1+\im\eta_1) (D_2+\im\eta_2)}
+\frac{1}{(D_1-\im\eta_1) (D_2+\im\eta_2)}\right.
\breq &\quad  \left.
+\frac{1}{(D_1-\im\eta_1) (D_2-\im\eta_2)}\right] \equiv \mathcal{F}_\text{pp,hh,ph}^{\text{reduced},\ast\ast} ,
\end{align}
which is real.
Substituting this for $\mathcal{F}_\text{pp,hh,ph}^\text{cyclic}$ in 
Eqs.~\eqref{Omega3pp}, \eqref{Omega3hh}, and~\eqref{Omega3ph} and performing the functional derivative one 
obtains the third-order contribution to $U^\text{BdD}$.

\vspace*{0mm}
\begin{figure}[h] 
\centering
\vspace*{0mm}
\hspace*{-0.05cm}
\includegraphics[width=0.48\textwidth]{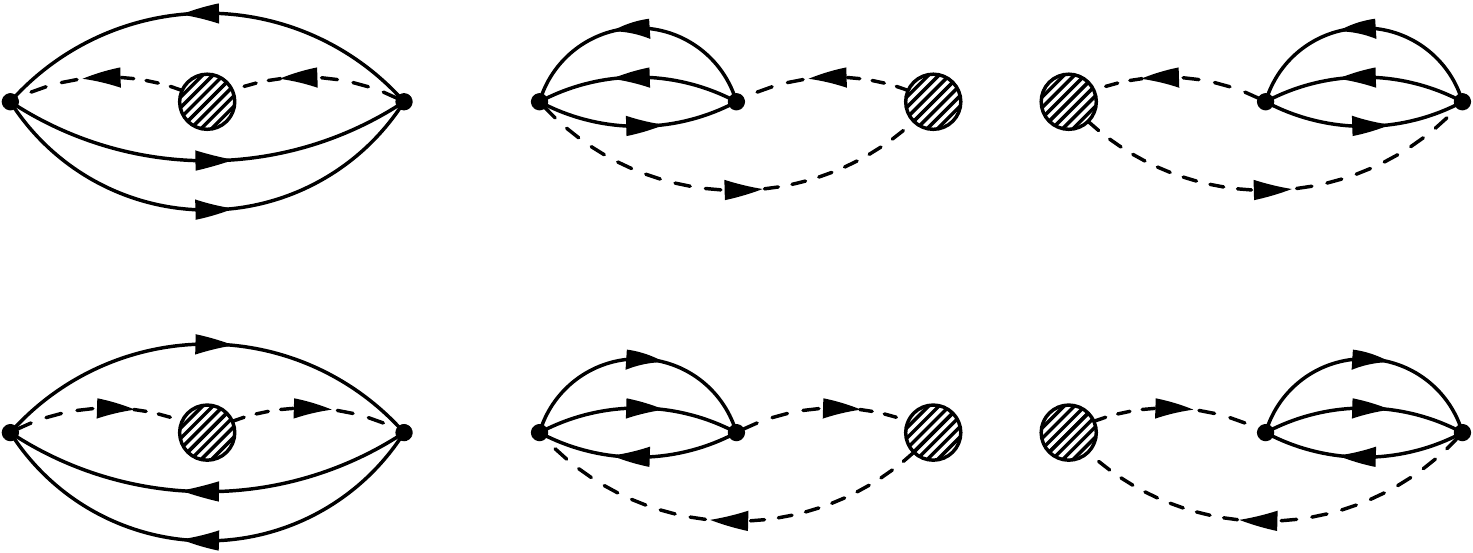} 
\vspace*{-4mm}
\caption{The six third-order two-particle reducible diagrams composed of one second-order and one first-order piece. 
Articulation lines are shown as dashed lines.
The shaded blobs represent vertices with loops (first-order pieces). 
In each row, the diagram on the left is a normal diagram, and the other two are anomalous. 
The diagrams in each row transform into each other under cyclic vertex permutations. 
The set of all six diagrams is closed under general vertex permutations.}
\label{fig3red}
\end{figure}

The normal third-order two-particle reducible diagrams are shown in Fig.~\ref{fig3red}.
Also shown are the cyclically related anomalous diagrams.
The cyclic expression for the sum of these diagrams is given by
\begin{align}\label{Omega3_21}
\Omega_{3,(21)}^\text{cyclic}&= -\frac{1}{4} \sum_{ijab}
\zeta^{ijab} n_{ij}\bar n_{ab} \mathcal{F}_{(21)}^\text{cyclic}
\big(n_i U_{1,i}-\bar n_a U_{1,a}\big),
\end{align}
where
\begin{align}
\mathcal{F}_{(21)}^\text{cyclic}
= \frac{1-\e^{-\beta D}-\beta D\e^{-\beta D}}{D^2}
,
\end{align}
with $D=D_{ab,ij}$.
In Hartree-Fock MBPT, the contribution from these diagrams is (as is well known) canceled 
by the corresponding diagrams where the first-order pieces are replaced by $-U_{1}$ vertices.
Nevertheless, it will be still be useful to regularize these contributions. We will then find that, if $U_1$ were left out, the anomalous part of these diagrams
can still be canceled via $U_2^\text{BdD}$.\footnote{It should be noted that, while 
the complete cancellation of two-particle reducible diagrams (with first-order pieces) is specific to $U_1$,
including $U_2^\text{BdD}$, $U_3^\text{BdD}$, etc. does not only eliminate anomalous contributions but also partially cancels normal contributions. 
Note also that the reduced contributions from 
normal two-particle reducible diagrams with single-vertex loops can be resummed as geometric series; 
in zero-temperature MBPT this is equivalent to the change from ${U=0}$ to ${U=U_1}$ for isotropic systems (only).}

Substituting $D\rightarrow D+\im\eta$ and applying Eqs.~\eqref{ndouble} and \eqref{Dexp}
and the relabeling $(i,j)\leftrightarrow (a,b)$, we can 
separate $\Omega_{3,(21)}^\text{cyclic}$ into the three contributions
\begin{align}\label{Omega3_21a}
\Omega_{3,(21),\text{normal}}^{\text{reduced,}\circ\circ,\div}&= -\frac{1}{4} \sum_{ijab}
\zeta^{ij,ab} n_{ij}\bar n_{ab} \frac{1}{(D+\im\eta)^2}
\big(U_{1,i}-U_{1,a}\big),
\\ \label{Omega3_21b}
\Omega_{3,(21),\text{anom.}}^{\text{reduced,}\circ\circ,\div}&= \frac{\beta}{4} \sum_{ijab}
\zeta^{ij,ab} n_{ij}\bar n_{ab} \frac{1}{D+\im\eta}
\big(\bar n_iU_{1,i}- n_aU_{1,a}\big),
\\ \label{Omega3_21c}
\Omega_{3,(21),\text{pseudo-a.}}^{\text{reduced,}\circ\circ,\div}&= \frac{1}{4} \sum_{ijab}
\zeta^{ij,ab} n_{ij}\bar n_{ab} \left[\frac{1}{(D+\im\eta)^2}-\frac{1}{(D-\im\eta)^2}\right]
\breq &\quad \times
\big(n_iU_{1,i}-\bar n_aU_{1,a}\big).
\end{align}
Here, ${\circ\circ}$ refers to an incomplete (in fact, incorrect) regularization:
none of the three contributions 
given by Eq.~\eqref{Omega3_21a}, \eqref{Omega3_21b} and \eqref{Omega3_21c} is real, and (more severely) also their sum is not real.  
As explained below Eq.~\eqref{Omega2dirc}, the reason for this deficiency is that 
we have applied Eq.~\eqref{Dexp} and relabeled indices 
without applying the Sokhotski-Plemelj-Fox formula first.
To repair this 
we have to average over the signs of the imaginary parts, which 
leads to
\begin{align} \label{T0canc}
\Omega_{3,(21),\text{normal}}^{\text{reduced,}\ast\ast,\div}&= -\frac{1}{8} \sum_{ijab}
\zeta^{ij,ab} n_{ij}\bar n_{ab} 
\breq &\quad 
\times \left[\frac{1}{(D+\im\eta)^2}+\frac{1}{(D-\im\eta)^2}\right]
 \big(U_{1,i}- U_{1,a}\big),
\\
\Omega_{3,(21),\text{anom.}}^{\text{reduced,}\ast\ast,\div}&= \frac{\beta}{8} \sum_{ijab}
\zeta^{ij,ab} n_{ij}\bar n_{ab} 
\left[\frac{1}{(D+\im\eta)}+\frac{1}{(D-\im\eta)}\right],
\breq &\quad 
\times (\bar n_iU_{1,i}- n_aU_{1,a}),
\\ \label{321pseudovanish}
\Omega_{3,(21),\text{pseudo-a.}}^{\text{reduced,}\ast\ast,\div}&= 0.
\end{align}
The pseudoanomalous contribution has vanished:
this feature, which is essential to obtain the Fermi-liquid relations at $T\neq 0$ (but not ${T=0}$), 
holds to all orders (see Sec.~\ref{sec4}).
Note that
the vanishing of the pseudoanomalous contributions holds only if 
all vertex permutations are included, i.e., 
it holds not separately for cyclically closed sets (in the present case, the two rows in Fig.~\ref{fig3red}).

\begin{figure}[h] 
\centering
\vspace*{0mm}
\hspace*{-0.0cm}
\includegraphics[width=0.14\textwidth]{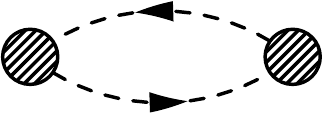} 
\vspace*{-2mm}
\caption{The anomalous diagram composed of two pieces of the mean-field or single-vertex loop type.}
\label{fig4UU}
\end{figure}

The anomalous contribution has the factorized form given by Eq.~\eqref{reducedfactorized1} (with $\ast\ast$ instead of $\ast$), i.e.,
\begin{align}
\Omega_{3,(21),\text{anom.}}^{\text{reduced,}\ast\ast,\div}&= -\beta \,  U_{2,i}^{\text{reduced,}\ast\ast,(\div)} n_i \bar n_i
\, U_{1,i}.
\end{align}
Thus, the anomalous contribution from the diagrams of Fig.~\ref{fig3red} gets 
canceled by the 
contribution from the diagram shown in Fig.~\ref{fig4UU} where
one piece is a first-order diagram and the other one either a $-U_1$ vertex or a $-U_{2}^\text{BdD}$ vertex.
The same cancellation occurs between the rotated diagram and the one with two mean-field vertices, and similar for the case where
both $U_1$ and $U_{2}^\text{BdD}$ are included.

\vspace*{-0mm}

\subsection{Integration variables}\label{sec32}
We now discuss how the formulas derived in Sec.~\ref{sec31} can be evaluated in numerical 
calculations.
Nonvanishing contributions with poles of even degree appear 
first at fourth order in the BdD renormalization scheme. 
These have to be evaluated in terms of the Hadamard finite part, which obviously represents 
a major difficulty in the numerical application of the  BdD scheme at high orders.
We leave out the discussion of methods to evaluate the Hadamard finite part numerically, and 
defer numerical applications of the BdD scheme (and the other schemes) to future research.

For an isotropic system and MBPT without a mean-field potential (${U_\bs{k}=0}$) where $\en_\bs{k}=\en_{0,\bs{k}}=\bs{k}^2/(2M)$, using as integration variables 
relative momenta $\bs{p}=(\bs{k}_i-\bs{k}_j)/2$ and $\bs{A}=(\bs{k}_a-\bs{k}_b)/2$ as well as the average momentum
$\bs{K}=(\bs{k}_i+\bs{k}_j)/2=(\bs{k}_a+\bs{k}_b)/2$, one obtains from Eq.~\eqref{Omega2c}
the following expression for the second-order normal contribution:
\begin{align}\label{Omega2num1}
\Omega_{2,\text{normal}}^\text{BdD}&=
-2 M \sum_{\bs{K},\bs{p},\bs{A}}
\zeta^{ijab} n_{ij} \bar n_{ab} \frac{P}{A^2-p^2}.
\end{align}
The functional derivative of this expression with respect to $n_{\bs{k}_i}$ is given by
\begin{align}\label{U2num1}
U_{2,\bs{k}_i}^\text{BdD}[\en_{0,\bs{k}_i}]=
-4 M \sum_{\bs{p},\bs{A}}
\zeta^{ijab} \big(n_{j} \bar n_{ab}+n_{ab} \bar n_{j}\big) \frac{P}{A^2-p^2}.
\end{align}
For truncation order ${N=2}$, the single-particle energies
in the BdD scheme are obtained 
from the self-consistent equation
\begin{align}
\en_\bs{k}=\frac{\bs{k}^2}{2M}+ U_{1,\bs{k}}+ U_{2,\bs{k}}^\text{BdD}[\en_\bs{k}],
\end{align}
where one may use for $U_{2,\bs{k}}^\text{BdD}[\en_\bs{k}]$
the expression obtained by substituting in Eq.~\eqref{U2num1} the term $D_{ab,ij}=\en_{\bs{k}_a}+\en_{\bs{k}_b}-\en_{\bs{k}_i}-\en_{\bs{k}_j}$
for $M/(A^2-p^2)$ if
this substitution does not introduce additional poles; 
otherwise one must go back to the expression with infinitesimal imaginary parts, Eq.~\eqref{Omega2d}.
This issue can be seen also in the ${U_\bs{k}=0}$ case 
if $\bs{k}_a$, $\bs{k}_i$ and $\bs{k}_j$ are used as integration variables to evaluate Eq.~\eqref{Omega2c}.
Considering a one-dimensional system for simplicity, we have
\begin{align}\label{Omega2num2a}
\Omega_{2,\text{normal}}^\text{BdD}&=
-\frac{ M}{4}\!\! \sum_{k_a,k_i,k_j}\!\! \zeta^{ijab} n_{ij} \bar n_{ab} \,
\left[\frac{1}{\kappa+\im\eta}
+\frac{1}{\kappa-\im\eta} \right],
\end{align}
with $k_b=k_i+k_j-k_a$ and $\sum_k=\int dk/(2\pi)$. Moreover,  
$\kappa=(k_a-k_i)(k_a-k_j)$, i.e., now there are two poles.
To bring Eq.~\eqref{Omega2num2a} into a form where the Sokhotski-Plemelj theorem can be applied, we note that
\begin{align}
&(k_a-k_i+ \im\eta)(k_a-k_j+\im\eta)
\breq
&\quad \quad=
(\kappa +\im\eta) \, \theta(2k_a-k_i-k_j)
+(\kappa - \im\eta) \, \theta(k_i+k_j-2k_a),
\end{align}
so
\begin{align}
\sum_{\text{sgn}(\eta)}\frac{1}{\kappa+\im\eta}
=
\sum_{\text{sgn}(\eta)}
\frac{1}{(k_a-k_i+\im\eta)(k_a-k_j+\im\eta)}.
\end{align}
The Sokhotski-Plemelj theorem can now be applied (assuming that $k_a$ is integrated after $k_i$ or $k_j$), which leads to
\begin{align}\label{Omega2num2b}
\Omega_{2,\text{normal}}^\text{BdD}&=
-\frac{ M}{2} \sum_{k_a,k_i,k_j} \zeta^{ijab} n_{ij} \bar n_{ab} 
\breq &\quad \times
\bigg[\frac{P}{k_a-k_i}\frac{P}{k_a-k_j}
+
\pi^2 \delta(k_a-k_i)\delta(k_a-k_j) \bigg],
\end{align}
where the integration order is fixed. Changing the integration order such that 
$k_a$ is integrated first would lead to an incorrect result, as evident from the 
Poincar\'{e}-Bertrand transformation 
formula \cite{hardy,poincare,bertrand,Muskheli,dispersions}
\begin{align} \label{bertrand}
&\int \!\! dx \!\int \!\! dy 
\,\,\varphi(x,y)
\frac{P}{x-y}\frac{P}{x-z} 
\breq
&\quad = 
\int \!\! dy \!\int \!\! dx 
\,\,\varphi(x,y)
\frac{P}{x-y}\frac{P}{x-z} 
+
\pi^2 \varphi(z,z).
\end{align}
Since it has only one pole, the expression given by Eq.~\eqref{Omega2num1} is however preferable compared
to the one where
$\bs{k}_a$, $\bs{k}_i$ and $\bs{k}_j$ are used as integration variables.

At third order the issue manifested by the Poincar\'{e}-Bertrand transformation formula becomes unavoidable.
For an isotropic system and ${U_\bs{k}=0}$, using relative and average momenta as integration variables one obtains 
for $\Omega_{3,\text{pp}}^\text{BdD}$ the expression (see also Refs.~\cite{Kondo,YosidaHirosi})
\begin{align}\label{Omega3ppnum}
\Omega_{3,\text{pp}}^\text{BdD}&=
\frac{M^2}{3} \!
\sum_{\bs{K},\bs{p},\bs{A},\bs{B}}
\zeta^{ijabcd}_\text{pp} n_{ij} \bar n_{abcd} 
\breq & \quad \times
\left[3\frac{P}{A^2-p^2}\frac{P}{B^2-p^2}+\pi^2 \frac{\delta(A-p)\delta(B-p)}{(A+p)(B+p)}\right],
\end{align}
where $\bs{p}=(\bs{k}_i-\bs{k}_j)/2$, $\bs{A}=(\bs{k}_a-\bs{k}_b)/2$, $\bs{B}=(\bs{k}_c-\bs{k}_d)/2$
and
$\bs{K}=(\bs{k}_i+\bs{k}_j)/2$.
In Eq.~\eqref{Omega3ppnum}, the integration order is such that $p$ is integrated after $A$ or $B$.\footnote{Notably, the same expression results 
if one naively introduces principal values in $\Omega_{3,\text{pp}}^\text{reduced}$ and 
averages over three different integration orders (where in one case $p$ is integrated before $A$ or $B$); 
for Eq.~\eqref{Omega2num2a} this procedure would, however, lead to an incorrect result.}
The expression for $\Omega_{3,\text{hh}}^\text{BdD}$ is similar to Eq.~\eqref{Omega3ppnum}.
For $\Omega_{3,\text{ph}}^\text{BdD}$, however,
using relative and average momenta as integration variables leads to
\begin{align}\label{Omega3phnum}
\Omega_{3,\text{ph}}^\text{BdD}&=
\frac{8 M^2}{3} 
\sum_{\bs{K},\bs{p},\bs{A},\bs{Y}}
\zeta^{ijkabc}_\text{ph} n_{ijk} \bar n_{abc}
\breq & \quad \times
\left[\mathcal{F}_{[\eta_1,\eta_2]}+\mathcal{F}_{[-\eta_1,\eta_2]}+\mathcal{F}_{[-\eta_1,-\eta_2]} \right],
\end{align}
where $\bs{p}=(\bs{k}_i-\bs{k}_j)/2$, $\bs{A}=(\bs{k}_a-\bs{k}_b)/2$, $\bs{Y}=(\bs{k}_a-\bs{k}_c)/2$
and
$\bs{K}=(\bs{k}_i+\bs{k}_j)/2$, and
\begin{align}
\mathcal{F}_{[\eta_1,\eta_2]}=
\frac{1}{\big[A^2-p^2+\im\eta_1\big]\, \big[(\bs{p}-\bs{A})\cdot(\bs{A}-2\bs{Y}+\bs{p})+\im\eta_2\big]}.
\end{align}
From here one would have to proceed similar to the steps that lead from Eq.~\eqref{Omega2num2a} to Eq.~\eqref{Omega2num2b}.

\section{Factorization to All Orders}\label{sec4}
Here, 
we prove to all orders that the BdD renormalization scheme implies the 
thermodynamic relations associated with Fermi-liquid theory and (consequently)
leads to a perturbation series
that manifests the concistency of the adiabatic zero-temperature formalism, for both isotropic and anisotropic systems.

First, in Sec.~\ref{sec41}, we examine more closely how the linked-cluster theorem manifests itself.
Second, in Sec.~\ref{sec42} we systematize the disentanglement ($\div$) of the grand-canonical perturbation series. 
These two steps provide the basis for Sec.~\ref{sec43}, where we prove to all orders the reduced
factorization property for finite systems, Eq.~\eqref{reducedfactorized1}.
In Sec.~\ref{sec44} we then infer that the reduced factorization property holds also for the BdD renormalization scheme. This implies the Fermi-liquid relations and the consistency of the adiabatic formalism.
Finally, in Sec.~\ref{sec45} we point out that the BdD renormalization scheme maintains the 
cancellation of the divergencies (at ${T=0}$) from
energy denominator poles and
discuss the minimal renormalization requirement 
for the consistency of the adiabatic formalism
with the modified perturbation series 
for the free energy, $F(T,\mu_\text{ref})$, in the anisotropic case.

\subsection{Linked-cluster theorem}\label{sec41}

\begin{figure}[t] 
\centering
\vspace*{0mm}
\hspace*{-0.0cm}
\includegraphics[width=0.25\textwidth]{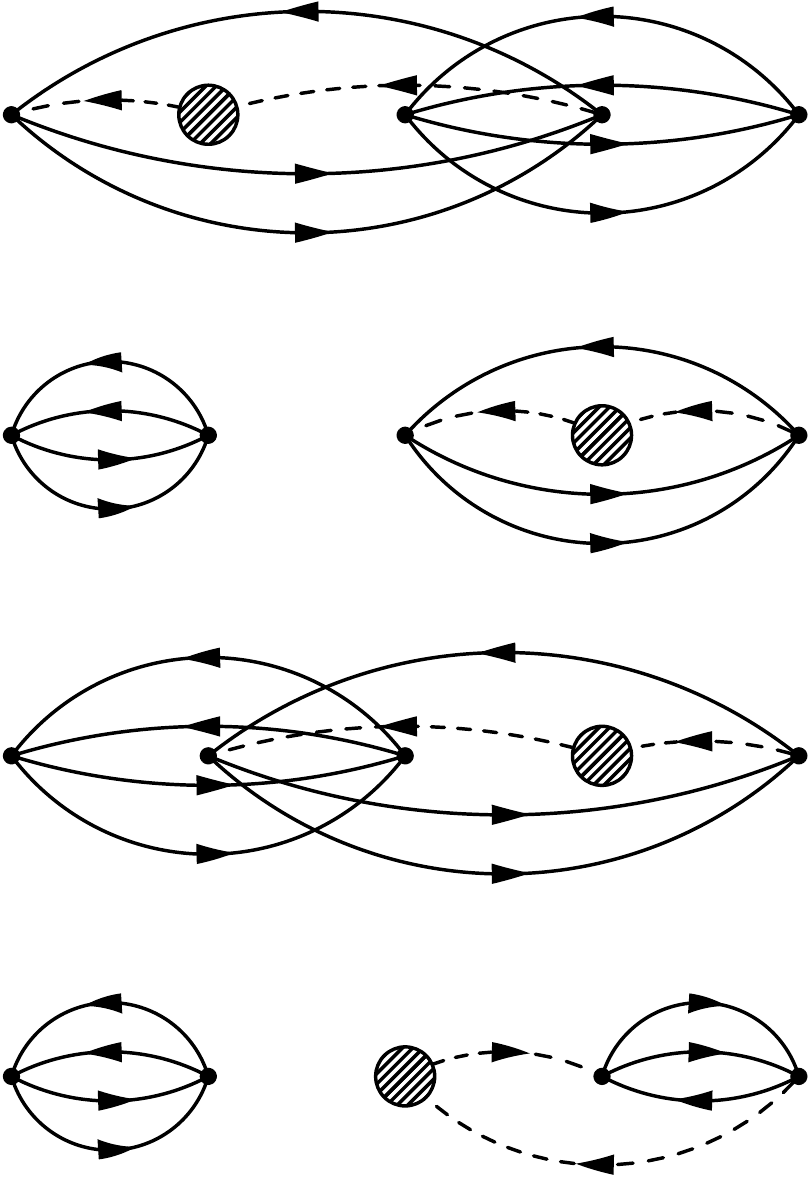} 
\vspace*{-2mm}
\caption{Vertex permutations for an unlinked diagram with two linked parts (subdiagrams).
If the diagram in the first row represents the original vertex ordering, then the second and third diagram correspond to (nonoverlapping and overlapping, respectively) orderings $\in O$, and
the fourth diagram to an ordering $\in P/O$.}
\label{figx}
\end{figure}

Letting the truncation order (formally) go to infinity, 
the sum of all perturbative contributions to $\Omega(T,\mu)$ can be written as
\begin{align} \label{OmegaY}
\Delta \Omega = \sum_{n=1}^\infty \Omega_n
=-\frac{1}{\beta}\ln \left[1-\beta\sum_{n=1}^\infty \Upsilon_n \right],
\end{align}
where $\Upsilon_n$ denotes the contribution of order $n$ from both linked and unlinked diagrams.
We refer to the various linked parts of an unlinked diagram as subdiagrams.
Further, we denote the contribution---evaluated via a given \textit{time-independent} ($\aleph$) formula (i.e., direct, cyclic, or reduced with $\ast$ or $\ast\ast$)---to $\Upsilon_n$ from 
a diagram composed of
$K=k\sum_{i=0}^{k} \alpha_i$
linked parts involving
$k$ different subdiagram species $\Gamma_1\neq \Gamma_2 \neq \ldots \neq \Gamma_k$ where each $\Gamma_i$ appears $\alpha_i$ times in the complete diagram,
by 
$\Upsilon^\aleph_{[\Gamma_{1}^{\alpha_1}\cdots\Gamma_{k}^{\alpha_k}]_n}$.
In this notation, Eq.~\eqref{OmegaY} reads
\begin{align} \label{redfac1}
\Delta\Omega =
-\frac{1}{\beta} \ln \Big[1-\beta  \sum_{n=1}^\infty
\sum_{[\Gamma_{1}^{\alpha_1}\cdots\Gamma_{k}^{\alpha_k}]_n} 
\sum_O
\Upsilon^\aleph_{[\Gamma_{1}^{\alpha_1}\cdots\Gamma_{k}^{\alpha_k}]_n} \Big],
\end{align} 
where $\sum_{[\Gamma_{1}^{\alpha_1}\cdots\Gamma_{k}^{\alpha_k}]_n}$ is the sum over all possible (i.e., those consistent with order $n$) combinations of subdiagrams (including repetitions), and
$\sum_O$ denotes the sum over all distinguishable vertex permutations of the unlinked diagram
that leave the subdiagrams invariant. 
This is illustrated in Fig.~\ref{figx}.
We write
\begin{align} \label{Gammasum}
\sum_{n=1}^\infty   \sum_{[\Gamma_{1}^{\alpha_1}\cdots\Gamma_{k}^{\alpha_k}]_n}
\sum_O
\Upsilon^\aleph_{[\Gamma_{1}^{\alpha_1}\cdots\Gamma_{k}^{\alpha_k}]_n}
=
\sum_{\Gamma_{1}^{\alpha_1}\cdots\Gamma_{k}^{\alpha_k}}
\sum_O
\Upsilon^\aleph_{\Gamma_{1}^{\alpha_1}\cdots\Gamma_{k}^{\alpha_k}}.
\end{align} 
It is
\begin{align} \label{redfac1b}
\sum_{\Gamma_{1}^{\alpha_1}\cdots\Gamma_{k}^{\alpha_k}} 
\sum_O
\Upsilon^\aleph_{\Gamma_{1}^{\alpha_1}\cdots\Gamma_{k}^{\alpha_k}} 
=
\sum_{\widetilde{\Gamma}_{1}^{\alpha_1}\cdots\widetilde{\Gamma}_{k}^{\alpha_k}} 
\sum_P
\Upsilon^\aleph_{\widetilde{\Gamma}_{1}^{\alpha_1}\cdots\widetilde{\Gamma}_{k}^{\alpha_k}} ,
\end{align}
where $\sum_P$ denotes the sum over all distinguishable vertex  orderings, and 
$\sum_{\widetilde{\Gamma}_{1}^{\alpha_1},
\ldots,\widetilde{\Gamma}_{k}^{\alpha_k}}$ sums over 
all combinations of subdiagrams where in the underlying set of linked diagrams $\{\Gamma_i\}$ only one (arbitrary) element is included for  
each set of diagrams that is closed under vertex permutations. For example, among the first two diagrams of Fig.~\ref{fig3ppph} only one is included, and only one of the six diagrams of Fig.~\ref{fig3red}.

The generalization of Eq.~\eqref{OmegaT} for 
$\Upsilon_n$ is given by
\begin{align} \label{OmegadirectT}  
\Upsilon_n^{\text{direct}[P]}=
-\frac{1}{\beta}\frac{(-1)^{n}}{n!} 
\int\limits_{0}^\beta \!
d \tau_n \cdots d \tau_1 \;
\Braket{  \mathcal{T}\big[
\mathcal{V}(\tau_n) \cdots
\mathcal{V}(\tau_1) \big]
}.
\end{align}
We denote the expressions obtained from
Eq.~\eqref{OmegadirectT} for the contribution from a given permutation invariant set of (linked or unlinked) diagrams by 
$\Upsilon^{\text{direct}[P]}_{\widetilde{\Gamma}_{1}^{\alpha_1}\cdots
\widetilde{\Gamma}_{k}^{\alpha_k}}$.
As noted in Sec.~\ref{sec22}, these expressions
are equivalent 
to the summed expressions obtained from any of the time-independent formulas (direct, cyclic, or reduced with $\ast$ or $\ast\ast$), i.e., 
\begin{align} \label{sumuppi}
\Upsilon^{\text{direct}[P]}_{\widetilde{\Gamma}_{1}^{\alpha_1}\cdots
\widetilde{\Gamma}_{k}^{\alpha_k}}
&=
\sum_P\Upsilon^\aleph_{\widetilde{\Gamma}_{1}^{\alpha_1}\cdots
\widetilde{\Gamma}_{k}^{\alpha_k}}.
\end{align}
Now, the number of ways the $n$ perturbation operators in Eq.~\eqref{OmegadirectT} can be partitioned into the subgroups specified by 
$\Upsilon^{\text{direct}[P]}_{\widetilde{\Gamma}_{1}^{\alpha_1}\cdots
\widetilde{\Gamma}_{k}^{\alpha_k}}$ is given by 
\cite{Abrikosov,Fetter}
\begin{align}\label{partitioningcount}
\frac{1}{\alpha_1!
\cdots 
\alpha_k!}\frac{n!}{(n_1!)^{\alpha_1}
\cdots (n_k!)^{\alpha_k}},
\end{align}
where $n_i$ are the orders of the respective subdiagrams.
From Eq.~\eqref{OmegadirectT}, this leads to
\begin{align} \label{redfac2}
\sum_P\Upsilon^\aleph_{\widetilde{\Gamma}_{1}^{\alpha_1}\cdots
\widetilde{\Gamma}_{k}^{\alpha_k}}
&= 
-\frac{1}{\beta} \prod_{i=1}^k
\frac{\left(-\beta \,
\Upsilon^{\text{direct}[P]}_{\widetilde{\Gamma}_{i}}\right)^{\alpha_i}}{\alpha_i!}
\breq &= 
-\frac{1}{\beta} \prod_{i=1}^k
\frac{\left(-\beta \,\sum_{P_i}
\Upsilon^\aleph_{\widetilde{\Gamma}_{i}}\right)^{\alpha_i}}{\alpha_i!},
\end{align}
where in the second step we have applied Eq.~\eqref{sumuppi}.
In Sec.~\ref{sec43} we will see that Eq.~\eqref{redfac2} implies the (direct, cyclic, and reduced) factorization properties 
for anomalous diagrams.

It is now straightforward to verify by explicit comparison with Eq.~\eqref{redfac2} that
\begin{align} \label{redfac3}
 \sum_{\widetilde{\Gamma}_{1}^{\alpha_1}\cdots\widetilde{\Gamma}_{k}^{\alpha_k}} 
\sum_P
\left(-\beta\Upsilon^\aleph_{\widetilde{\Gamma}_{1}^{\alpha_1}\cdots\widetilde{\Gamma}_{k}^{\alpha_k}}\right)
&= 
\sum_{\alpha=1}^\infty \frac{1}{\alpha!}
\bigg[  -\beta 
\sum_{\Gamma}
\Upsilon^\aleph_{\Gamma}
\bigg]^\alpha 
\breq 
&=
-1+\exp\Big( -\beta \sum_{\Gamma} \Upsilon^\aleph_{\Gamma}\Big).
\end{align}
Applying this to Eq.~\eqref{redfac1b} leads to
\begin{align} 
\Delta \Omega &=
\sum_{\Gamma} \Upsilon^\aleph_{\Gamma},
\end{align} 
which constitutes the linked-cluster theorem.
Now, expanding the logarithm in Eq.~\eqref{OmegaY} we find 
\begin{align}\label{Insert2}
\Delta \Omega&=
\sum_{n=1}^\infty  \, \Bigg[ \sum_{\nu,\{n_i\},\{k_i\}}
\beta^{k_1+\ldots+k_\nu-1}
\binom{k_1+\ldots+k_\nu}{k_1,\ldots,k_\nu}
\frac{(\Upsilon_{n_1})^{k_1}\cdots (\Upsilon_{n_\nu})^{k_\nu}}{k_1 +\ldots +k_\nu}\Bigg],
\end{align}
where the inner sum is subject to the constraints 
$\sum_{i=1}^\nu n_i k_i=n$, $1\leq n_1<n_2<\ldots< n_\nu$, $k_i\geq 1$, and $\nu\geq 1$. 
The linked-cluster theorem implies that, 
if evaluated in terms of the usual Wick contraction formalism~\cite{Fetter}, in Eq.~\eqref{Insert2} the contributions 
with $\nu=1$ and $k_1=1$
from unlinked diagrams
are all canceled by 
the contributions with $\nu>1$ or $k_1>1$.
The individual expressions from these canceling terms are not size extensive, i.e., in the thermodynamic limit they
diverge with higher powers of the confining volume.

\subsection{Disentanglement}\label{sec42}

Here, we first introduce the cumulant formalism,\footnote{In the context of MBPT for Fermi systems this method was 
introduced by Brout and Englert~\cite{brout2,PhysRev.115.824} (see also Ref.~\cite{horwitz2}).}
which allows systematizing
the disentanglement ($\div$). 
Then, we show that this formalism provides a new representation and evaluation method for the
contributions associated (in the usual Wick contraction formalism) with anomalous diagrams and the subleading parts of 
Eqs.~\eqref{ndouble1}--\eqref{ndouble4}, etc. 
Finally, we construct and discuss the modified perturbation series for the free energy $F(T,\mu_\text{ref})$.

\subsubsection{Cumulant formalism} 
We define $\mathcal{C}_{i_1 \ldots i_n}$ as the unperturbed ensemble average of a fully-contracted (indicted by paired indices) but not necessarily linked sequence of creation and annihilation operators, i.e.,
\begin{align}\label{chi1def}
\mathcal{C}_{i_1 \ldots i_n}=\braket{a_{i_1}^\dagger a_{i_1} \cdots a_{i_n}^\dagger a_{i_n}},
\end{align}
where some of the index tuples may be identical (articulation lines). 
In Eq.~\eqref{chi1def}, all contractions are of the hole type.
For the case where there are also particles we introduce the notation
\begin{align}\label{chi1defpart}
\mathcal{C}_{i_1 \cdots i_n}^{a_1 \cdots a_m}=\braket{a_{i_1}^\dagger a_{i_1} \cdots a_{i_n}^\dagger a_{i_n}
a_{a_1} a_{a_1}^\dagger \cdots a_{a_m} b_{a_m}^\dagger }.
\end{align}
This can be expressed in terms of functional derivatives of the unperturbed partition function $\mathcal{Y}_{\!\text{ref}}=\Tr[\e^{-\beta(\mathcal{H}_\text{ref}-\mu \mathcal{N})}]$, 
i.e.,\footnote{The number operator is given by $\mathcal{N}=\sum_\bs{k} a_\bs{k}^\dagger a_\bs{k}$.}
\begin{align} \label{chi1defpart2}
\mathcal{C}_{i_1 \cdots i_n}^{a_1 \cdots a_m}&=\frac{1}{\mathcal{Y}_{\!\text{ref}}} 
\frac{\delta}{\delta[-\beta \eps_{i_1}]} \cdots 
\frac{\delta}{\delta[-\beta \eps_{i_n}]}
\breq &\quad\times
\bigg(1-\frac{\delta}{\delta[-\beta \eps_{a_1}]}\bigg)
\cdots 
\bigg(1-\frac{\delta}{\delta[-\beta \eps_{a_m}]}\bigg)\;\mathcal{Y}_{\!\text{ref}}.
\end{align}
This shows that the upper indices can be lowered iteratively, i.e.,
$\mathcal{C}_{i_1 \cdots i_n}^{a_1 \cdots a_m}=\mathcal{C}_{i_1 \cdots i_n}^{a_1 \cdots a_{m-1}}-\mathcal{C}_{i_1 \cdots i_n a_m}^{a_1 \cdots a_{m-1}}$,
which leads to
\begin{align} \label{chi1defpart3}
\mathcal{C}_{i_1 \cdots i_n}^{a_1 \cdots a_m}=
\sum_{\mathcal{P}\subset\{1,\ldots,m\}} (-1)^{|P|}
\mathcal{C}_{i_1 \cdots i_n \{a_k\}_{k \in P}  }.
\end{align}
The cumulants $\mathcal{K}_{i_1 \ldots i_n}$ are defined by
\begin{align}\label{Maverage}
\mathcal{K}_{i_{1}\ldots i_{n}}= \frac{\delta^n \ln \mathcal{Y}_{\!\text{ref}}}{\delta[-\beta \eps_{i_1}] \cdots \delta[-\beta \eps_{i_n}]}.
\end{align}
The relation between the $\mathcal{K}$'s and the $\mathcal{C}$'s is given by \cite{sokal}
\begin{align}  \label{generalG}
\mathcal{C}_{i_1 \cdots i_n}=\sum_{\substack{\mathcal{P}\in\,\text{partitions}\\\text{of}\,\{1,\ldots,n\}}} \prod_{I\in \mathcal{P}}
\mathcal{K}_{  \{i_k\}_{k\in I}  }.
\end{align}
These formulas provide an alternative way (compared 
to the Wick contraction formalism) to evaluate the various contributions $\Upsilon_{[\Gamma_{1}^{\alpha_1}\cdots\Gamma_{k}^{\alpha_k}]_n}$ in Eq.~\eqref{Insert2}.

\subsubsection{Simply connected unlinked diagrams} 

For linked diagrams without articulation lines (i.e., two-particle irreducible diagrams) the contributions from higher 
cumulants have measure zero for infinite systems.
For such diagrams, the (sums of the) contributions from higher cumulants vanish also in the finite case, via exchange antisymmetry. 
This is clear, since these (nonextensive) contributions are absent in the Wick contraction formalism.
For instance, for the first-order diagram $\mathcal{C}_{ij}=\mathcal{K}_{i}\mathcal{K}_{j}+\delta_{ij}\mathcal{K}_{ii}$
the part $\delta_{ij}\mathcal{K}_{ii}$ gives no contribution (by antisymmetry).
Overall, it is
\begin{align}\label{highK1}
i_1\neq i_2\neq  \ldots \neq  i_n:\;\;\;
\mathcal{C}_{i_1\cdots i_n}=\prod_{\nu=1}^n \mathcal{K}_{i_\nu}
\end{align}
for linked diagrams. This means that 
for linked two-particle irreducible diagrams the cumulant formalism leads to the same expressions as the Wick contraction formalism.
For two-particle reducible diagrams, however, there 
are additional size extensive contributions from higher cumulants corresponding to articulation lines with identical three-momenta.
This has the effect that 
for each set of normal articulation lines with identical three-momenta there is only a single distribution function, i.e.,
\begin{align} \label{Gnormal}
\mathcal{C}_{i_1 \cdots i_n j \cdots j}^{a_1 \cdots a_m}&=\mathcal{C}_{i_1 \cdots i_n j}^{a_1 \cdots a_m},
\\ \label{Gnormal2}
\mathcal{C}_{i_1 \cdots i_n}^{a_1 \cdots a_m  b \cdots b}&=\mathcal{C}_{i_1 \cdots i_n }^{a_1 \cdots a_m b},
\end{align}
see Ref.~\cite{Wellenhofer:2017qla}.
For example, $\mathcal{C}_{i_1\ldots i_n jj}=\mathcal{C}_{i_1\ldots i_n}(\mathcal{K}_{j}\mathcal{K}_{j}+\mathcal{K}_{jj})=\mathcal{C}_{i_1\ldots i_n j}$, since 
$\mathcal{K}_{j}\mathcal{K}_{j}+\mathcal{K}_{jj}=n_jn_j+n_j\bar n_j=n_j$.
Equations~\eqref{Gnormal} and \eqref{Gnormal2} together with Eq.~\eqref{chi1defpart3} imply that 
the contributions from
anomalous diagrams are zero:
\begin{align}\label{anomzero}
\mathcal{C}_{i_1 \cdots i_n j \cdots j}^{a_1 \cdots a_m j \cdots j}=0.
\end{align}
The contributions from anomalous diagrams and from the subleading parts of Eqs.~\eqref{ndouble1}--\eqref{ndouble4}, etc. ,
now arise instead from unlinked diagrams (with normal subdiagrams).
That is, for unlinked diagrams composed of $N$ subdiagrams the contributions from higher cumulants 
connecting $N$ lines with distinct three-momenta are size extensive. For example,  
for the case of three first-order subdiagrams with indices $(i,j)$, $(k,l)$, and $(m,n)$ 
one has the contributions
\begin{align}
\delta_{ik}\delta_{jm}
\mathcal{K}_{i i}\mathcal{K}_{j j} \mathcal{K}_{l} \mathcal{K}_{n}, \ldots, \;\; 
\delta_{ik}\delta_{im}
\mathcal{K}_{i i i} \mathcal{K}_{j} \mathcal{K}_{l} \mathcal{K}_{n},\ldots,
\end{align}
where $\mathcal{K}_{i i i}=n_i\bar n_i \bar n_i-n_i n_i \bar n_i$ 
and the ellipses represent terms with other index combinations.
By virtue of the linked-cluster theorem, 
the size extensive contributions from unlinked diagrams 
where not all higher-cumulant indices correspond to different subdiagrams
cancel against the corresponding
terms with $\nu>1$ or $k_1>1$ in Eq.~\eqref{Insert2}.
The remaining size extensive contributions 
from unlinked diagrams are exactly those
where the different (normal) subdiagrams are simply connected via higher cumulants.
This provides 
a new representation for the 
contributions associated (in the Wick contraction formalism) with anomalous diagrams and the contributions not included 
in Eqs.~\eqref{Gnormal} and \eqref{Gnormal2}.

\subsubsection{Modified thermodynamic perturbation series}

There are two methods for the construction of the modified perturbation series for the free energy $F(T,\mu_\text{ref})$.
The first, introduced by Kohn and Luttinger~\cite{Kohn:1960zz}, is based on grand-canonical MBPT; it constructs $F(T,\mu_\text{ref})$ in terms of a 
truncated formal expansion\footnote{The 
mean field is not expanded; i.e., the expansion is performed after $U(T,\mu)$ is replaced by $U(T,\mu_\text{ref})$.
This (and the truncation of the expansion) makes evident that at a given order the modified and the unmodified perturbation series lead to different results; see also 
Sec.\ref{sec23} and Refs.~\cite{PhysRevC.89.064009,Wellenhofer:2017qla}.}
of 
$F(T,\mu)$ about $\mu_\text{ref}$, see Refs.~\cite{Kohn:1960zz,PhysRevC.89.064009,Wellenhofer:2017qla} for details.
The second method, due to Brout and Englert~\cite{brout2}, starts from the canonical ensemble.
In canonical perturbation theory \cite{PhysRev.115.1374,Parry}, Eq.~\eqref{chi1def} is replaced by
\begin{align}
\mathscr{C}_{i_1 \ldots i_n}=\braket{a_{i_1}^\dagger a_{i_1} \cdots a_{i_n}^\dagger a_{i_n}}_{\!\varrho},
\end{align}
where $\braket{\ldots}_{\!\varrho}$ denotes the
unperturbed canonical ensemble average which involves only Fock states $\ket{\Psi_{\!\varrho}}$ with fixed $\varrho=\Braket{\Psi_{\!\varrho}|\mathcal{N}|\Psi_{\!\varrho}}$.
From this, we proceed analogously to the grand-canonical case, with $\mathcal{Y}_{\!\text{ref}}$ replaced by the unperturbed canonical partition function 
$\mathcal{Z}_\text{ref}=\sum_{\Psi_{\!\varrho}}\braket{\Psi_{\!\varrho}|\e^{-\beta \mathcal{H}_\text{ref}}|\Psi_{\!\varrho}}$, i.e., 
the cumulants are now given by
\begin{align}
\mathscr{K}_{i_{1}\ldots i_{n}}= \frac{\delta^n \ln \mathcal{Z}_\text{ref}}{\delta[-\beta \eps_{i_1}] \cdots \delta[-\beta \eps_{i_n}]}.
\end{align}
The decisive new step is now to evaluate the cumulants not directly (which would be practically impossible) but using the Legendre transformation
\begin{align} \label{legendre}
\ln \mathcal{Z}_\text{ref}(T,\varrho)&=\ln \mathcal{Y}_{\!\text{ref}}(T,\mu_\text{ref})- \mu_\text{ref} \frac{\partial \ln \mathcal{Y}_{\!\text{ref}}(T,\mu_\text{ref})}{\partial \mu_\text{ref}},
\end{align}
where $\mu_\text{ref}$ is the chemical potential of an unperturbed grand-canonical system with the same mean fermion number as the fully interacting canonical system, i.e., 
\begin{align}
\varrho=-\frac{1}{\beta} \frac{\partial \ln \mathcal{Y}_{\!\text{ref}}(T,\mu_\text{ref})}{\partial \mu_\text{ref}}
=-\frac{\partial \Omega_\text{ref}(T,\mu_\text{ref})}{\partial \mu_\text{ref}}
=\sum_\bs{k} \tilde{n}_\bs{k},
\end{align}
where $\tilde{n}_\bs{k}$ denotes the Fermi-Dirac  distribution with $\mu_\text{ref}$ as the chemical potential.
With $\varrho$ being fixed, $\varrho=\sum_\bs{k} \tilde{n}_\bs{k}$ determines $\mu_\text{ref}$ as a functional of the spectrum $\varepsilon_\bs{k}$.
From this and Eq.~\eqref{legendre},
the expression for $\mathscr{K}_{i}$ is given by
\begin{align}
\mathscr{K}_{i}
&=
\frac{\delta
\ln \mathcal{Y}_{\!\text{ref}}}{\delta[-\beta \varepsilon_i]}
-\frac{1}{\beta}\frac{\partial 
\ln \mathcal{Y}_{\!\text{ref}}}{\partial \mu_\text{ref}}
\Bigg(\frac{\delta \mu_\text{ref}}{\delta \varepsilon_i}\Bigg)_{\!\varrho}
-\varrho\,\Bigg( \frac{\delta\mu_\text{ref}}{\partial\varepsilon_i}\Bigg)_{\!\varrho}
=
\tilde{n}_{i}. 
\end{align}
The higher $\mathscr{K}$'s can then be determined iteratively, i.e.,
\begin{align}
\mathscr{K}_{i_1 i_2}
&=\Bigg(\frac{\delta \mathscr{K}_{i_1}}{\delta[-\beta \varepsilon_{i_2}]}\Bigg)_{\!\varrho}
\breq 
&=
\delta_{i_1i_2}
\frac{\partial \mathscr{K}_{i_1}}{\partial[-\beta \varepsilon_{i_1}]}
-
\frac{\partial \mathscr{K}_{i_1}}{\partial \mu_\text{ref}}\,
\Bigg[\frac{\partial \varrho}{\partial\mu_\text{ref}}\Bigg]^{-1}\!\!
\frac{\delta \varrho}{\delta[-\beta \varepsilon_{i_2}]}
\breq 
&=
\delta_{i_1i_2} \tilde{n}_{i_1} (1-\tilde{n}_{i_1})
-\frac{\tilde{n}_{i_1} (1-\tilde{n}_{i_1})\tilde{n}_{i_2} (1-\tilde{n}_{i_2})}{\sum_{i} \tilde{n}_{i} (1-\tilde{n}_{i})}.
\end{align}
For $\mathscr{K}_{i_1 i_2 i_3}$ and beyond
there is also a contribution 
where the energy derivative acts on 
$[\partial \varrho/\partial\mu_\text{ref}]^{-1}=\sum_{i} \tilde{n}_{i} (1-\tilde{n}_{i})$, i.e.,\footnote{Note that Eq.\ (B.12) of Ref.~\cite{brout2} is not valid; e.g., it misses the second part of Eq.~\eqref{Kscr3}.} 
\begin{align} \label{Kscr3}
\big[\mathscr{K}_{i_1 i_2 i_3}\big]_{i_1\neq i_2 \neq i_3}
&=
-
\frac{\partial \mathscr{K}_{i_1i_2}}{\partial \mu_\text{ref}}
\Bigg[\frac{\delta \varrho}{\delta\mu_\text{ref}}\Bigg]^{-1}
\frac{\delta \varrho}{\delta[-\beta \varepsilon_{i_3}]}
\breq 
&\quad 
-
\frac{\partial \mathscr{K}_{i_1}}{\beta \mu_\text{ref}}
\frac{\delta \varrho}{\delta[-\beta \varepsilon_{i_2}]}
\frac{\partial}{\partial[-\beta\varepsilon_{i_3}]} \Bigg[\frac{\partial \varrho}{\partial\mu_\text{ref}}\Bigg]^{-1}.
\end{align}
One can show that $\big[\mathscr{K}_{i_1 \ldots i_n}\big]_{i_a \neq i_b \,\forall a,b\in[1,n]}=\mathcal{O}(1/\varrho^{n-1})$, 
see Ref.~\cite{Wellenhofer:2017qla}, 
so the size
extensive contributions from unlinked diagrams
are again given by simply connected diagrams.
For isotropic systems the anomalous parts of these contributions cancel 
at each order in the zero-temperature limit,\footnote{This feature
is expected from indirect arguments~\cite{Luttinger:1960ua,Wellenhofer:2017qla}. The cancellation has been shown 
explicitly to all orders for certain subclasses of diagrams~\cite{Wellenhofer:2017qla}, but no direct proof to all orders 
exists.} thus
\begin{align}
\text{isotropy:}\;\;
F(T,\mu_\text{ref}) \xrightarrow{T\rightarrow 0} E^{(0)}(\en_\F),
\end{align}
with $\mu_\text{ref} \xrightarrow{T\rightarrow 0} \en_\F$.
By construction, within each of the order-by-order renormalization schemes (direct, cyclic, BdD), at each order the 
modified perturbation series $F(T,\mu_\text{ref})$
matches the
grand-canonical perturbation series for the free energy $F(T,\mu)=\Omega(T,\mu)-\mu \,\partial \Omega(T,\mu)/\partial \mu$.
The zero-temperature limit exists however only for the BdD scheme (see Sec.~\ref{sec2}).

\subsection{Factorization theorem(s)}\label{sec43}
Using the direct formula, the cyclic formula, or the reduced formula for finite systems ($\ast$) and
applying the cumulant formalism to Eq.~\eqref{redfac2} leads to 
\begin{align} \label{redfac2cumulant}
\sum_{P/A}\Upsilon^{\text{direct},\div}_{[\widetilde{\Pi}_{1}^{\alpha_1}\cdots
\widetilde{\Pi}_{k}^{\alpha_k}]_n}
&= 
-\frac{1}{\beta} \prod_{i=1}^k
\frac{\left(-\beta \,\sum_{P_i/A_i}
\Upsilon^{\text{direct},\div}_{\widetilde{\Pi}_{i}}\right)^{\alpha_i}}{\alpha_i!},
\\
\label{redfac2cumulantcyc}
\sum_{P/A}\Upsilon^{\text{cyclic},\div}_{[\widetilde{\Pi}_{1}^{\alpha_1}\cdots
\widetilde{\Pi}_{k}^{\alpha_k}]_n}
&= 
-\frac{1}{\beta} \prod_{i=1}^k
\frac{\left(-\beta \,\sum_{P_i/A_i}
\Upsilon^{\text{cyclic},\div}_{\widetilde{\Pi}_{i}}\right)^{\alpha_i}}{\alpha_i!},
\end{align}
\begin{align}\label{redfac7}
\sum_{P/A} \Upsilon^{\text{reduced,}\ast,(\div)}_{[\widetilde{\Pi}_{1}^{\alpha_1}\cdots
\widetilde{\Pi}_{k}^{\alpha_k}]_n}
&= 
-\frac{1}{\beta} \prod_{i=1}^k
\frac{\left(-\beta \,\sum_{P_i/A_i}
\Upsilon^{\text{reduced,}\ast,\div}_{\widetilde{\Pi}_{i}}\right)^{\alpha_i}}{\alpha_i!},
\end{align}
where the $\widetilde{\Pi}_i$ are all normal diagrams, and
$P/A$ excludes those permutations that lead to anomalous diagrams.
The combinatorics (and sign factors) of the higher-cumulant connections matches the combinatorics of the functional derivatives that 
generate the mean-field contributions from 
the perturbative contributions to the grand-canonical potential.
Hence, Eqs.~\eqref{redfac2cumulant} and \eqref{redfac2cumulantcyc} prove the direct and the cyclic version of the factorization property given by Eq.~\eqref{directfactorized} and its cyclic analog, and 
Eq.~\eqref{redfac7} proves the reduced factorization property for finite systems $(\ast)$, Eq.~\eqref{reducedfactorized1}.
Note that Eq.~\eqref{redfac7} implies that in the reduced finite case the pseudoanomalous contributions 
vanish at each order.

The reduced version of the factorization theorem can also be proved as follows.
For a given unlinked diagram where none of the linked parts are overlapping (see Fig.~\ref{figx}), 
the reduced formula has the form
\begin{align} 
\Upsilon^{\text{reduced,}\ast,(\div)}_{[{\Pi}_{1}^{\alpha_1}\cdots
{\Pi}_{k}^{\alpha_k}]_n} 
&\sim  
\underset{z=0}{\text{Res}}  \frac{\e^{-\beta z}}{z} \frac{1}{(-z)^K}
\prod_i\frac{1}{\En_i-z} 
\breq
& \sim 
\beta^{K-1}
\prod_i\frac{1}{\En_i} + \text{extra terms},
\end{align}
where the extra terms are proportional to $\beta^{K-n}$, with $n\in\{2,\ldots,K\}$.
The reduced expressions for unlinked diagrams with overlapping linked parts are composed entirely of such extra terms. 
These extra terms are incompatible with the linked-cluster theorem:
they do not match the temperature dependence of (the 
disentangled reduced expressions) for the corresponding
contributions with $\nu>1$ or $k_1>1$ in Eq.~\eqref{Insert2}.
The extra terms must therefore cancel each other at each order in the sum $\sum_{P/A}$.\footnote{This cancellation is not
always purely algebraic, see Eqs.~\eqref{4pseudoav2} and \eqref{4pseudoav3}.} 
Thus, symbolically we have
\begin{align} \label{redfac7b}
\sum_{P/A}
\Upsilon^{\text{reduced,}\ast,(\div)}_{[\widetilde{\Pi}_{1}^{\alpha_1}\cdots
\widetilde{\Pi}_{k}^{\alpha_k}]_n}
& \sim 
\beta^{K-1} \sum_{P/A}\left[
\prod_i\frac{1}{\En_i} \right],
\end{align}
which is equivalent to Eq.~\eqref{redfac7}.

\subsection{Statistical quasiparticles}\label{sec44}
The energy denominator regularization maintains the linked-cluster theorem.
From the proof of the (reduced) factorization theorem it can be inferred that this suffices to establish
that
\begin{align} \label{redfac7BdD}
\sum_{P/A} \Upsilon^{\text{reduced,}\ast\ast,\div}_{[\widetilde{\Pi}_{1}^{\alpha_1}\cdots
\widetilde{\Pi}_{k}^{\alpha_k}]_n}
&= 
-\frac{1}{\beta} \prod_{i=1}^k
\frac{\left(-\beta \,\sum_{P_i/A_i}
\Upsilon^{\text{reduced,}\ast\ast,\div}_{\widetilde{\Pi}_{i}}\right)^{\alpha_i}}{\alpha_i!},
\end{align}
which (by virtue of the cumulant formalism) implies the BdD factorization property
\begin{align}\label{reducedfactorizedBdD}
\Omega_{n_1+n_2,\text{anomalous}}^{\text{reduced,}\ast\ast,\div}
&= -\frac{\beta}{2} \sum_\bs{k} 
U^{\text{reduced,}\ast\ast,\div}_{n_1,\bs{k}} n_\bs{k} \bar n_\bs{k} \,
U^{\text{reduced,}\ast\ast,\div}_{n_2,\bs{k}}
\breq & \quad \times 
(2-\delta_{n_1,n_2})
,
\end{align}
and similar [i.e., as specified by Eq.~\eqref{redfac7BdD}] for anomalous contributions with several pieces (subdiagrams, in the cumulant formalism).

It is now clear 
how the cancellation between the contributions
from simply connected diagrams composed of $V$ vertices 
and those where also $-U$ vertices are present
works.
For a given simply connected diagram, only the subdiagrams 
with \textit{single} higher-cumulant connections can be replaced by $-U$ vertices, so 
at truncation orders $2N+1$ and $2N+2$ all anomalous contributions are removed 
if the mean field includes all contributions $U^{\aleph,\div}_{n,\bs{k}}$ with $n\leq N$.
However, this does \textit{not} imply consistency with the adiabatic formalism for $U^{\aleph,\div}_{n,\bs{k}}=U^{\text{reduced},\ast\ast,\div}_{n,\bs{k}}$ (irrespective of isotropy), 
since the relation between chemical potential $\mu$ and the fermion number $\varrho$ does not match 
the adiabatic relation $\varrho=\sum_\bs{k} \theta(\en_\F-\en_\bs{k})$.
For the consistency of the grand-canonical and the adiabatic formalism, the BdD mean field must 
include all contributions up to the truncation order; 
only then one preserves the 
thermodynamic relations of the
pure mean-field theory (where ${H=H_0+U}$, with $U\equiv U[n_\bs{k}]$), i.e., the Fermi-liquid relations
\begin{align} \label{StatQP1aFERMILIQ}
\varrho &= 
\sum_\bs{k} n_\bs{k},
\\ \label{StatQP2aFERMILIQ}
S &=-\sum_\bs{k} \big( n_\bs{k}\ln n_\bs{k} +  
\bar n_\bs{k}\ln \bar n_\bs{k}\big) , 
\\  \label{StatQP3aFERMILIQ}
\frac{\delta E}{\delta n_\bs{k}} &=  \en_{\bs{k}} .
\end{align}
These relations are valid for all temperatures.

\subsection{Zero-temperature limit}\label{sec45}

At zero temperature, the energy denominator poles 
are at the boundary of the integration region,\footnote{For an interesting implication of this feature, i.e., 
the singularity at fourth order and ${T=0}$ of the Maclaurin expansion in terms of $x=\mu_\uparrow-\mu_\downarrow$ (or, $x=\varrho_\uparrow-\varrho_\downarrow$) 
for a system of spin one-half fermions with spins $\uparrow$ and $\downarrow$, see Refs.~\cite{PhysRevC.91.065201,PhysRevC.93.055802,Wellenhofer:2017qla}. 
Note however that the statement in Refs.~\cite{PhysRevC.93.055802,Wellenhofer:2017qla} that the convergence radius of the expansion is still zero (instead of just very small) near (but not at) the degenerate limit
appears somewhat questionable. In particular, 
Fig. 6\break of Ref.~\cite{PhysRevC.93.055802} 
should be interpreted not in terms of the radius of convergence but in terms of convergence at $x=\pm 1$.} 
which implies that the contributions 
from two-particle reducible diagrams with several identical energy denominators diverge~\cite{Wellenhofer:2018dwh,Feldman1996}.
For MBPT with ${U=0}$ or ${U=U_1}$ one finds that
the divergent contributions cancel each other at each order.\footnote{See Ref.~\cite{Wellenhofer:2018dwh} [and Eq.~\eqref{T0canc}] for an example of this. We defer a more detailed analysis of these 
cancellations to a future publication.}
This cancellation is maintained in the BdD renormalization scheme:
the cancellation occurs separately for normal contributions, and for the sum 
of the matching contributions the Sokhotski-Plemelj-Fox formula is consistent with the 
${T\rightarrow 0}$ limit.

Notably, the energy denominator regularization is not required 
to construct a thermodynamic perturbation series that is consistent with the adiabatic formalism in the anisotropic case:  
at ${T=0}$, the BdD factorization theorem takes the form 
\begin{align}\label{reducedfactorizedT0}
T=0:\;\;\Omega_{n_1+n_2,\text{anomalous}}
&= -\frac{1}{2} \sum_\bs{k} 
U^{\text{reduced},\ast\ast,\div}_{n_1,\bs{k}} 
\delta(\en_\bs{k}-\mu) 
\breq &\quad \times U^{\text{reduced},\ast\ast,\div}_{n_2,\bs{k}}
(2-\delta_{n,m}),
\end{align}
and similar for anomalous diagrams with several pieces.
(At ${T=0}$, the symbols $\ast\ast$ and $\div$ (and the specification of $\aleph$ to reduced) are not needed for
the separation of normal and anomalous contributions to the grand-canonical potential.)
Thus, as recognized by Feldman \textit{et al.}~\cite{Feldman1999}, to cancel the anomalous contributions 
to $\Omega(T=0,\mu)$
the following mean field is sufficient (for truncation orders below $2N+2$):
\begin{align} \label{Ufermi}
U^{L_\F}_{\bs{k}} = U_{1,\bs{k}} + \sum_{n=2}^N L_\F \left[ U^{\text{reduced,}\ast\ast,\div}_{n,\bs{k}}(T=0,\mu)\right],
\end{align}
where $L_\F $ 
satisfies
$L_\F [g(\bs{k})] = g(\bs{k})$ for $\eps_\bs{k}=\mu$
and is smoothed off away from $\eps_\bs{k}=\mu$. 
There are still anomalous contributions to the particle number, so the adiabatic series is not reproduced.
The renormalization given by Eq.~\eqref{Ufermi} (with $\mu$ replaced by $\mu_\text{ref}$)
is however sufficient for the consistency of the adiabatic formalism
with the
modified perturbation series $F(T,\mu_\text{ref})$ in the anisotropic case.

\section{Conclusion}\label{summary}

In the present paper, we have, substantiating the outline by Balian and de 
Dominicis (BdD)~\cite{statquasi3,statquasi1},\footnote{Other studies regarding the derivation of 
statistical quasiparticle relations can be found in 
Refs.~\cite{nort1,nort2,article,PhysRevA.1.1243,RevModPhys.39.771,PhysRev.153.263,TUTTLE1966510,1964quasip,PhysRev.121.957,PhysRev.124.583}.}
derived a thermodynamic perturbation series for infinite Fermi systems
that (1) is consistent with the adiabatic zero-temperature formalism for both isotropic and anisotropic systems and (2)
satisfies 
at each order and for all temperatures the thermodynamic relations associated with Fermi-liquid theory. 
This result arises, essentially, as a corollary of the linked-cluster theorem.
The proof of (2) [which implies (1)] given here relies, apart from the earlier analysis of the disentanglement ($\div$) conducted by
Balian, Bloch, and de Dominicis \cite{Balian1961529}
and the
outline provided by Balian and de 
Dominicis, on the application of the cumulant formalism (as a systematic method to perform $\div$) introduced by Brout and Englert~\cite{brout2,PhysRev.115.824}.
The statistical quasiparticles associated 
with the thermodynamic Fermi-liquid relations 
are distinguished from the dynamical quasiparticles
associated with the asymptotic stability of the low-lying excited states; in particular, 
the energies of dynamical and statistical quasiparticles are different.

In the perturbation series derived in the present paper the\break 
reference Hamiltonian is renormalized at each order in terms 
of additional contributions to the self-consistent mean-field potential. 
Conceptually, such an order-by-order renormalization is appealing: 
At each new order, not only is new information about interaction effects included, but this information automatically improves the reference point.
Nevertheless, the 
relevance of this perturbation series depends on its convergence rate compared to
the modified perturbation series for the free energy $F(T,\mu_\text{ref})$ with a
fixed reference Hamiltonian; e.g., Hartree-Fock, or the (modified) second-order BdD mean field.
In addition to the complete removal of anomalous contributions, 
the higher-order mean-field contributions lead also to partial
cancellations of normal two-particle reducible contributions. This
suggests that the convergence rate may indeed improve by renormalizing the mean field at each 
order.\footnote{Note also that somewhat similar methods
have been applied with considerable success for (certain) finite systems~\cite{Tichai:2018qge,PhysRevLett.121.032501,PhysRevLett.75.2787}.}
Apart from the question of convergence, beyond second order the practicality of the BdD renormalization scheme is
impeded by the increasingly complicated regularization procedure required for its numerical application.

An alternative renormalization scheme, the direct scheme, was introduced by Balian, Bloch, and de Dominicis \cite{Balian1961529} (and rederived in the present paper, together with 
yet another scheme, the cyclic one). 
The thermodynamic relations resulting from the direct scheme however deviate from the Fermi-liquid relations. 
More severely, for the direct (and the cyclic) scheme the zero-temperature limit does not exist.
The direct scheme
may however still
be useful for numerical calculations close to the 
classical limit. In particular, the corresponding perturbation series reproduces the virial expansion in the classical limit~\cite{Balian1961529b}.
The BdD renormalization scheme is thus mainly targeted at calculations not too far from the degenerate limit, in particular perturbative
nuclear matter calculations (see, e.g., 
Refs.~\cite{BOGNER200559,PhysRevC.82.014314,PhysRevC.83.031301,PhysRevLett.110.032504,PhysRevC.88.025802,PhysRevC.89.025806,
PhysRevC.93.054314,PhysRevC.94.054307,Drischler:2017wtt,
PhysRevC.87.014322,PhysRevC.89.044321,PhysRevC.91.054311,Sammarruca:2018bqh}.
Notably, the statistical quasiparticle relations may be useful 
for the application of the Sommerfeld expansion~\cite{prakash} and
to connect with phenomenological parametrizations~\cite{Yasin:2018ckc}.

In conclusion, future research in the many-fermion problem will 
investigate the perturbation series derived in the present paper.\footnote{More generally, the effect on convergence 
of higher-order contributions to the mean field (in the modified perturbation series for the free energy) will be investigated.}
\\

I thank A. Carbone, C. Drischler, K. Hebeler, J. W. Holt, F. Hummel, N. Kaiser, R. Lang, M. Prakash, S. Reddy, A. Schwenk and W. Weise for useful discussions.
Moreover, I thank the referees for helpful comments.
Finally, I thank for their warm hospitality the group T39 (TU M\"unchen), the INT (Seattle), the CEA (Saclay) and the ECT* (Trento), where parts of this work have been presented.
This work is supported by the  Deutsche Forschungsgemeinschaft (DFG, German Research Foundation) -- Projektnummer 279384907-- SFB 1245 as well as the DFG and NSFC through the CRC 110 ``Symmetries and the Emergence of Structure in QCD''.

\appendix
\section{Two-particle reducible diagrams at fourth order}\label{app1}
\begin{figure}[t] 
\centering
\vspace*{0mm}
\hspace*{-0.0cm}
\includegraphics[width=0.48\textwidth]{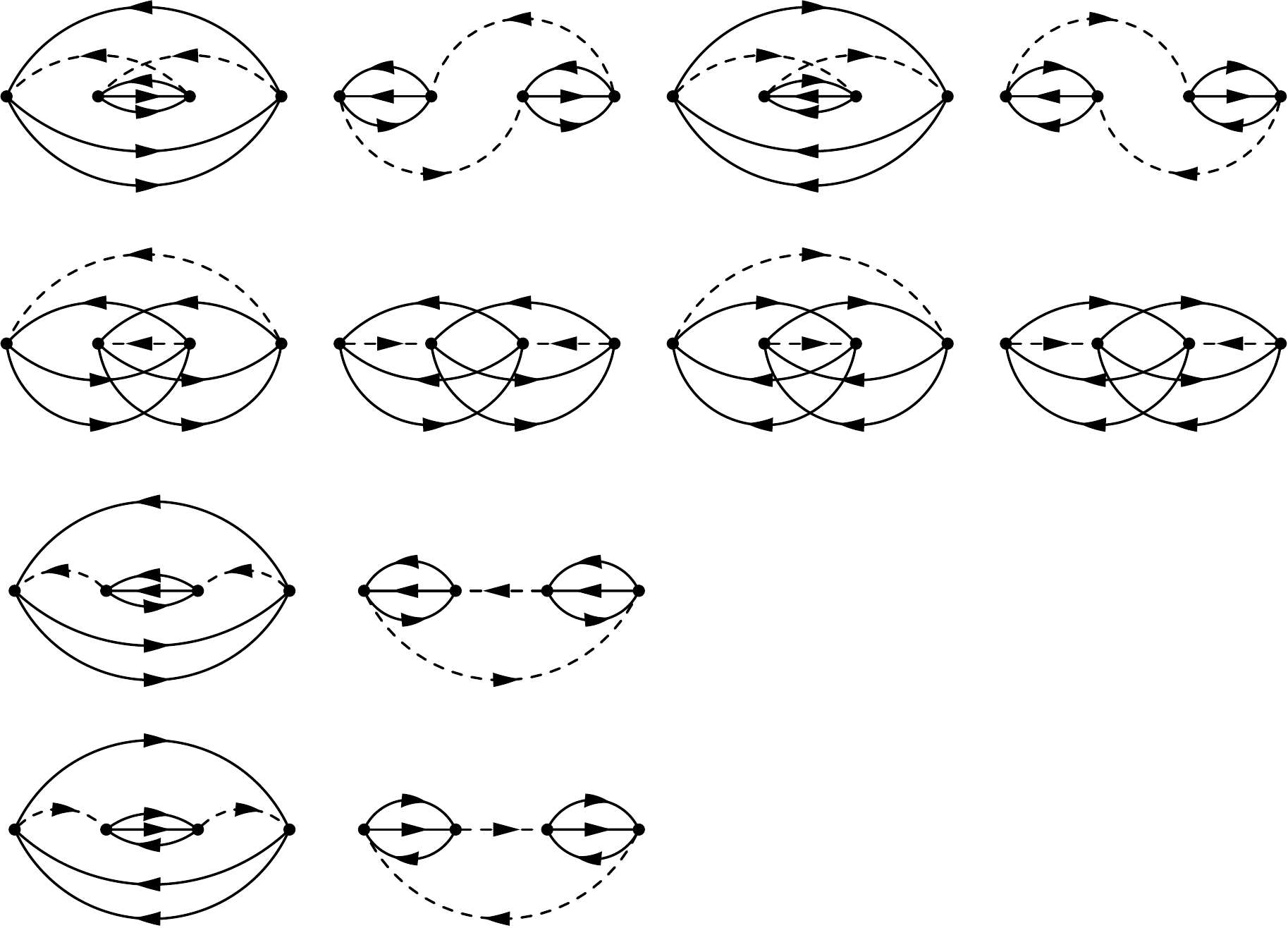} 
\vspace*{-4mm}
\caption{The 12 fourth-order two-particle reducible diagrams composed of two second-order normal pieces.
In each of the four rows (1, 2, 3, 4), the first (and third) diagram is a normal diagram, the other anomalous.
The diagrams in each row
transform into each other under cyclic vertex permutations.
The set of all 12 diagrams is closed under general vertex permutations.
}
\label{fig4}
\end{figure}

Here we derive explicitly 
the regularized ($\ast\ast$) disentangled ($\div$) reduced expressions for the contributions from two-particle reducible diagrams at fourth order. 
Diagrams with single-vertex loops are canceled by the ones with $-U_1$ vertices; 
the remaining diagrams with $V$ vertices only are shown in Fig.~\ref{fig4}.
One can choose indices such that 
for each diagram
the matrix elements are given by
\begin{align}
\zeta=
V^{ij,ab}V^{ik,cd}V^{cd,ki}V^{ab,ij},
\end{align}
and
the energy denominators corresponding to the two second-order pieces are given by
$D_1=D_{ab,ij}$ and $D_2=D_{cd,ik}$.
The cyclic expression for the sum of the diagrams in each row $\nu \in\{1,2,3,4\}$ can then be written as
\begin{align}
\Omega^\text{cyclic}_{4,\nu}
=
\xi_{\nu} \sum_{ijkabcd}
\zeta \;
\mathcal{N}_{\nu} \, 
\mathcal{F}_{\nu}^\text{cyclic},
\end{align}

\noindent
where $\xi_{1,2,3,4}=(-1/4,-1/4,1/8,1/8)$.
For the chosen indices the $\mathcal{N}_\nu$ are fixed as
$\mathcal{N}_{1}=\mathcal{N}_{2}= n_{iijk} \bar{n}_{abcd}$,\break
$\mathcal{N}_{3}= n_{iijcd} \bar{n}_{abk}$, and
$\mathcal{N}_{4}= n_{abk} \bar{n}_{iijcd}$.
Finally, from
Eq.~\eqref{cyclic}, the $\mathcal{F}^\text{cyclic}_\nu$ are given by
\begin{align}
\mathcal{F}_{1}^\text{cyclic}
&=
\frac{1}{D_1^2 D_{1+2}}
-\frac{\e^{-\beta (D_{1+2})}}{D_2^2 D_{1+2}}
+\frac{\e^{-\beta D_1}D_{1-2}}{D_1^2 D_2^2}
-\beta\frac{\e^{-\beta D_1}}{D_1 D_2},
\\
\mathcal{F}_{2}^\text{cyclic}
&=
\frac{1}{D_1D_2 D_{1+2}}
-\frac{\e^{-\beta (D_{1+2})}}{D_1D_2 D_{1+2}}
+\frac{\e^{-\beta D_1}}{D_1 D_2 D_{1-2}}
\breq &\quad
-\frac{\e^{-\beta D_2}}{D_1 D_2 D_{1-2}},
\\
\mathcal{F}_{3}^\text{cyclic}
&=
\frac{1}{D_1^2 D_{1-2}}
-\frac{\e^{-\beta (D_{1-2})}}{D_2^2 D_{1-2}}
+\frac{\e^{-\beta D_1}D_{1+2}}{D_1^2 D_2^2}
+\beta\frac{\e^{-\beta D_1}}{D_1 D_2},
\\
\mathcal{F}_{4}^\text{cyclic}
&=
-\frac{1}{D_1^2 D_{1-2}}
+\frac{\e^{\beta (D_{1-2})}}{D_2^2 D_{1-2}}
-\frac{\e^{\beta D_1}D_{1+2}}{D_1^2 D_2^2}
+\beta\frac{\e^{\beta D_1}}{D_1 D_2},
\end{align}
where $D_{1\pm 2}=D_1 \pm D_2$. Although their individual parts have poles, the $\mathcal{F}^\text{cyclic}_{\nu}$ are regular for any zero of
$D_1 D_2 D_{1+2} D_{1-2}$.
To separate the various parts, we 
add
infinitesimal imaginary parts
to the energy denominators, i.e.,
\begin{align}
\En_{1} \rightarrow \,& \Ens_{1,\eta_{1}}=\En_{1} +\im\eta_{1},
\\
\En_{2} \rightarrow \,& \Ens_{2,\eta_{2}}=\En_{2} +\im\eta_{2},
\end{align}
where $|\eta_1| \neq |\eta_2|$, since otherwise $\Ens_{1+2,\eta_{1+2}}$ or $\Ens_{1-2,\eta_{1-2}}$ has zeros.
Averaging over the signs of the imaginary parts and
applying Eqs.~\ref{ndouble} and \eqref{Dexp}, we can reorganize the sum of the 12 diagrams according to
\begin{align}
\sum_{\nu=1}^4\Omega^\text{cyclic}_{4,\nu}
=
\Omega_{4,\text{normal}}^{\text{reduced,}\ast\ast,\div} + 
\Omega_{4,\text{anom.}}^{\text{reduced,}\ast\ast,\div} + 
\Omega_{4,\text{pseudo-a.}}^{\text{reduced,}\ast\ast,\div},
\end{align}

\noindent
where 
\begin{align} \label{4normal}
\Omega_{4,\text{normal}}^{\text{reduced,}\ast\ast,\div} = & 
\sum_{\alpha=1}^4 
\bigg[
\frac{1}{8} \sum_{ijkabcd} \!\!
\zeta \; 
\mathcal{N}^\text{normal}_{\alpha}\;
\mathcal{F}^{\text{reduced},\ast\ast}_{\alpha,\text{normal}} 
\bigg]
,
\\ \label{4anom}
\Omega_{4,\text{anom.}}^{\text{reduced,}\ast\ast,\div} = & 
\sum_{\alpha=1,3,4} 
\bigg[
\frac{\beta}{8} \sum_{ijkabcd} \!\!
\zeta \; 
\mathcal{N}^\text{anom.}_{\alpha} \;
\mathcal{F}^{\text{reduced},\ast\ast}_{\alpha,\text{anom.}} 
\bigg]
, 
\\ \label{4pseud}
\Omega_{4,\text{pseudo-a.}}^{\text{reduced,}\ast\ast,\div} = & 
\sum_{\alpha=1,3,4}  
\bigg[
\frac{1}{8} \sum_{ijkabcd} \!\!
\zeta \; 
\mathcal{N}^\text{anom.}_{\alpha}\;
\mathcal{R}^{\text{reduced},\ast\ast}_{\alpha}
\bigg]
,
\end{align}
with 
\begin{align}
\mathcal{F}^{\text{reduced},\ast\ast}_{\alpha,\text{normal}}
=
\!\!\!\!\!  \sum_{\text{sgn}(\eta_{1}),\text{sgn}(\eta_{2})} \!\!
\mathcal{F}^{\text{reduced},\ast\ast}_{\alpha,\text{normal},[\eta_1,\eta_2]}
,
\end{align}
and similar for the anomalous and pseudoanomalous contributions.
The correspondence $\alpha \cong \nu$ holds only 
for the anomalous contributions, and the normal ones with $\alpha=3,4$.
For the normal contributions with $\alpha=1,2$, we combine the (disentangled) contributions from the first two ($\alpha=1$) and the third two ($\alpha=2$) diagrams 
in the first two rows.
Regarding the pseudoanomalous contributions, each $\alpha$ 
corresponds to several $\nu$'s, 
by virtue of the application of Eq.~\eqref{ndouble}.
In the anomalous contribution
\begin{align} \label{Nanom1}
\mathcal{N}^\text{anom.}_1&= n_{iabk} \bar{n}_{ijcd},\\ \label{Nanom2}
\mathcal{N}^\text{anom.}_3&= n_{iabcd} \bar{n}_{ijk},\\ \label{Nanom3}
\mathcal{N}^\text{anom.}_4&= n_{ijk} \bar{n}_{iabcd},
\end{align}

\vspace*{-0.2cm}

\noindent
and
\begin{align}
\mathcal{F}^{\text{reduced},\ast\ast}_{1,\text{anom.}}
&=\!\!\!\!\!\!
\sum_{\text{sgn}(\eta_{1}),\text{sgn}(\eta_{2})} \!\!
\left[-\frac{2}{\Ens_{1,\eta_1}\Ens_{2,\eta_2}}\right]
,
\\
\mathcal{F}^{\text{reduced},\ast\ast}_{3,\text{anom.}}
&=\!\!\!\!\!\!
\sum_{\text{sgn}(\eta_{1}),\text{sgn}(\eta_{2})} \!\!\frac{1}{\Ens_{1,\eta_1}\Ens_{2,\eta_2}},
\\
\mathcal{F}^{\text{reduced},\ast\ast}_{4,\text{anom.}}
&=\!\!\!\!\!\!
\sum_{\text{sgn}(\eta_{1}),\text{sgn}(\eta_{2})} \!\!\frac{1}{\Ens_{1,\eta_1}\Ens_{2,\eta_2}}.
\end{align}
Suitably relabeling indices, we obtain the BdD factorization property
\begin{align} \label{4thorderfactorized}
\Omega_{4,\text{anom.}}^{\text{reduced,}\ast\ast,\div}
=-\frac{\beta}{2}\sum_i  U_{2,i}^{\text{reduced,}\ast\ast,(\div)}   n_i \bar n_i \, U_{2,i}^{\text{reduced,}\ast\ast,(\div)},
\end{align}
with $U_{2,i}^{\text{reduced,}\ast\ast,(\div)}$ given by Eq.~\eqref{Omega2d}. 
Relabeling indices
according to Eqs.~\eqref{Nanom1}, \eqref{Nanom2}, and \eqref{Nanom3},
the energy denominators in the pseudoanomalous contribution
are
\begin{align}
\mathcal{R}^{\text{reduced},\ast\ast}_{1,[\eta_1,\eta_2]}
&=
-\frac{2}{\Ens_{1,\eta_1}(\Ens_{2,\eta_2})^2} 
+\frac{2}{(\Ens_{1,\eta_1})^2\Ens_{2,\eta_2}}
\breq & \quad
-\frac{2}{\Ens_{1,\eta_1}\Ens_{2,\eta_2}\Ens_{1-2,\eta_{1-2}}}
-\frac{2}{\Ens_{1,\eta_2}\Ens_{2,\eta_1}\Ens_{1-2,\eta_{2-1}}}
\breq & \quad
+\frac{1}{(\Ens_{2,\eta_1})^2\Ens_{1-2,\eta_{2-1}}}
+\frac{1}{(\Ens_{2,\eta_2})^2\Ens_{1-2,\eta_{1-2}}}
\breq & \quad
+\frac{1}{(\Ens_{1,\eta_1})^2\Ens_{1-2,\eta_{1-2}}}
+\frac{1}{(\Ens_{1,\eta_2})^2\Ens_{1-2,\eta_{2-1}}}
,
\\
\mathcal{R}^{\text{reduced},\ast\ast}_{3,[\eta_1,\eta_2]}
&=
-\frac{2}{(\Ens_{2,\eta_2})^2\Ens_{1+2,\eta_{1+2}}}
-\frac{2}{\Ens_{1,\eta_1}\Ens_{2,\eta_2}\Ens_{1+2,\eta_{1+2}}}
\breq & \quad
+\frac{1}{\Ens_{1,\eta_1}(\Ens_{2,\eta_2})^2} 
+\frac{1}{(\Ens_{1,\eta_1})^2\Ens_{2,\eta_2}}
,
\\
\mathcal{R}^{\text{reduced},\ast\ast}_{4,[\eta_1,\eta_2]}
&=
\frac{2}{(\Ens_{2,\eta_2})^2\Ens_{1+2,\eta_{1+2}}}
+\frac{2}{\Ens_{1,\eta_1}\Ens_{2,\eta_2}\Ens_{1+2,\eta_{1+2}}}
\breq & \quad
-\frac{1}{\Ens_{1,\eta_1}(\Ens_{2,\eta_2})^2} 
-\frac{1}{(\Ens_{1,\eta_1})^2\Ens_{2,\eta_2}}.
\end{align}
In these expressions, the parts with 
three different denominators require special attention: 
the formal application of the Sokhotski-Plemelj-Fox formula
assumes that each energy denominator is 
used as an explicit integration variable, but 
this is not possible for terms with denominators of the form $\Ens_{1,\eta_1}\Ens_{2,\eta_2}\Ens_{1\pm 2,\eta_{1\pm 2}}$.
To evaluate these terms, we use the relations
\begin{align} \label{combineD}
\frac{1}{(\Ens_{1,\eta_1})^2\Ens_{1\pm 2,\eta_{1\pm 2}}}\pm
\frac{1}{\Ens_{1,\eta_1}\Ens_{2,\eta_2}\Ens_{1\pm 2,\eta_{1\pm 2}}}
&=
\pm \frac{1}{(\Ens_{1,\eta_1})^2\Ens_{2,\eta_{2}}},
\\ \label{combineD2}
\frac{1}{(\Ens_{2,\eta_2})^2\Ens_{1\pm 2,\eta_{1\pm 2}}}\pm
\frac{1}{\Ens_{1,\eta_1}\Ens_{2,\eta_2}\Ens_{1\pm 2,\eta_{1\pm 2}}}
&=
\pm\frac{1}{\Ens_{1,\eta_1}(\Ens_{2,\eta_{2}})^2}.
\end{align}
This leads to

\begin{align}
\mathcal{R}^{\text{reduced},\ast\ast}_{1}
&=0
,
\\ \label{4pseudoav2}
\mathcal{R}^{\text{reduced},\ast\ast}_{3}
&=\!\!\!
\sum_{\text{sgn}(\eta_{1}),\text{sgn}(\eta_{2})} 
\left[
-\frac{1}{\Ens_{1,\eta_1}(\Ens_{2,\eta_2})^2} 
+\frac{1}{(\Ens_{1,\eta_1})^2\Ens_{2,\eta_2}}
\right]
,
\\ \label{4pseudoav3}
\mathcal{R}^{\text{reduced},\ast\ast}_{4}
&=\!\!\!
\sum_{\text{sgn}(\eta_{1}),\text{sgn}(\eta_{2})}
\left[
\frac{1}{\Ens_{1,\eta_1}(\Ens_{2,\eta_2})^2} 
-\frac{1}{(\Ens_{1,\eta_1})^2\Ens_{2,\eta_2}}
\right]
.
\end{align}
One sees that
$\mathcal{R}^{\text{reduced},\ast\ast}_{3}$ and $\mathcal{R}^{\text{reduced},\ast\ast}_{4}$ are antisymmetric under $D_1 \leftrightarrow D_2$.
In each case, the remaining 
part of the integrand is symmetric under $D_1 \leftrightarrow D_2$.
Thus, the pseudoanomalous contribution is zero:
\begin{align}
\Omega_{4,\text{pseudo-a.}}^{\text{reduced,}\ast\ast,\div}
=0.
\end{align}
Finally, in the normal contribution
\begin{align} \label{Nnorm1}
\mathcal{N}^\text{normal}_1&= n_{ijk} \bar{n}_{abcd},\\ \label{Nnorm2}
\mathcal{N}^\text{normal}_2&= n_{abcd} \bar{n}_{ijk},\\ \label{Nnorm3}
\mathcal{N}^\text{normal}_3&= n_{ijcd} \bar{n}_{abk},\\ \label{Nnorm4}
\mathcal{N}^\text{normal}_4&= n_{abk} \bar{n}_{ijcd},
\end{align}
and

\begin{align} \label{Enorm1}
\mathcal{F}^{\text{reduced},\ast\ast}_{1,\text{normal},[\eta_1,\eta_2]}
&=
-\frac{2}{(\Ens_{1,\eta_1})^2\Ens_{1+2,\eta_{1+2}}}
\breq & \quad
-\frac{2}{\Ens_{1,\eta_1}\Ens_{2,\eta_2}\Ens_{1+2,\eta_{1+2}}}
,
\\ \label{Enorm2}
\mathcal{F}^{\text{reduced},\ast\ast}_{2,\text{normal},[\eta_1,\eta_2]}
&=
\frac{2}{(\Ens_{2,\eta_2})^2\Ens_{1+2,\eta_{1+2}}}
+\frac{2}{\Ens_{1,\eta_1}\Ens_{2,\eta_2}\Ens_{1+2,\eta_{1+2}}}
,
\\ \label{Enorm3}
\mathcal{F}^{\text{reduced},\ast\ast}_{3,\text{normal},[\eta_1,\eta_2]}
&=
\frac{1}{(\Ens_{1,\eta_1})^2\Ens_{1-2,\eta_{1-2}}}
+\frac{1}{(\Ens_{1,\eta_2})^2\Ens_{1-2,\eta_{2-1}}}
,
\\ \label{Enorm4}
\mathcal{F}^{\text{reduced},\ast\ast}_{4,\text{normal},[\eta_1,\eta_2]}
&=-\frac{1}{(\Ens_{1,\eta_1})^2\Ens_{1-2,\eta_{1-2}}}
-\frac{1}{(\Ens_{1,\eta_2})^2\Ens_{1-2,\eta_{2-1}}}
,
\end{align}
where we have suitably relabeled indices.
Applying Eqs.~\eqref{combineD} and \eqref{combineD2}, the averaged expressions are given by
\begin{align} \label{Enorm1av}
\mathcal{F}^{\text{reduced},\ast\ast}_{1,\text{normal}}
&=\!\!\!
\sum_{\text{sgn}(\eta_{1}),\text{sgn}(\eta_{2})}
\left[-\frac{2}{(\Ens_{1,\eta_1})^2\Ens_{2,\eta_{2}}}\right]
,
\\ \label{Enorm2av}
\mathcal{F}^{\text{reduced},\ast\ast}_{2,\text{normal}}
&=\!\!\!
\sum_{\text{sgn}(\eta_{1}),\text{sgn}(\eta_{2})}
\frac{2}{\Ens_{1,\eta_{1}}(\Ens_{2,\eta_2})^2}
,
\\ \label{Enorm3av}
\mathcal{F}^{\text{reduced},\ast\ast}_{3,\text{normal}}
&=\!\!\!
\sum_{\text{sgn}(\eta_{1}),\text{sgn}(\eta_{2})}
\frac{1}{(\Ens_{1,\eta_1})^2\Ens_{1-2,\eta_{1-2}}}
,
\\ \label{Enorm4av}
\mathcal{F}^{\text{reduced},\ast\ast}_{4,\text{normal}}
&=\!\!\!
\sum_{\text{sgn}(\eta_{1}),\text{sgn}(\eta_{2})} 
\left[-\frac{1}{(\Ens_{1,\eta_1})^2\Ens_{1-2,\eta_{1-2}}}\right]
.
\end{align}
In addition to the contribution from the 12 diagrams shown in Fig.~\ref{fig4}, in the BdD scheme
the two-particle reducible contribution at fourth order 
involves the six diagrams of Fig.~\ref{fig3red} with 
the first-order subdiagrams replaced by $-U_{2}^\text{BdD}$ vertices, and 
the diagram composed of two $-U_{2}^\text{BdD}$ vertices (Fig.~\ref{fig4UU}).
The anomalous contributions from these 19 diagrams cancel each other (as a consequence of Eq.~\eqref{4thorderfactorized}).
Notably, there is also a partial analytic cancellation between the contributions 
from Eqs.~\eqref{Enorm1av} and \eqref{Enorm2av}
and the normal contribution from the third-order diagrams with one $-U_{2}^\text{BdD}$ vertex (Fig.~\ref{fig3red}), see Refs.~\cite{Becker:1971asg,Jones:1970hwh,Wellenhofer:2017qla}.
Such partial cancellations
can be found also at higher orders for the normal contribution from certain (normal) two-particle reducible diagrams, i.e., for those where cutting the articulation lines and closing them 
such that an unlinked diagram (with two linked parts) is generated leaves
the number of holes invariant.\footnote{We defer a more detailed analysis of these partial analytic cancellations
to a future publication.}

Finally, the two-particle reducible fourth-order contribution to $U^\text{BdD}$ is 
given by the functional derivative of the regularized disentangled reduced normal contributions
from the diagrams of Fig.~\ref{fig4} and the ones of Fig.~\ref{fig3red} with 
the first-order subdiagrams replaced by $-U_{2}^\text{BdD}$ vertices.\footnote{Here, the functional derivative is supposed to 
disregard the implicit dependence on $n_\bs{k}$ of $U_{2}^\text{BdD}$; see Sec.~\ref{sec23}.}

\section{Self-energy, mass function, mean field and all that}\label{app2}

Here, we discuss the various forms of the self-energy, and their relation to the grand-canonical potential, the
mean occupation numbers, and the (various forms of the) mean field.

\subsection{Analytic continuation(s) of the Matsubara self-energy}\label{app20}

Although it is defined in terms of the self-consistent Dyson equation, the proper Matsubara self-energy $\Xi_\bs{k}(z_l)$ can 
also be calculated using bare propagators; in that case, also two-particle reducible self-energy diagrams
contribute to $\Xi_\bs{k}(z_l)$; see, e.g., Ref.~\cite{PLATTER2003250}.
From $\Xi_\bs{k}(z_l)$
the frequency-space self-energy $\Sigma_\bs{k}(z)$ is 
obtained as the analytic continuation of $\Xi_\bs{k}(z_l)$ that has the following properties:\footnote{See the second paragraph of part \ref{app21} of this Appendix.}
\begin{enumerate}[start=1,label={(\arabic*)}]
\item $\Sigma_\bs{k}(z)$ is analytic 
in both the upper and lower half plane, vanishes at infinity, and has a branch cut along the real axis 
where $\text{Im}[\Sigma_\bs{k}(z)]$ changes sign, with ${\text{Im}[\Sigma_\bs{k}(z)]\lessgtr 0}$ for ${\text{Im}[z] \gtrless 0}$.
\end{enumerate}
With these properties, $\Sigma_\bs{k}(z)$ leads to the spectral representation of the mean occupation number, see Eq.~\eqref{fspectralrep} below.

Now, as shown below, in bare MBPT another analytic continuation of the Matsubara self-energy $\Xi_\bs{k}(z_l)$ can be defined, here referred to as the mass function $\mathcal{M}_\bs{k}(z)$.
It has the following properties:
\begin{enumerate}[start=1,label={(\arabic*)}]
\item $\mathcal{M}_\bs{k}(z)$ is entire and real on the real axis for ${T\neq 0}$.
\item It vanishes at infinity, except for $1/\text{Re}[z]=0^{+}$ where it has an essential singularity.
\item For $\text{Re}[z]>\mu$ it has an essential singularity at ${T=0}$. 
\end{enumerate}
If $z=\en_\bs{k}+\mathcal{M}_\bs{k}(z)$ has 
no solutions off the real axis, then 
one can obtain from $\mathcal{M}_\bs{k}(z)$ another simple
expression for the mean occupation numbers:
the mass function representation $f_\bs{k}=n(\mathscr{E}_\bs{k})$, see Eq.~\eqref{fMass}. The ${T\rightarrow 0}$ limit of this representation
is singular for $\mathscr{E}_\bs{k}>\mu$, and gives $f_\bs{k}=n_\bs{k}$ for $\mathscr{E}_\bs{k}<\mu$.
The issue whether $z=\en_\bs{k}+\mathcal{M}_\bs{k}(z)$ 
may have nonreal solutions is discussed further below. 
We did not see an argument that guarantees that nonreal solutions exist.

The functional forms of the bare perturbative contributions to $\Xi_\bs{k}(z_l)$, $\Sigma_\bs{k}(z)$, and $\mathcal{M}_\bs{k}(z)$ are 
related to ones of the different time-independent formulas ($\aleph$) for the perturbative contributions
to the grand-canonical potential, $\Omega_n^\aleph$. 
This is examined in part \ref{app23} of this Appendix.

\subsection{Mean occupation numbers from Dyson equation}\label{app21}

Here, we first derive
the mass function representation 
for the mean occupation number,
Eq.~\eqref{fMass}. 
Then, we derive the spectral representation Eq.~\eqref{fspectralrep}.
Only the spectral representation
can be derived also from the real-time propagator.\footnote{In that sense, the mass function representation 
(as well as the direct representation of part \ref{app22})
represents a purely statistical result, 
while the spectral representation corresponds to a statistical-dynamical result.
Only the statistical-dynamical result has a well-behaved ${T\rightarrow 0}$ limit.}
Last, we examine the relation between the collision self-energy $\Sigma^{\text{coll}}_\bs{k}(\omega)$ and the frequency-space self-energy $\Sigma_\bs{k}(z)$ at ${T=0}$.

\subsubsection{Mass function}
\begin{figure}[t] 
\centering
\vspace*{0mm}
\hspace*{-0.0cm}
\includegraphics[width=0.45\textwidth]{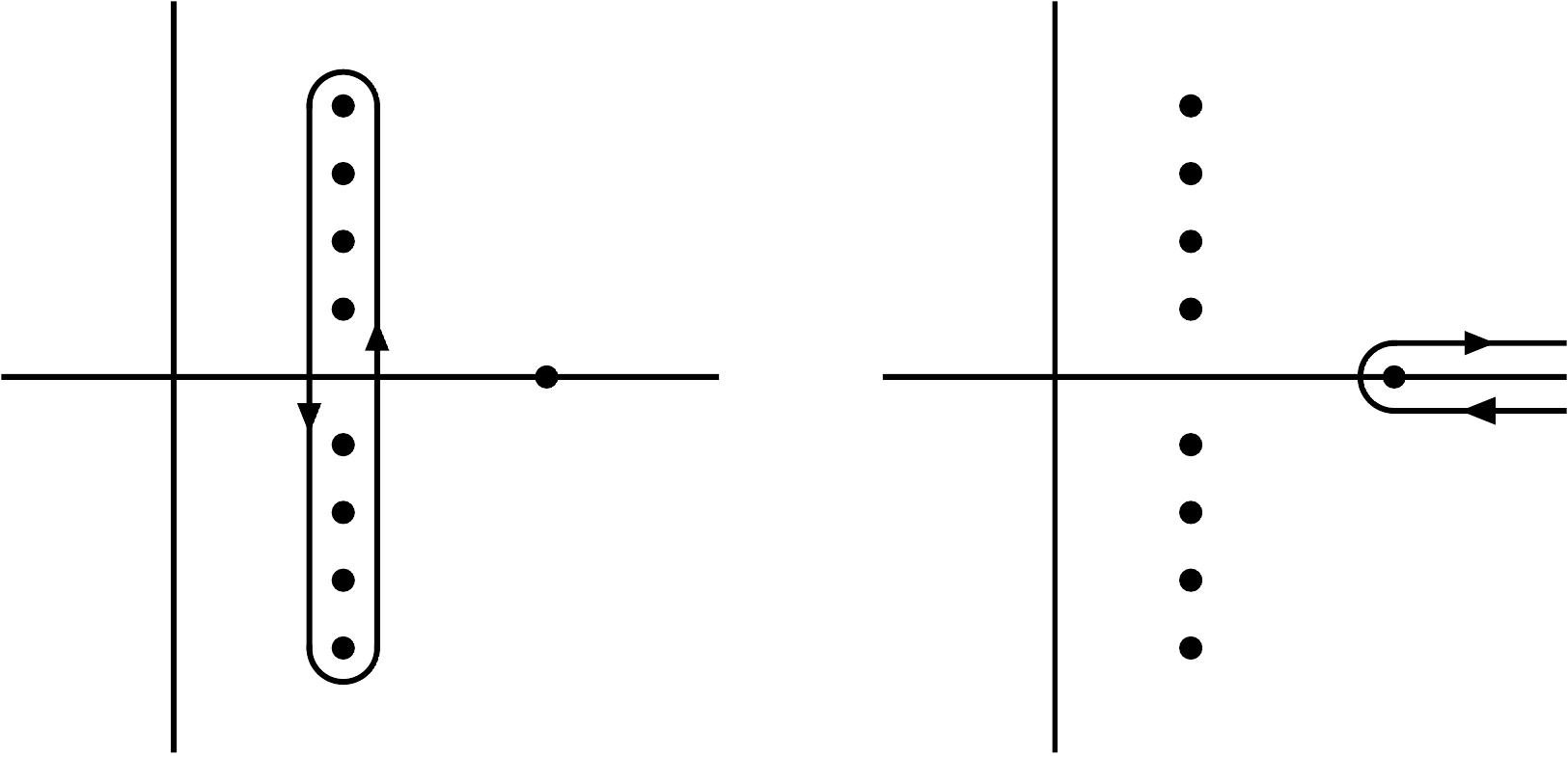} 
\vspace*{-2mm}
\caption{Contours $C_0[l_\text{max}]$ (left panel) and $C_1$ (right panel).}
\label{figc1}
\end{figure}

The
imaginary-time propagator is given by
\begin{align}
\mathscr{G}_\bs{k}(\tau-\tau')=-\Braket{\! \Braket{ \mathcal{T}\left[a_\bs{k}(\tau) a_\bs{k}^\dagger(\tau') \right] } \!},
\end{align}
where $a_\bs{k}(\tau)=a_\bs{k}\e^{-\en_\bs{k} \tau}$ and $a^\dagger_\bs{k}(\tau)=a^\dagger_\bs{k}\e^{\en_\bs{k} \tau}$,
and $\braket{\!\braket{\ldots}\!}$ is the true ensemble average.
Its Fourier series is
\begin{align} \label{Gfourierser}
\mathscr{G}_\bs{k}(\tau)=\frac{1}{\beta} 
\lim_{l_\text{max} \rightarrow \infty}
\sum_{l\in \mathcal{L}(l_\text{max}) } \mathscr{G}_\bs{k}(z_l) \e^{-z_l\tau},
\end{align}
where $\mathcal{L}(l_\text{max})=\{-l_\text{max},\ldots,l_\text{max}\}$, and $z_l$ are the Matsubara frequencies (see Eq.~\eqref{Matsubarafreq}).
The Fourier coefficients are given by
\begin{align}\label{Gfourierserinv}
\mathscr{G}_\bs{k}(z_l)=\int \limits_{0}^\beta d\tau  \,\mathscr{G}_\bs{k}(\tau)\e^{z_l\tau}.
\end{align}
The 
Dyson equation in Fourier (Matsubara) space is given by
\begin{align}
\mathscr{G}_\bs{k}(z_l) 
&=
g_\bs{k}(z_l) + g_\bs{k}(z_l)\, \Xi_\bs{k}(z_l)\, \mathscr{G}_\bs{k}(z_l),
\end{align}
where $\Xi_\bs{k}(z_l)$ is the Matsubara self-energy and
$g_{\bs{k}}(\omega_l)$ is the unperturbed propagator in Matsubara space, i.e.,
\begin{align} \label{g0fourier}
g_{\bs{k}}(z_l)
=\frac{1}{z_l-\en_\bs{k}}.
\end{align}
Iterating the Dyson equation and summing the resulting geometric series  leads to
\begin{align} \label{Dysonsummed}
\mathscr{G}_\bs{k}(z_l) 
&=
\frac{1}{z_l-\en_\bs{k}-\Xi_\bs{k}(z_l)}.
\end{align}
Inserting this into the Fourier series Eq.~\eqref{Gfourierser} 
and replacing the discrete frequency sums by
a contour integral leads to 
\begin{align} \label{Gim0C0}
\mathscr{G}_\bs{k}(\tau)
&= \!
\oint\limits_{C_0[l_\text{max}]} \!\!\!
\frac{d z}{2\pi \im} \, 
\e^{-z\tau}n(z)\,
\frac{1}{z-\en_\bs{k}-\mathcal{M}_\bs{k}(z)}  ,
\end{align}
where $l_\text{max}\rightarrow \infty$ is implied, and $n(z)=[1+\e^{\beta(z-\mu}]^{-1}$.
The
contour $C_0[l_\text{max}]$ encloses all the Matsubara poles $z=z_{l\in\mathcal{L}(l_\text{max})}$ but not the pole at $z=\en_\bs{k}+\mathcal{M}_\bs{k}(z)$, see Fig.~\ref{figc1}.
Note 
that by construction $C_0[l_\text{max}]$ crosses the real axis. 
Thus, for Eq.~\eqref{Gim0C0} to be equivalent to
Eq.~\eqref{Gfourierser}, the mass function $\mathcal{M}_\bs{k}(z)$ must be an analytic continuation of 
the Matsubara self-energy $\Xi_\bs{k}(z_l)$ that is analytic on the real axis (and near the Matsubara poles).
This is easy to get:
for the second-order two-particle irreducible contribution to $\mathcal{M}_\bs{k}(z)$ we obtain from Eq.~\eqref{sigma2} the expression
\begin{align} \label{sigma2bMass}
\mathcal{M}_{2,\bs{k}}(z) &= 
\frac{1}{2}
\sum_{\bs{k}_2, \bs{k}_3, \bs{k}_4} \!
|\braket{\psi_{\bs{k}}\psi_{\bs{k}_2}| V|\psi_{\bs{k}_3} \psi_{\bs{k}_4}}|^2  n_{\bs{k}_2} \bar n_{\bs{k}_3} \bar n_{\bs{k}_4}
\breq & \quad \times
\frac{e^{-\beta(\en_{\bs{k}_3}+\en_{\bs{k}_4} - \en_{\bs{k}_2} - z)}-1}{\en_{\bs{k}_3}+\en_{\bs{k}_4} - \en_{\bs{k}_2} - z},
\end{align}
i.e., in contrast to Eq.~\eqref{sigma2b}, we do \emph{not}
substitute\break $\e^{\beta(z_l-\mu)}=-1$ before
performing the analytic continuation, and similar for higher-order contributions.
Since with this prescription there are no poles,
$\mathcal{M}_\bs{k}(z)$ is entire and real on the real axis, and regular everywhere except for $\text{Re}[z]\rightarrow \infty$.
However, the ${T\rightarrow 0}$ limit of $\mathcal{M}_\bs{k}(z)$ is singular for $\text{Re}[z]>\mu$, due to terms $\e^{\beta (z-\mu)}$ as in Eq.~\eqref{sigma2bMass}.

\vspace*{2mm}

In Eq.~\eqref{Gim0C0}, for $\tau<0$ the term $\e^{-z\tau} n_\bs{k}(z)$ is regular 
at infinity, and the term $[z-\en_\bs{k}-\mathcal{M}_\bs{k}(z)]^{-1}$ 
vanishes at infinity. 
Hence, 
if we assume that $z=\en_\bs{k}+\mathcal{M}_\bs{k}(z)$ has no solutions off the real axis, 
for $\tau<0$ the contour $C_0[l_\text{max}]$ can be deformed
into the contour $C_1$ (see Fig.~\ref{figc1})
that encloses the pole on the real axis at 
\begin{align} \label{massSC}
\mathscr{E}_\bs{k}=\en_\bs{k}+\mathcal{M}_\bs{k}(\mathscr{E}_\bs{k}).
\end{align} 
For nonreal $z=x+iy$, $z=\en_\bs{k}+\mathcal{M}_\bs{k}(z)$ is equivalent
to the two coupled nonlinear equations $x=\en_\bs{k}+\text{Re}[\mathcal{M}_\bs{k}(x+iy)]$ and $y=\text{Im}[\mathcal{M}_\bs{k}(x+iy)]$.
At second order, this is given by
\begin{align} \label{Mcoup1}
x &= \en_\bs{k} - U_{\bs{k}} + U_{1,\bs{k}} + \mathcal{M}^{\ddagger}_{2,\bs{k}}
\breq & \quad  +
\frac{1}{2}
\sum_{\bs{k}_2, \bs{k}_3, \bs{k}_4} \!
|\braket{\psi_{\bs{k}}\psi_{\bs{k}_2}| V|\psi_{\bs{k}_3} \psi_{\bs{k}_4}}|^2 
\frac{n_{\bs{k}_2} \bar n_{\bs{k}_3} \bar n_{\bs{k}_4}}{\left[D(x)\right]^2+y^2}
\breq & \quad \times
\{ D(x)[\cos(\beta y)e^{-\beta D(x)}-1] -y \sin(\beta y) e^{-\beta D(x)} \}
,
\\ \label{Mcoup2}
y &= 
\frac{1}{2}
\sum_{\bs{k}_2, \bs{k}_3, \bs{k}_4} \!
|\braket{\psi_{\bs{k}}\psi_{\bs{k}_2}| V|\psi_{\bs{k}_3} \psi_{\bs{k}_4}}|^2 
\frac{n_{\bs{k}_2} \bar n_{\bs{k}_3} \bar n_{\bs{k}_4}}{\left[D(x)\right]^2+y^2}
\breq & \quad \times
\{ y[ \cos(\beta y)e^{-\beta D(x)}-1] +D(x) \sin(\beta y) e^{-\beta D(x)} \},
\end{align}
with $D(x)=\en_{\bs{k}_3}+\en_{\bs{k}_4} - \en_{\bs{k}_2} - x$. Here, $-U_{\bs{k}}$ corresponds to the diagram composed of a single $-U$ vertex, 
$U_{1,\bs{k}}$ to the one with a single $V$ vertex,
and $\mathcal{M}^{\ddagger}_{2,\bs{k}}$ denotes the second-order two-particle reducible contribution to $\mathcal{M}_{2,\bs{k}}(z)$.
We did not see an argument that guarantees that Eqs.~\eqref{Mcoup1} and \eqref{Mcoup2} have solutions for ${y\neq 0}$.

Assuming that there is only the pole given by Eq.~\eqref{massSC}, we get
\begin{align} \label{fMass0}
\mathscr{G}_\bs{k}(\tau<0)
&=n\left(\mathscr{E}_\bs{k}\right)\e^{-\mathscr{E}_\bs{k}\tau},
\end{align}
and the expression for the mean occupation numbers is given by
\begin{align} \label{fMass}
f_\bs{k}
&=
\mathscr{G}_\bs{k}(0^{-})
=n\left(\mathscr{E}_\bs{k}\right),
\end{align}
i.e., the exact mean occupation numbers are given by the Fermi-Dirac distribution with the reference spectrum renormalized in terms 
of the on-shell mass function $\mathcal{M}_\bs{k}(\mathscr{E}_\bs{k})$ 
defined via Eq.~\eqref{massSC}
and the analytic continuation of the Matsubara self-energy $\Xi_\bs{k}(z_l)$ that is real analytic on the real axis. 
As discussed, the ${T\rightarrow 0}$ limit of $\mathcal{M}_\bs{k}(\mathscr{E}_\bs{k})$ is ill-behaved.

\subsubsection{Frequency-space self-energy}
The real-time propagator is\footnote{In this paragraph we follow for the most part Kadanoff and Baym~\cite{kadanoffbaym}, Fetter and Walecka~\cite{Fetter}, and 
Ref.~\cite{ThesisRios}.}
\begin{align}
\im G_\bs{k}(t-t')= \Braket{\! \Braket{ \mathcal{T}\left[a_\bs{k}(t) a_\bs{k}^\dagger(t') \right] } \!},
\end{align}
with $a_\bs{k}(t)=a_\bs{k}\e^{-\im \en_\bs{k} t}$ and $a^\dagger_\bs{k}(t)=a^\dagger_\bs{k}\e^{\im\en_\bs{k}t}$.
It can be decomposed as
\begin{align}
\im G_\bs{k}(t-t')= 
\theta(t-t') \underbrace{\braket{\! \braket{a_\bs{k}(t) a_\bs{k}^\dagger(t')  }\!} }_{ \im G^{>}_\bs{k}(t-t')}
-
\theta(t'-t)\underbrace{\braket{\!  \braket{a_\bs{k}^\dagger(t') a_\bs{k}(t)  }\!} }_{-\im G^{<}_\bs{k}(t-t')},
\end{align}
where we have defined the correlation functions $\im G^{>}_\bs{k}(t-t')$ and $\im G^{<}_\bs{k}(t-t')$.
The Fourier transforms of the real-time propagator and the correlation functions are given by
\begin{align}
G_\bs{k}(\omega)&=\int \limits_{-\infty}^\infty \!\! dt \, \im G_\bs{k}(t)\e^{\im\omega t},
\\
G^{>}_\bs{k}(\omega)&=\int \limits_{-\infty}^\infty \!\! dt \, \im G_\bs{k}^{>}(t)\e^{\im\omega t},
\\
G^{<}_\bs{k}(\omega)&=\int \limits_{-\infty}^\infty \!\! dt \, \left(-\im G_\bs{k}^{<}(t)\e^{\im\omega t}\right),
\end{align}
with inverse transforms
\begin{align} \label{Gfouriertransform}
\im G_\bs{k}(t)=\int \limits_{-\infty}^\infty \!\! \frac{d \omega}{2 \pi} \, G_\bs{k}(\omega)  \e^{-\im\omega t},
\\
\im G^{>}_\bs{k}(t)=\int \limits_{-\infty}^\infty \!\! \frac{d \omega}{2 \pi} \, G^{>}_\bs{k}(\omega)  \e^{-\im\omega t},
\\
-\im G^{<}_\bs{k}(t)=\int \limits_{-\infty}^\infty \!\! \frac{d \omega}{2 \pi} \, G^{<}_\bs{k}(\omega)  \e^{-\im\omega t}.
\end{align}
The Fourier transforms of the correlation functions satisfy the KMS relation~\cite{kadanoffbaym,ThesisRios} (see also Refs.~\cite{PhysRev.115.1342,Haag1967,doi:10.1143/JPSJ.12.570})
\begin{align}
G^{<}_\bs{k}(\omega) = \e^{-\beta(\omega-\mu)} G^{>}_\bs{k}(\omega).
\end{align}
From this relation it follows that we can write
\begin{align}
G^{>}_\bs{k}(\omega)
&=
\bar n(\omega) \mathcal{A}_\bs{k}(\omega) ,
\\
G^{<}_\bs{k}(\omega)
&=
n(\omega) \mathcal{A}_\bs{k}(\omega),
\end{align}
with
\begin{align}
\mathcal{A}_\bs{k}(\omega) 
&=
G^{>}_\bs{k}(\omega)+G^{<}_\bs{k}(\omega),
\end{align}
and $\bar n(\omega)=1-n(\omega)$.
From the Lehmann representations 
of $G^{>}_\bs{k}(t)$ and $G^{<}_\bs{k}(t)$ 
it can be seen that 
the spectral function is semipositive, i.e., 
\begin{align} \label{Ageq0}
\mathcal{A}_\bs{k}(\omega) & \geq 0 ,
\end{align}
and satisfies the sum rule
\begin{align} \label{sumrule}
\int \limits_{-\infty}^\infty \!\! \frac{d \omega}{2\pi}\mathcal{A}_\bs{k}(\omega) & =1,
\end{align}
see e.g., Refs.~\cite{Fetter,ThesisRios}.
Consider now the function $\Gamma_\bs{k}(z)$
defined by
\begin{align} \label{Gammadef}
\Gamma_\bs{k}(z)
&=
\int \limits_{-\infty}^\infty \!\! \frac{d \omega}{2\pi}
\frac{\mathcal{A}_\bs{k}(\omega)}{z-\omega}.
\end{align}
From the Lehmann representation of the 
imaginary-time propagator $\mathscr{G}_\bs{k}(\tau)$
it can be seen that~\cite{Fetter}
\begin{align}
\mathscr{G}_\bs{k}(z_l) 
&=
\int \limits_{-\infty}^\infty \!\! \frac{d \omega}{2\pi}
\frac{\mathcal{A}_\bs{k}(\omega)}{z_l-\omega}.
\end{align}
From the sum rule for $\mathcal{A}_\bs{k}(\omega)$, Eq.~\eqref{sumrule}, it then follows that
$\Gamma_\bs{k}(z)$ corresponds to the (unique \cite{baym}) analytic continuation of $\mathscr{G}_\bs{k}(z_l)$ that 
satisfies $\Gamma_\bs{k}(z)\sim z^{-1}$ for $|z|\rightarrow \pm\infty$.
From Eq.~\eqref{Dysonsummed}, this can be obtained via\footnote{Note 
that Eqs.~\eqref{Ageq0} and \eqref{Gammadef} imply that $\text{Im}[\Sigma_\bs{k}(z)]\lessgtr 0$ for $\text{Im}[z] \gtrless 0$.}
\begin{align}\label{Sigmadef}
\Gamma_\bs{k}(z)
&=
\frac{1}{z-\en_\bs{k}-\Sigma_\bs{k}(z)},
\end{align}
where the frequency-space self-energy $\Sigma_\bs{k}(z)$ 
is defined as the analytic continuation of the Matsubara self-energy $\Xi_\bs{k}(z_l)$
that satisfies $\Sigma_\bs{k}(z)\rightarrow 0$ for $|z|\rightarrow \pm\infty$.
In bare MBPT, this is given by the prescription noted before Eq.~\eqref{sigma2b} in Sec.~\ref{sec24}, i.e., 
one first substitutes $\e^{\beta(\omega_l-\mu)}=-1$ and then 
performs the analytic continuation.
For convenience, we give again the irreducible part of the bare second-order contribution to $\Sigma_\bs{k}(z)$, i.e.,
\begin{align} \label{sigma2again}
\Sigma_{2,\bs{k}}(z) &= 
-\frac{1}{2}
\sum_{\bs{k}_2, \bs{k}_3, \bs{k}_4} \!
|\braket{\psi_{\bs{k}}\psi_{\bs{k}_2}| V|\psi_{\bs{k}_3} \psi_{\bs{k}_4}}|^2  
\frac{
n_{\bs{k}_2} \bar n_{\bs{k}_3} \bar n_{\bs{k}_4} +
n_{\bs{k}_3} n_{\bs{k}_4} \bar n_{\bs{k}_2}}
{\en_{\bs{k}_3}+\en_{\bs{k}_4} - \en_{\bs{k}_2} - z}.
\end{align}
From Eqs.~\eqref{Gammadef} and \eqref{Sigmadef}, we obtain for the spectral function the expression
\begin{align} \label{spectralfundef}
\mathcal{A}_\bs{k}(\omega)
&=\im \Big[\Gamma_\bs{k}(\omega+\im\eta) - \Gamma_\bs{k}(\omega-\im\eta) \Big].
\breq 
&=\im \left[\frac{1}{\omega-\en_\bs{k}-\Sigma_\bs{k}(\omega+\im\eta)+\im\eta} - \text{c.c.} \right],
\end{align}
where c.c. denotes the complex conjugate.
Note that inserting Eq.~\eqref{dynquasi1a} into
Eq.~\eqref{spectralfundef} leads to the Breit-Wigner form of the spectral function, Eq.~\eqref{spectral}.
The relation between $\mathcal{A}_\bs{k}(\omega)$ and the Fourier transform of the real-time propagator $G_\bs{k}(\omega)$
is obtained as follows:
\begin{align} \label{Gomega}
G_\bs{k}(\omega) 
&=
\int \limits_{-\infty}^\infty \!\! dt \,\e^{\im\omega t} \Big[\theta(t) \im G^{>}_\bs{k}(t)+\theta(-t) \im G^{<}_\bs{k}(t) \Big]
\breq
&=
-\int \limits_{-\infty}^\infty \!\! dt \,\e^{\im\omega t} 
\left[
\int \limits_{-\infty}^\infty \!\! \frac{d \xi}{2 \pi\im} \,\frac{\e^{-\im\xi t} }{\xi+\im\eta} \im G^{>}_\bs{k}(t)
\right.
 \left.
-
\int \limits_{-\infty}^\infty \!\! \frac{d \xi}{2 \pi\im} \,\frac{\e^{-\im\xi t} }{\xi-\im\eta}  \im G^{<}_\bs{k}(t)
\right]
\breq
&= 
-\int \limits_{-\infty}^\infty \!\! \frac{d \xi}{2\pi \im} 
\left[ 
\frac{G^{>}_\bs{k}(\omega-\xi)}{\xi+\im \eta}+\frac{G^{<}_\bs{k}(\omega-\xi)}{\xi-\im \eta}
\right]
\breq
&= 
-\int \limits_{-\infty}^\infty \!\! \frac{d \xi}{2\pi\im} 
\left[ 
\frac{\bar n(\xi) \mathcal{A}_\bs{k}(\xi)}{\omega-\xi+\im\eta} +\frac{n(\xi) \mathcal{A}_\bs{k}(\xi)}{\omega-\xi-\im\eta} 
\right],
\end{align}
where we have used the relation~\cite{Fetter}
\begin{align}
\theta(\pm t) = \mp\int\limits_{-\infty}^\infty \!\!  \frac{d \xi}{2 \pi\im} \, \frac{\e^{-\im\xi t} }{\xi\pm \im\eta}.
\end{align}
From Eq.~\eqref{Gfouriertransform} we then have
\begin{align}
\im G_\bs{k}(t)= 
-\int \limits_{-\infty}^\infty \!\!  
\frac{d \omega}{2\pi}
\e^{-\im\omega t}\!\!
\int \limits_{-\infty}^\infty \!\! \frac{d \xi}{2\pi\im} 
\left[ 
\frac{\bar n(\xi) \mathcal{A}_\bs{k}(\xi)}{\omega-\xi+\im\eta} +\frac{n(\xi) \mathcal{A}_\bs{k}(\xi)}{\omega-\xi-\im\eta} 
\right].
\end{align}
For $t<0$
we can 
close the $\omega$ integral in the upper half plane.
Interchanging the integration order, we then get
\begin{align} \label{Gtless0}
\im G_\bs{k}(t<0)=
-\int \limits_{-\infty}^\infty  \!\! \frac{d \xi}{2\pi} \,
 \e^{-\im\xi t} n(\xi)  \mathcal{A}_\bs{k}(\xi).
\end{align}
Thus, the expression for the exact mean occupation numbers is
\begin{align} \label{fspectralrep}
f_\bs{k}
&=
-\im G_\bs{k}(0^{-})
=\int \limits_{-\infty}^\infty \!\! \frac{d \xi}{2\pi} \,
n(\xi)  \mathcal{A}_\bs{k}(\xi).
\end{align}
Here, in contrast to Eq.~\eqref{fMass}, the ${T\rightarrow 0}$ limit is well-behaved, and its analysis 
reveals that $f_\bs{k}(T=0,\mu)$ has a discontinuity at $\bs{k}=\bs{k}_\F$, see Ref.~\cite{PhysRev.119.1153} and Sec.~\ref{sec24}.

\vspace*{2mm}
\begin{figure}[h] 
\centering
\vspace*{0mm}
\hspace*{-0.0cm}
\includegraphics[width=0.45\textwidth]{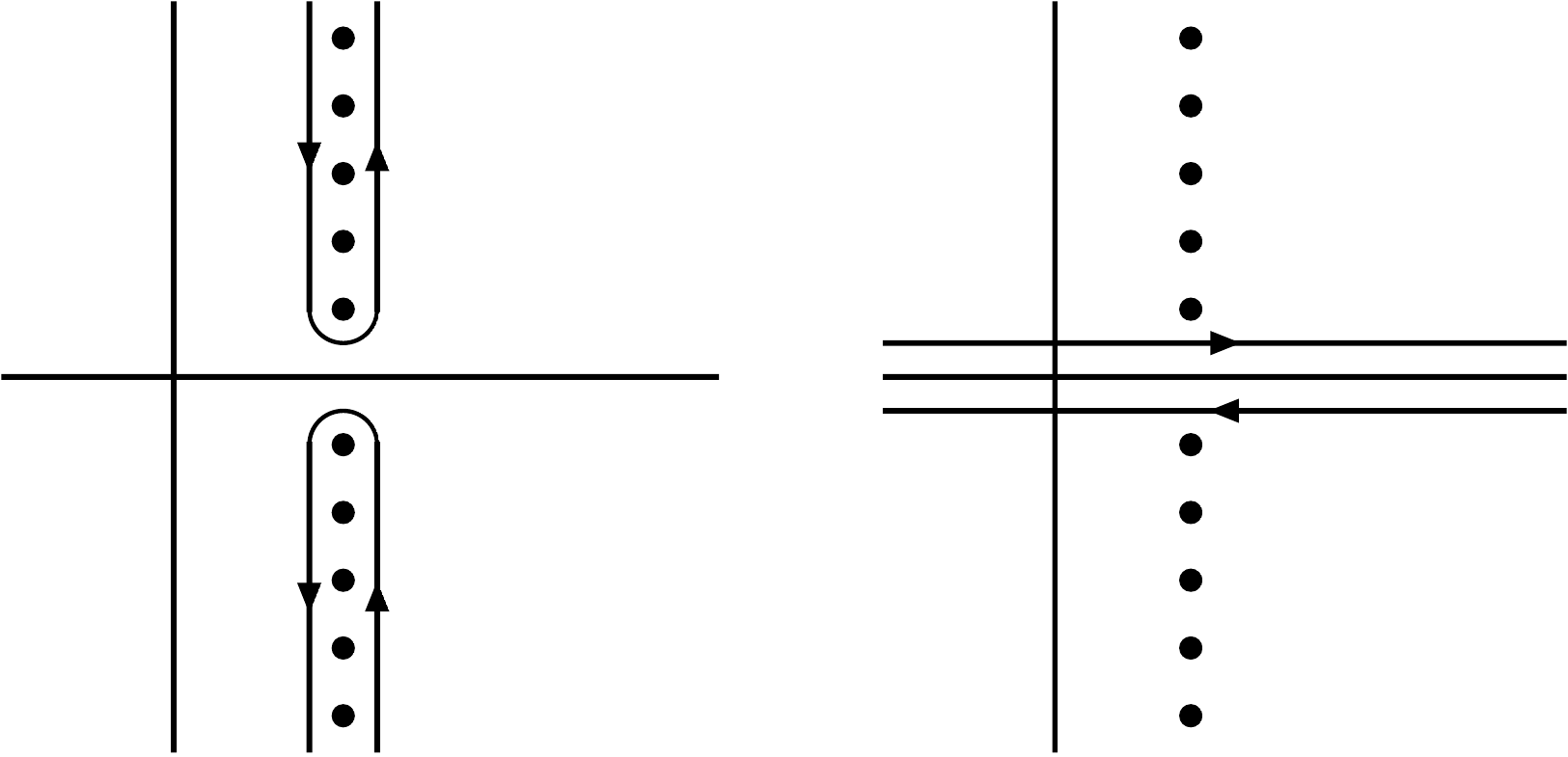} 
\vspace*{-2mm}
\caption{Contours $C^{\pm}_0$ (left panel) and $C_2$ (right panel).}
\label{figc2}
\end{figure}

The result 
given by Eq.~\eqref{fspectralrep}
can also be obtained directly from the Fourier expansion of the imaginary-time propagator, Eq.~\eqref{Gfourierser}. 
That is, taking first the limit ${l_\text{max}\rightarrow\infty}$ and then
performing the analytic continuation 
of $\Xi_\bs{k}(z_l)$ to $\Sigma_\bs{k}(z)$
we get
\begin{align} 
\mathscr{G}_\bs{k}(\tau)
&=
\oint\limits_{C^{\pm}_0} \frac{d z}{2\pi \im} \, 
\e^{-z\tau} n(z) 
\frac{1}{z-\en_\bs{k}-\Sigma_\bs{k}(z)} ,
\end{align}
with $C^{\pm}_0=C^{+}_0+C^{-}_0$, where $C^{+}_0$ encloses the Matsubara poles in the upper half plane without crossing the real axis, and $C^{-}_0$ the poles in the lower half plane.
Since $\Sigma_\bs{k}(z)$ is analytic in the two half planes and vanishes at complex infinity, 
and $\e^{-z\tau} n_\bs{k}(z)$ is regular at infinity for $\tau<0$, 
for $\tau<0$
these two contours can be deformed into the contour $C_2$ that encloses the real axis, see Fig.~\ref{figc2}, i.e.,
\begin{align} \label{Gim0C0again}
\mathscr{G}_\bs{k}(\tau<0)
&= \!\!
\int\limits_{-\infty}^\infty \!\! \frac{d \xi}{2\pi \im}  \,
\e^{-\xi\tau} n(\xi) 
\underbrace{\left[\frac{1}{\xi-\en_\bs{k}-\Sigma_\bs{k}(\xi+\im\eta)+\im\eta}
-\text{c.c.}\right]}_{-\im\mathcal{A}_\bs{k}(\xi)},
\end{align}
which is just the Wick rotation of Eq.~\eqref{Gtless0}.

\subsubsection{Collision self-energy at zero temperature}
The self-energy 
corresponding to the real-time propagator, here referred to as the collision self-energy
$\Sigma_\bs{k}^\text{coll}(\omega)$, can be defined by~\cite{Fetter}\footnote{Note that 
Fetter and Walecka omit the factor $\im$ in the definition of $G_\bs{k}(\omega)$, so no $\im$ appears in their 
version of our Eq.~\eqref{Ecoll}, i.e., in Eq.~(9.33) of Ref.~\cite{Fetter}.}
\begin{align} \label{Ecoll}
G_\bs{k}(\omega) = \im \frac{1}{\omega-\en_\bs{k}- \Sigma_\bs{k}^\text{coll}(\omega) }.
\end{align}
In the following, we examine how at ${T=0}$ the collision self-energy 
relates to the frequency-space self-energy $\Sigma_\bs{k}(z)$. 
For this, using the Sokhotski-Plemelj theorem we rewrite Eq.~\eqref{Gomega} as
\begin{align}
G_\bs{k}(\omega) = 
\int \limits_{-\infty}^\infty \!\! \frac{d \xi}{2\pi\im}  \mathcal{A}_\bs{k}(\xi) \frac{P}{\omega-\xi} 
-
\frac{\bar n(\omega) \mathcal{A}_\bs{k}(\omega)}{2}
+
\frac{ n(\omega) \mathcal{A}_\bs{k}(\omega)}{2}
.
\end{align}
From $n(\omega)\xrightarrow{T\rightarrow 0}\theta(\mu-\omega)$, 
at zero temperature we have
\begin{align}
G_\bs{k}(\omega) = 
\theta(\omega-\mu)\,G_\bs{k}^\text{R}(\omega)
+
\theta(\mu-\omega)\,G_\bs{k}^\text{A}(\omega)
,
\end{align}
with the Fourier transforms of the retarded and advanced propagators given by
\begin{align}
G_\bs{k}^R(\omega)&= 
\int \limits_{-\infty}^\infty \!\! \frac{d \xi}{2\pi\im}  \mathcal{A}_\bs{k}(\xi) \frac{P}{\omega-\xi} 
-
\frac{\mathcal{A}_\bs{k}(\omega)}{2},
\\
G_\bs{k}^A(\omega)&= 
\int \limits_{-\infty}^\infty \!\! \frac{d \xi}{2\pi\im}  \mathcal{A}_\bs{k}(\xi) \frac{P}{\omega-\xi} 
+
\frac{\mathcal{A}_\bs{k}(\omega)}{2}.
\end{align}
Comparing with Eq.~\eqref{Gammadef} we see that
\begin{align}
G_\bs{k}^\text{R}(\omega) = 
\im\Gamma_\bs{k}(\omega+ \im \eta),
\;\;\;\;\;\;\;\;
G_\bs{k}^\text{A}(\omega) = 
\im\Gamma_\bs{k}(\omega- \im \eta).
\end{align}
Thus, from Eq.~\eqref{Sigmadef} we can at ${T=0}$ make the identification
\begin{align}
\Sigma_\bs{k}^\text{coll}(\omega)=
\theta(\omega-\mu)\,\Sigma_\bs{k}(\omega+ \im \eta)
+
\theta(\mu-\omega)\,\Sigma_\bs{k}(\omega- \im \eta).
\end{align}
From 
Eq.~\eqref{dynquasi1a}, i.e.,
\begin{align} \label{dynquasi1aagain}
\Sigma_\bs{k}(\omega\pm\im\eta) = \mathcal{S}_\bs{k}(\omega) \mp \im \mathcal{J}_\bs{k}(\omega),
\end{align}
we have (at ${T=0}$)
\begin{align}
\Sigma_\bs{k}^\text{coll}(\omega) &= 
\theta(\omega-\mu)\,\big[\mathcal{S}_\bs{k}(\omega)-\im\mathcal{J}_\bs{k}(\omega) \big]
\breq & \quad 
+\theta(\mu-\omega)\,\big[\mathcal{S}_\bs{k}(\omega)+\im\mathcal{J}_\bs{k}(\omega) \big].
\end{align}
In particular, (as discussed in Sec.~\ref{sec24}), at zero temperature it is 
${\mathcal{J}_\bs{k}(\omega) \xrightarrow{\omega \rightarrow \mu} C_\bs{k}(\mu) (\omega-\mu)^2}$, with $C_\bs{k}(\mu)\geq 0$, so (at ${T=0}$)
\begin{align}
\text{Im}\big[\Sigma_\bs{k}^\text{coll}(\omega)\big] &= 
-\theta(\omega-\mu)\,\mathcal{J}_\bs{k}(\omega)
+\theta(\mu-\omega)\,\mathcal{J}_\bs{k}(\omega) 
\breq
&\xrightarrow{\omega \rightarrow \mu}
-C_\bs{k}(\mu)\, (\omega-\mu)|\omega-\mu|.
\end{align}
Finally, for the on-shell collision self-energy this leads to
\begin{align}
\text{Im}\big[\Sigma_\bs{k}^\text{coll}(\en_\bs{k})\big] 
&\xrightarrow{\en_\bs{k} \rightarrow \mu}
-C_\bs{k}(\mu)\, (\en_\bs{k}-\mu)|\en_\bs{k}-\mu|,
\end{align}
which is the property 
quoted in Refs.\cite{Holt13a,Kaiser:2001ra}.\footnote{In the 
adiabatic formalism only real-time propagators appear, so it is the collisional self-energy that is calculated.}

\subsection{Mean occupation numbers from direct mean-field renormalization}\label{app22}

In the direct renormalization scheme the exact mean occupation numbers are identified with the Fermi-Dirac distributions (i.e., with the mean occupation numbers in the reference system), i.e.,
\begin{align} \label{fisn}
\text{direct scheme:}\;\;\;\;
f_\bs{k}=n_\bs{k}.
\end{align}
From this one may conclude that
in the direct scheme the mass function is zero, $\mathcal{M}_\bs{k}(z)=0$, and 
the spectral function is given by the unperturbed one, $\mathcal{A}_\bs{k}(\omega)=2\pi \delta(\omega-\en_\bs{k})$.
More generally, one may conclude that the Matsubara self-energy is zero, $\Xi_\bs{k}(z_l)=0$.

However, these conclusions come with two caveats:
\begin{enumerate}[start=1,label={(\arabic*)}]
\item 
The cancellations that 
lead to Eq.~\eqref{fisn} are not available in Matsubara space.
That is, the result $\Xi_\bs{k}(z_l)=0$ is obtained 
only from the Fourier expansion of the direct expression for the propagator, Eq.~\eqref{Gdirect}.
If one instead Fourier expands the (unperturbed) propagators (cf.~Appendix~\ref{app23}) in 
the time-integral representation,
Eq.~\eqref{Gmbpt},
then one obtains the usual result, i.e., $\Xi_\bs{k}(z_l)\neq 0$, also in the direct scheme.
\item The (proper) Matsubara self-energy $\Xi_\bs{k}(z_l)$ is defined in terms of the Dyson equation, Eq.~\eqref{Dysonsummed}.
The Dyson equation is inconsistent with a perturbative truncation order. 
In contrast, Eq.~\eqref{fisn} relies on a finite truncation order $N$.
\end{enumerate}
Caveat (1) implies that the ${T\neq 0}$ part\footnote{We note again that in the direct (and cyclic) scheme the ${T\rightarrow 0}$ limit is nonexistent.} of the general results of Sec.~\ref{sec24} can be obtained also in the direct scheme, and caveat (2) makes evident that these results and Eq.~\eqref{fisn} do not contradict each other; they correspond to 
different partial summations of a divergent asymptotic series.

\subsubsection{Proof of Eq.~(\ref{fisn})}
The perturbation series for the imaginary-time propagator $\mathscr{G}_{\bs{k}}(\tau-\tau')$ is given by
\begin{align}
\mathscr{G}_{n,\bs{k}}(\tau-\tau')
&=
g_{\bs{k}}(\tau-\tau')
+\sum_{n=1}^N \mathscr{G}_{n,\bs{k}}(\tau-\tau').
\end{align}
Here, the unperturbed propagator $g_{\bs{k}}(\tau-\tau')$ is given by
\begin{align} \label{smallg}
g_{\bs{k}}(\tau-\tau')
&= - \Braket{ \mathcal{T}\left[a_\bs{k}(\tau) a_\bs{k}^\dagger(\tau') \right] }
\breq 
&= \theta(\tau-\tau')\, n_\bs{k} \e^{\en_\bs{k} (\tau-\tau')} 
-
\theta(\tau'-\tau)\, \bar n_\bs{k} \e^{\en_\bs{k} (\tau-\tau')},
\end{align}
and its Matsubara coefficients are given by Eq.~\eqref{g0fourier}.
The perturbative contributions $\mathscr{G}_{n,\bs{k}}(\tau-\tau')$ are given by the expression~\cite{Fetter}
\begin{align} \label{Gmbpt}
\mathscr{G}_{n,\bs{k}}(\tau-\tau')
&=
\frac{(-1)^{n+1}}{n!}
\int \limits_{0}^{\beta} d\tau_n\cdots d\tau_1 
\breq &\quad \times
\Braket{  \mathcal{T}\big[ a_\bs{k}(\tau) \,
\mathcal{V}(\tau_n) \cdots
\mathcal{V}(\tau_1) \,
a_\bs{k}^\dagger(\tau') \big]
}_{L}
\breq 
&\equiv\mathscr{G}^{\text{direct}(P)}_{n,\bs{k}}(\tau-\tau').
\end{align}
This can be written as 
\begin{align} 
\mathscr{G}_{n,\bs{k}}(\tau-\tau')
&=
(-1)^{n+1}
\int \limits_{0}^{\beta} d\tau_{n} \int \limits_{0}^{\tau_n} d\tau_{n-1}\cdots \int \limits_{0}^{\tau_2} d\tau_1 
\breq &\quad \times
\Braket{  \mathcal{T}\big[ a_\bs{k}(\tau) \,
\mathcal{V}(\tau_{n}) \cdots
\mathcal{V}(\tau_1) \,
a_\bs{k}^\dagger(\tau') \big]
}_{\!L}
\breq 
&\equiv\mathscr{G}^\text{direct}_{n,\bs{k}}(\tau-\tau')
.
\end{align}
From here, we can follows the steps that lead Bloch and de Dominicis~\cite{BDDnuclphys7} to the 
direct formula\footnote{Because of the two external lines no cyclic and reduced versions of Eq.~\eqref{Gdirect} are available; see 
Ref.~\cite{BDDnuclphys7} for details on the derivation of the 
cyclic formula and the reduced formula.} for 
the perturbative contributions to the grand-canonical potential (see Sec.~\ref{sec22}).
Because $\mathscr{G}_{n,\bs{k}}(\tau)$ is antiperiodic with period $\beta$ we can, without loss of generality, set $\tau<0$ and $\tau'<0$.
For $\tau-\tau'<0$, this leads to
\begin{align} \label{Gdirect}
\mathscr{G}^\text{direct}_{n,\bs{k}}(\tau<0)&=
\e^{\en_\bs{k}\tau}
\frac{(-1)^{n}}{2\pi \im} 
\oint_{C}  dz \frac{\e^{-\beta z}}{z^2}
\breq &\quad \times
\Braket{ 
\mathcal{V} \frac{1}{\En_n-z} \cdots
\mathcal{V} \frac{1}{\En_1-z}
\mathcal{V} \,a_\bs{k}^\dagger\, a_\bs{k}
}_{\!L}.
\end{align}
For truncation order $N$, the contributions to $\mathscr{G}_{\bs{k}}(\tau)$ are given by 
all linked (one-particle irreducible and reducible) propagator diagrams that satisfy Eq.~\eqref{Ncounting}.
Applying the cumulant formalism,
the contributions to Eq.~\eqref{Gdirect} are given by
normal propagator diagrams\footnote{We use the notion normal propagator diagrams to refer to 
diagrams that have no anomalous articulation lines and are either
(i) one-particle irreducible propagator diagrams 
or (ii) one-particle reducible propagator diagrams where all cuttable propagator lines go in the same direction.} with normal Hugenholtz 
diagrams attached via higher-cumulant connections, plus 
diagrams composed of multiple normal propagator diagrams 
simply-connected via higher-cumulant connections attached to normal Hugenholtz 
diagrams.
With the mean field given by
\begin{align}
U_\bs{k}=\sum_{n=1}^N U_{n,\bs{k}}^{\text{direct},\div},
\end{align}
the contributions with higher-cumulant connections are removed. 
\textit{Furthermore}, because propagator diagrams 
involve all possible orderings of the vertices, (an analog of)
the direct factorization theorem applies also for the remaining contributions; e.g., 
for a one-particle irreducible propagator diagram with non $-U$ self-energy part (i.e., at least one $V$ vertex is involved) we have
\begin{align} \label{Gdirectfactorized1}
\mathscr{G}_{n,\bs{k}}^{\div}(\tau)&= g_{\bs{k}}(\tau) \;
U^{\text{direct,}\div}_{n,\bs{k}},
\end{align}
for a one-particle reducible diagram with two non $-U$ self-energy parts we have
\begin{align} \label{Gdirectfactorized2}
\mathscr{G}_{n_1+n_2,\bs{k}}^{\div}(\tau)&= g_{\bs{k}}(\tau) \;
U^{\text{direct,}\div}_{n_1,\bs{k}} \;  U^{\text{direct,}\div}_{n_2,\bs{k}},
\end{align}
etc.
Hence, in the direct scheme these contributions 
are canceled by the diagrams where the self-energy parts are replaced 
by $-U^{\text{direct},\div}$ vertices.
Thus,\footnote{In the cyclic and the BdD scheme only the contributions with Hugenholtz diagrams attached via higher-cumulant connections 
can be canceled.
The 
remaining propagator contributions in these schemes are then given by
Eqs.~\eqref{Gdirectfactorized1}, \eqref{Gdirectfactorized2}, etc., with the $-U$ vertices (but not the self-energy parts) given by $-U^{\text{cyclic},\div}_n$ and $-U^{\text{BdD}}_n$, respectively,
plus
diagrams that have self-energy parts consisting of $-U^{\text{cyclic},\div}_n$ and $-U^{\text{BdD}}_n$ vertices (with $2\leq n\leq N$), respectively.}
\begin{align} 
\text{direct scheme:}\;\;\;\;
\mathscr{G}_{\bs{k}}(\tau)=
g_{\bs{k}}(\tau),
\end{align}
and Eq.~(\ref{fisn}) is proved.

\subsection{Self-energy, mass function, and grand-canonical potential}\label{app23}
The (proper) Matsubara self-energy $\Xi_\bs{k}(z_l)$ can be calculated 
using self-consistent propagators or using bare propagators (or, anything in between).
In the bare case, also two-particle reducible self-energy diagrams contribute to $\Xi_\bs{k}(z_l)$; see, e.g., Ref.~\cite{PLATTER2003250}.

Below, we first explain how the bare perturbative 
contributions to the improper Matsubara self-energy $\Xi^\star_\bs{k}(z_l)$ can be obtained. 
From this, the bare contributions to $\Xi_\bs{k}(z_l)$
are obtained via the restriction to one-particle irreducible diagrams.

Second, we
derive the functional relations between the 
bare perturbative contributions to the (various forms of the) improper self-energy and the grand-canonical potential.\footnote{For the self-consistent functional relations between the proper
self-energy and the grand-canonical potential, see, e.g., Refs.~\cite{Abrikosov,Luttinger:1960ua,Baym:1962sx}.}
In particular, we find the simple relation for the proper frequency-space self-energy $\Sigma_\bs{k}(z)$ given by Eq.~\eqref{dynquasi4}.

\subsubsection{Matsubara self-energy}

The improper Matsubara self-energy $\Xi^\star_{\bs{k}}(z_l)$ is defined by~\cite{Luttinger:1960ua}
\begin{align} \label{Xistar0}
\mathscr{G}_{\bs{k}}(z_l)
&=
g_{\bs{k}}(z_l)
+
g_{\bs{k}}(z_l)\,
\Xi^\star_{\bs{k}}(z_l)\,
g_{\bs{k}}(z_l),
\end{align}
i.e., 
the perturbative contributions to $\Xi^\star_{\bs{k}}(z_l)$ are defined by
\begin{align} \label{Xistar1}
\mathscr{G}_{n,\bs{k}}(z_l)
&=
g_{\bs{k}}(z_l)\,
\Xi^\star_{n,\bs{k}}(z_l)\,
g_{\bs{k}}(z_l).
\end{align}
For example, from Eq.~\eqref{Gmbpt} the second-order irreducible contribution to $\mathscr{G}_{\bs{k}}(\tau)$ is given by
\begin{align} 
\mathscr{G}_{2,\bs{k}}(\tau)
&=
-\frac{1}{2} \sum_{\bs{k}_2,\bs{k}_3,\bs{k}_4}
|\braket{\psi_{\bs{k}}\psi_{\bs{k}_2}| V|\psi_{\bs{k}_3}\psi_{\bs{k}_4}}|^2 
\int \limits_{0}^{\beta} \! d\tau_1 \int \limits_{0}^{\beta} \!d\tau_2
\breq &\quad \times
g_\bs{k}(\tau-\tau_1)g_\bs{k}(\tau_2-0)
g_{\bs{k}_2}(\tau_{21})g_{\bs{k}_3}(\tau_{12}) g_{\bs{k}_5}(\tau_{12}),
\end{align}
with $\tau_{ij}=\tau_i-\tau_j$. 
Inserting 
the Fourier series of the unperturbed propagators $g_{\bs{k}}(\tau)=\beta^{-1}\sum_l g_{\bs{k}}(z_l) e^{-z_l \tau}$ we obtain the expression
\begin{align}  \label{Xistar2}
\mathscr{G}_{2,\bs{k}}(\tau) 
&=
-\frac{1}{2} 
\sum_{\bs{k}_2,\bs{k}_3,\bs{k}_4} \!\!
|\braket{\psi_{\bs{k}}\psi_{\bs{k}_2}| V|\psi_{\bs{k}_3}\psi_{\bs{k}_4}}|^2 \sum_{l,l_2,l_3,l_4}
\breq &\quad \times
\frac{1}{\beta^4} \e^{-z_{l}\,\tau} 
\int \limits_{0}^{\beta} \! d\tau_1
\e^{-(z_{l_3}+z_{l_4}-z_{l_2}-z_{l})\,\tau_1} 
\breq &\quad \times
\left[g_{\bs{k}}(z_{l}) \right]^2\,g_{\bs{k}_2}(z_{l_2})g_{\bs{k}_3}(z_{l_3}) g_{\bs{k}_4}(z_{l_4}),
\end{align}
where we have eliminated the $\tau_2$ integral and one Matsubara sum via the relation
\begin{align} \label{kroneck}
\frac{1}{\beta}\int \limits_{0}^{\beta} \! d\tau_2 \e^{\pm(z_{l_3}+z_{l_4}-z_{l_2}-z_{l'})\,\tau_2}=\delta_{l_3+l_4,l_2+l'}.
\end{align}
From $\mathscr{G}_{2,\bs{k}}(\tau)=\beta^{-1}\sum_l \mathscr{G}_{2,\bs{k}}(z_l) e^{-z_l \tau}$ 
and Eq.~\eqref{Xistar1}
we then find that
\begin{align} \label{Xistar3}
\Xi_{2,\bs{k}}[g_\bs{k}(z_l),z_l]
&=
-\frac{1}{2} 
\sum_{\bs{k}_2,\bs{k}_3,\bs{k}_4} \!\!
|\braket{\psi_{\bs{k}}\psi_{\bs{k}_2}| V|\psi_{\bs{k}_3}\psi_{\bs{k}_4}}|^2 \;
\sum_{l_2,l_3,l_4}
\breq &\quad \times
\frac{1}{\beta^3}\int \limits_{0}^\beta \! d\tau \e^{-(z_{l_3}+z_{l_4}-z_{l_2}-z_{l})\,\tau} 
\breq &\quad \times
g_{\bs{k}_2}(z_{l_2})g_{\bs{k}_3}(z_{l_3}) g_{\bs{k}_4}(z_{l_3}),
\end{align}
i.e.,
\begin{align} \label{Xistar4}
\Xi_{2,\bs{k}}[g_\bs{k}(\tau),z_l]
&=
-\frac{1}{2} 
\sum_{\bs{k}_2,\bs{k}_3,\bs{k}_4} \!\!
|\braket{\psi_{\bs{k}}\psi_{\bs{k}_2}| V|\psi_{\bs{k}_3}\psi_{\bs{k}_4}}|^2 \;
\breq &\quad \times
\int \limits_{0}^\beta \! d\tau \e^{z_l\,\tau} 
g_{\bs{k}_2}(\tau)g_{\bs{k}_3}(\tau) g_{\bs{k}_4}(\tau).
\end{align}
Since $\tau>0$ in the time integral, from Eq.~\eqref{smallg} we have 
\begin{align} \label{Xi2calc}
\Xi_{2,\bs{k}}[n_\bs{k},z_l]
&=
-\frac{1}{2} 
\sum_{\bs{k}_2,\bs{k}_3,\bs{k}_4}
|\braket{\psi_{\bs{k}}\psi_{\bs{k}_2}| V|\psi_{\bs{k}_3}\psi_{\bs{k}_4}}|^2 
\breq &\quad \times 
\int \limits_{0}^{\beta} \! d\tau \e^{-(\en_{\bs{k}_3}+\en_{\bs{k}_4}-\en_{\bs{k}_2}-z_l)\,\tau} n_{\bs{k}_2}\bar n_{\bs{k}_3}\bar n_{\bs{k}_4},
\end{align}
and carrying out the time integral we get Eq.~\eqref{sigma2}.

\subsubsection{Functional relations}

The functional relations between the perturbative contributions to the improper Matsubara self-energy  
and the grand-canonical potential are given by (see, e.g., Ref.~\cite{Luttinger:1960ua})
\begin{align} \label{funcrel1a}
\Omega^{\aleph}_{n}[g_{\bs{k}}(z_l)] &= \frac{1}{2n\beta} \sum_\bs{k}  \sum_{l}
g_{\bs{k}}(z_l)\,
\Xi^\star_{n,\bs{k}}[g_{\bs{k}}(z_l),z_l],
\\ \label{funcrel1b}
\Xi^\star_{n,\bs{k}}[g_{\bs{k}}(z_l),z_l]
&=
\beta\frac{\delta \Omega^{\aleph}_{n}[g_{\bs{k}}(z_l)]}{\delta [g_{\bs{k}}(z_l)]},
\end{align}
and similar for $\Omega^{\aleph}_{n}[g_{\bs{k}}(\tau)]$ and $\Xi^\star_{n,\bs{k}}[g_{\bs{k}}(\tau),\tau]$.
The question is, what does $\aleph$ correspond to?

To find this out, we first evaluate the 
expression obtained from Eq.~\eqref{OmegaT} for the second-order normal contribution, i.e.,
\begin{align} \label{O2calc1}
\Omega^{\text{direct}(P)}_{2,\text{normal}}[g_{\bs{k}}(\tau)]
&=
\frac{1}{8 \beta} 
\sum_{\bs{k}_1,\bs{k}_2,\bs{k}_3,\bs{k}_4} \!\!
|\braket{\psi_{\bs{k}}\psi_{\bs{k}_2}| V|\psi_{\bs{k}_3}\psi_{\bs{k}_4}}|^2
W^{\text{direct}(P)}_{\bs{k}_1,\bs{k}_2,\bs{k}_3,\bs{k}_4},
\end{align}
where 
\begin{align} \label{Wcalc1}
W^{\text{direct}(P)}_{\bs{k}_1,\bs{k}_2,\bs{k}_3,\bs{k}_4}
&=
\int \limits^{\beta}_0 \!\!d\tau_1 \int \limits^{\beta}_0 \!\! d\tau_2 \,
g_{\bs{k}_1}(\tau_{21})g_{\bs{k}_2}(\tau_{21})g_{\bs{k}_3}(\tau_{12}) g_{\bs{k}_4}(\tau_{12})
\breq
&=
\int \limits^{\beta}_0 \!\!d\tau_1 \!\!\!\! \int \limits^{\beta-\tau_1}_{-\tau_1} \!\! d\tau' \,
g_{\bs{k}_1}(\tau')g_{\bs{k}_2}(\tau')g_{\bs{k}_3}(-\tau') g_{\bs{k}_4}(-\tau')
\breq
&=
\int \limits^{\beta}_0 \!\!d\tau_1 \!\!\ \int \limits^{0}_{-\tau_1} \!\! d\tau' \,
\bar n_{\bs{k}_1}\bar n_{\bs{k}_2}n_{\bs{k}_3} n_{\bs{k}_4}
\e^{-D\tau'}
\breq
& \quad +
\int \limits^{\beta}_0 \!\!d\tau_1 \!\! \int \limits^{\beta-\tau_1}_{0} \!\!\!\! d\tau' \,
n_{\bs{k}_1}n_{\bs{k}_2}\bar n_{\bs{k}_3} \bar n_{\bs{k}_4}
\e^{-D\tau'}
\breq
&=
\bar n_{\bs{k}_1}\bar n_{\bs{k}_2}n_{\bs{k}_3} n_{\bs{k}_4} \frac{\beta D-1+\e^{\beta D}}{D^2}
\breq
& \quad -
n_{\bs{k}_1}n_{\bs{k}_2}\bar n_{\bs{k}_3} \bar n_{\bs{k}_4} \frac{\beta D-1+\e^{-\beta D}}{D^2},
\end{align}
with $D=\en_{\bs{k}_3}+\en_{\bs{k}_4} - \en_{\bs{k}_2} - \en_{\bs{k}_1}$.\footnote{It can be seen by 
regularizing the energy denominators that 
the expression obtained from Eqs.~\eqref{O2calc1} and \eqref{Wcalc1} is equivalent to the direct, cyclic, and regularized reduced expressions for the
(permutation invariant) second-order normal diagram; see Sec.~\ref{sec31}.}
Now, we can evaluate $W_{\bs{k}_1,\bs{k}_2,\bs{k}_3,\bs{k}_4}$ also
by inserting 
in the first expression in Eq.~\eqref{Wcalc1}
the Fourier expansion of the unperturbed propagators. 
This leads to
\begin{align} \label{Wcalc2}
W^{\aleph}_{\bs{k}_1,\bs{k}_2,\bs{k}_3,\bs{k}_4}
&=
\frac{1}{\beta^3} 
\sum_{l_1,l_2,l_3,l_4} \delta_{l_3+l_4,l_2+l_1}
g_{\bs{k}_1}(z_{l_1})
g_{\bs{k}_2}(z_{l_2})g_{\bs{k}_3}(z_{l_3}) g_{\bs{k}_4}(z_{l_3}).
\end{align}
Note that this is the expression we get by substituting $\Xi^\star_{2,\bs{k}}[g_{\bs{k}}(z_l),z_l]$ 
into Eq.~\eqref{funcrel1a}.
Using Eq.~\eqref{kroneck} we find
\begin{align} \label{Wcalc3}
W^{\aleph}_{\bs{k}_1,\bs{k}_2,\bs{k}_3,\bs{k}_4}
&=
\beta
\int \limits_{0}^{\beta} \! d\tau \e^{-(\en_{\bs{k}_3}+\en_{\bs{k}_4}-\en_{\bs{k}_2}-\en_{\bs{k}_1})\,\tau} 
n_{\bs{k}_1} n_{\bs{k}_2}\bar n_{\bs{k}_3}\bar n_{\bs{k}_4}
\breq
&= \beta n_{\bs{k}_1} n_{\bs{k}_2}\bar n_{\bs{k}_3}\bar n_{\bs{k}_4} \frac{\e^{-\beta D}-1}{D},
\end{align}
which corresponds to the \emph{cyclic} expression, Eq.~\eqref{Omega2}.
However, we could have easily evaluated Eq.~\eqref{Wcalc2} such 
that the expression given by Eq.~\eqref{Wcalc1} would be obtained (i.e., by reversing the step that lead to Eq.~\eqref{Wcalc2}).
Thus, the $\aleph$ in Eq.~\eqref{funcrel2c} depends on how the Matsubara sums are carried out.
The identification of $\aleph$ with cyclic can however be fixed (formally) by substituting 
$\Xi^\star_{n,\bs{k}}[n_{\bs{k}},z_l]$
for 
$\Xi^\star_{n,\bs{k}}[g_{\bs{k}}(z_l),z_l]$, i.e.,
\begin{align}\label{funcrel1c}
\Omega^\text{cyclic}_{n}[g_{\bs{k}}(z_l)] &= \frac{1}{2n\beta} \sum_\bs{k}  \sum_{l}
g_{\bs{k}}(z_l)\,
\Xi^\star_{n,\bs{k}}[n_{\bs{k}},z_l].
\end{align}
Note that no functional derivative relation 
is available for $\Xi^\star_{n,\bs{k}}[n_\bs{k},z_l]$.

Now, from Eqs.~\eqref{funcrel1a} and \eqref{funcrel1b}, we obtain by analytic continuation the relations
\begin{align}\label{funcrel2a}
\Omega_{n}^\aleph[g_{\bs{k}}(z)] &= \frac{1}{2n} \sum_\bs{k} 
\oint\limits_{C_0} 
\frac{d z}{2\pi \im} \,
g_{\bs{k}}(z)\,n_{\bs{k}}(z)\,
\Xi^\star_{n,\bs{k}}[g_{\bs{k}}(z),z],
\\ \label{funcrel2b}
\Xi^\star_{n,\bs{k}}[g_{\bs{k}}(z')] 
&=
\frac{\delta \Omega_{n}^\aleph[g_{\bs{k}}(z)]}{\delta[ g_{\bs{k}}(z')]},
\end{align}
where $C_0 \in \{C_0[l_\text{max}],C_0^{\pm}\}$, with $C_0[l_\text{max}]$ from Fig.~\ref{figc1} and $C_0^{\pm}$ from Fig.~\ref{figc2}.
Note that these relations require that 
$\Xi^\star_{n,\bs{k}}$ is represented as a functional of $g_{\bs{k}}(z)$.
Replacing $\Xi^\star_{n,\bs{k}}[g_{\bs{k}}(z),z]$ 
by the mass function $\mathcal{M}^\star_{n,\bs{k}}[n_\bs{k},z]$ 
leads to
\begin{align}\label{funcrel2c}
\Omega_{n}^\text{cyclic}[n_\bs{k}] &= \frac{1}{2n} \sum_\bs{k} 
\oint\limits_{C_0[l_\text{max}]} \!\!\!\!\!
\frac{d z}{2\pi \im} \,
g_{\bs{k}}(z)\,n_{\bs{k}}(z)\,
\mathcal{M}^\star_{n,\bs{k}}[n_\bs{k},z].
\end{align}
Because $g_{\bs{k}}(z)\,n_{\bs{k}}(z)\,
\mathcal{M}^\star_{n,\bs{k}}[n_\bs{k},z]$ vanishes at infinity we can
deform the contour $C_0[l_\text{max}]$ into the contour $C_1$ from Fig.~\ref{figc1}. 
Since $\mathcal{M}^\star_{n,\bs{k}}(z)$ is entire, we get only the contributions from the pole at $1/g_{\bs{k}}(z)=0$, i.e., 
at $z=\en_\bs{k}$, so
\begin{align}\label{funcrel2d}
\Omega_{n}^\text{cyclic}[n_\bs{k}] &= \frac{1}{2n} \sum_\bs{k} n_{\bs{k}}
\mathcal{M}^\star_{n,\bs{k}}[n_\bs{k},\en_\bs{k}].
\end{align}
Finally, as discussed above, from the expressions for $\mathcal{M}^\star_{n,\bs{k}}[n_\bs{k},z]$
the ones for the perturbative contributions to the frequency-space self-energy $\Sigma^\star_{n,\bs{k}}[n_\bs{k},z]$ are obtained 
by substituting $\e^{\beta(z_l-\mu)}=-1$ and removing the remaining
energy denominator exponentials via Eq.~\eqref{Dexp}. 
From this, we find (analogous to zero-temperature MBPT~\cite{Kaiser:2001ra,Kaiser:2013bua}) that
\begin{align}\label{sigmastarred}
\Sigma^\star_{n,\bs{k}}[n_\bs{k},z] =  
\frac{\delta\Omega_n^{\text{reduced}}[n_\bs{k}]}{\delta n_\bs{k}}\bigg|_{\en_\bs{k}= z},
\end{align}
which implies the relation for the proper 
frequency-space self-energy $\Sigma_{n,\bs{k}}(z)$ given by Eq.~\eqref{dynquasi4}.

\bibliographystyle{apsrev4-1}		
\bibliography{refs}		

\end {document}